\documentclass[reqno,12pt]{amsart}

\usepackage{amsfonts}%
\usepackage{amsmath}%

\usepackage{amsthm}
\usepackage{mathrsfs}
\usepackage{epsfig}
\usepackage{graphicx}

\def\baselskip{19pt}
\newcommand{\DATUM}{September 16, 2007}               %                        %
\def\cHs{\mathring{\cH}^{out/in}}
\def\Lambdas{\kappa}

\def\HPs{H_{\vP}^{\sigma}}
\def\HPst{H_{\vP}^{\sigma_t}}

\def\Hw{K}
\def\HPsw{\Hw_{\vP}^{\sigma}}
\def\HPstw{\Hw_{\vP}^{\sigma_t}}

\def\Ksp{\cR}

\def\rf{\rho^-}

\def\sigmat{{\sigma_t}}
\newcommand{\change}%     % Puts Hash Mark # on the margin to indicate a %
{{\marginpar{\#}}}        % change or mistake in the manuscript.         %
\newcommand{\comma}{\: ,}              % Comma and Period to be used to   %
\newcommand{\Om}{\Omega}                %%%%%%%%%%%%%%%%%%%%%%%%%%%%%%%%%%%

\newcommand{\one}{{\bf 1}}
\newcommand{\cA}{{\mathcal{A}}}
\newcommand{\cB}{{\mathcal{B}}}

\newcommand{\cF}{{\mathcal{F}}}

\newcommand{\cH}{{\mathcal{H}}}
\newcommand{\cI}{{\mathcal{I}}}
\newcommand{\cJ}{{\mathcal{J}}}

\newcommand{\cO}{{\mathcal{O}}}         %                      %
\newcommand{\cR}{{\mathcal{R}}}
\newcommand{\cS}{{\mathcal{S}}}

\newcommand{\cW}{{\mathcal{W}}}

\newcommand{\RR}{\mathbb{R}}            %%%%%%%%%%%%%%%%%%%%%%%%%%%
\newcommand{\ZZ}{\mathbb{Z}}            %                         %
\newcommand{\NN}{\mathbb{N}}            % Blackboard Bold Letters %
\newcommand{\CC}{\mathbb{C}}            %                         %
\newcommand{\fh}{\mathfrak{h}}         %%% Gothic Letters %%%%%%%%
\newcommand{\vA}{{\vec{A}}}             %%%%%%%%%%%%%%%%%%%%%%%%%%%%%
\newcommand{\vF}{{\vec{F}}}             % Letters with Arrow on Top %
\newcommand{\vG}{{\vec{G}}}             %                           %
\newcommand{\vnabla}{{\vec{\nabla}}}
\newcommand{\veps}{{\vec{\varepsilon}}}

\newcommand{\vq}{{\vec{q}}}
\newcommand{\vk}{{\vec{k}}}

\newcommand{\vp}{{\vec{p}}}
\newcommand{\vP}{{\vec{P}}}
\newcommand{\vv}{{\vec{v}}}

\newcommand{\vx}{{\vec{x}}}

\newcommand{\vy}{{\vec{y}}}
\newcommand{\cirS}{\mathop{\bigcirc\kern -.73em {\scriptstyle{\rm S}}}}

\newcommand{\at}{{el}}

\def\Bra{\Big\langle}
\def\Ket{\Big\rangle}
\def\bra{\big\langle}
\def\ket{\big\rangle}
\newcommand{\Proof}{\noindent{\em Proof.}}
\newcommand{\QED}{\phantom{blablabla}\hfill\qed\newline}  % End of a Proof %
\renewcommand{\thesection}%            %%%%%%%%%%%%%%%%%%%%%%%%%%%%%%%%%%%%
{\Roman{section}}                      % To begin a section, write \secct %
\renewcommand{\theequation}%           % instead of \section. The section %
{\thesection.\arabic{equation}}        % number will then be displayed    %
\newcommand{\secct}[1]{\section{#1}%   % as a roman number, e.g., IV. for %
\setcounter{equation}{0}}              % the fourth section               %
\newtheorem{hypothesis}{Main Assumption}[section]
\newtheorem{theorem}{Theorem}[section]         %%%%%%%%%%%%%%%%%%%%%%%%%%%%%%
\newtheorem{lemma}[theorem]{Lemma}             % These Theoremlike environm.%
\theoremstyle{plain}
\renewcommand{\theequation}{\thesection.\arabic{equation}}
\newcommand{\resetequ}{\setcounter{equation}{0}}
\begin{document}
\bibliographystyle{plain}
%%%%%%%%%%%%%%%%%%%%%%%%%%%%%%%%%%%%%%%%%%%%%%%%%%%%%%%%%%%%%%%%%%%%%%%%%%%%
%\setcounter{page}{0}
\thispagestyle{empty}

\baselineskip=\baselskip

\title[Infraparticle states in QED I.]{Infraparticle Scattering States in Non-Relativistic QED:
I. The Bloch-Nordsieck Paradigm}

\author[T.Chen]{Thomas Chen}
\address{Department of Mathematics, University of Texas at Austin, Austin, TX 78712, USA}
\email{tc@math.utexas.edu}
\author[J.Fr\"ohlich]{J\"{u}rg Fr\"{o}hlich}
\address{Institut f\"ur Theoretische Physik, ETH H\"{o}nggerberg,
CH-8093 Z\"{u}rich, Switzerland, and IH\'ES, Bures sur Yvette, France}
\email{juerg@itp.phys.ethz.ch}
\author[A. Pizzo]{Alessandro Pizzo}
\address{Department of Mathematics, University of California Davis, Davis, CA 95616, USA}
\email{pizzo@math.ucdavis.edu}

%\date{\DATUM}

\begin{abstract}
  We construct infraparticle scattering states for Compton scattering
  in the standard model of non-relativistic QED.
  In our construction, an infrared cutoff initially introduced to regularize the model
  is removed completely.
  We rigorously establish the properties of
  infraparticle scattering theory
  predicted in the classic  work of Bloch and Nordsieck from the 1930's,
  Faddeev and Kulish, and others.
  Our results represent a basic step towards solving the infrared problem
  in (non-relativistic) QED.
\end{abstract}

\maketitle

\section{Introduction}
\label{sec-intro}
\resetequ

The construction of scattering states in Quantum Electrodynamics (QED) is an
old open problem. The main difficulties in solving this problem are linked to the infamous
\emph{infrared catastrophe} in QED: It became clear very early in the
development of QED that,
at the level of perturbation theory (e.g., for Compton scattering),
the transition amplitudes between formal scattering states with a {\em finite} number of photons are
{\em ill-defined}, because, typically, Feynman amplitudes containing
vertex corrections exhibit logarithmic infrared divergences; \cite{PauliFierz,JauchRohrlich}.

A pragmatic approach proposed by Jauch and Rohrlich, \cite{JauchRohrlich1,Rohrlich}, and by
Yennie, Frautschi, and Suura, \cite{YFS},
is to circumvent this difficulty by considering
\emph{inclusive cross sections}: One sums over all possible final states
that include photons whose total energy lies below an arbitrary threshold energy $\epsilon>0$.
Then the infrared divergences due to soft virtual photons
are \emph{formally} canceled by those corresponding to the emission of \emph{soft} photons of total energy
below $\epsilon$, order by order in perturbation theory
in powers of the finestructure constant $\alpha$.
A drawback of this approach becomes apparent when
one tries to formulate a scattering theory that is $\epsilon$-independent:
Because the transition probability $P^{\epsilon}$ for an inclusive process is estimated to be
$\cO(\epsilon^{const.\alpha})$, the threshold energy $\epsilon$ cannot be allowed to approach zero,
unless "Bremsstrahlungs"processes (emission of photons)
are properly incorporated in the calculation.

An alternative approach to solving the infrared problem
is to go beyond inclusive cross sections and
to define $\alpha$-dependent scattering states containing an infinite number of photons
(so-called \emph{soft-photon clouds}), which are expected to yield
finite transition amplitudes, order by order in perturbation theory.
The works of Chung \cite{Chung}, Kibble \cite{Kibble}, and Faddeev and Kulish \cite{FK},
between  1965 and
1970, represent promising, albeit incomplete progress in this direction. Their approaches are
guided by an \emph{ansatz} identified in the analysis of certain solvable models
introduced in early work by Bloch and Nordsieck,
\cite{BlochNordsieck}, and extended by Pauli and Fierz, \cite{PauliFierz}, in the late 1930's.
In a seminal paper \cite{BlochNordsieck} by Bloch and Nordsieck, it was shown
(under certain approximations that render their model solvable) that,
in the presence of asymptotic charged particles, the scattering representations of
the asymptotic photon field are {\em coherent non-Fock} representation, and that
formal scattering states with a
finite number of soft photons do not belong to the physical Hilbert space of a system
of asymptotically freely moving electrons interacting with the quantized radiation field.
These authors also showed that the coherent states describing the soft-photon cloud are
parameterized by the asymptotic velocities of the electrons.

The perturbative recipes for the construction of scattering states did not remove some of the major
conceptual problems. New puzzles appeared, some of which are related to the
problem that Lorentz boosts cannot be unitarily implemented on charged
scattering states; see \cite{FrMoSt}.
This host of problems was addressed in a fundamental analysis of the structural properties of QED,
and of the \emph{infrared problem} in the framework of \emph{general quantum field theory}; see \cite{WightmanStrocchi}.
Subsequent developments in \emph{axiomatic quantum field theory}
have led to results that are of great importance for the topics treated in the present paper:
\begin{itemize}
\item[i)] Absence of dressed one-electron states with a sharp mass;
see \cite{Schroer}, \cite{Buchholz2}.
\item[ii)] Corrections to the asymptotic dynamics, as compared to the
one in a theory with a positive mass
gap; see
%\cite{FMG},
\cite{Buchholz1}.
\item[iii)] Superselection rules pertaining to the space-like asymptotics of the quantized
electromagnetic field, and connections to Gauss' law; see \cite{Buchholz2}.
\end{itemize}
In the early 1970's, significant advances on the infrared problem were
made for Nelson's model,
which describes non-relativistic matter linearly coupled to a scalar field of massless bosons.
In \cite{Fr73}, \cite{Fr74},  the disappearance of a sharp mass shell
for the charged particles was established for
Nelson's model, in the limit where an infrared cut-off is removed.
(An infrared cutoff is introduced, initially, with the
purpose to eliminate the interactions between charged particles
and soft boson modes).
Techniques developed in \cite{Fr73,Fr74} have become standard tools in
more recent work on non-relativistic QED, and attempts made  in \cite{Fr73,Fr74}
have stimulated a deeper
understanding of the asymptotic dynamics of charged particles and photons.
The analysis of spectral and dynamical aspects of non-relativistic
QED and of Nelson's model constitutes an active branch of contemporary
mathematical physics. In questions relating to the infrared problem,
mathematical control of the removal of the infrared cutoff is a critical issue still
unsolved in many situations.

The construction of an \emph{infraparticle scattering theory} for Nelson's model,
after removal of the infrared cutoff, has recently been achieved in \cite{Pizzo2005}
by introducing a suitable scattering scheme.
This analysis involves spectral results substantially improving those in \cite{Fr74}.
It is based on a new \emph{multiscale technique}
developed in \cite{Pizzo2003}.

While the interaction in Nelson's model is linear in the creation- and
annihilation operators of the boson field, it is non-linear and of vector type in
non-relativistic QED.
For this reason, the methods developed in \cite{Pizzo2003,Pizzo2005} do not
directly apply to the latter. The main goal of the present
work is to construct an infraparticle scattering theory for non-relativistic
QED inspired by the methods of \cite{Pizzo2003,Pizzo2005}.
In a companion paper, \cite{ChFrPi2}, we derive those spectral properties of QED that
are crucial for our analysis of scattering theory and determine the mass shell structure
in the infrared limit. We will follow ideas developed in \cite{Pizzo2003}.
Bogoliubov transformations, proven in  \cite{ChFr} to characterize the soft photon clouds
in non-relativistic QED, represent an important element in our construction.
The proof in  \cite{ChFr} uses the uniform bounds on the renormalized electron mass
previously established in \cite{Chen}.

We present a detailed definition of the
model of non-relativistic QED in Section {\ref{sec-I.2}}.
Aspects of \emph{infraparticle scattering theory},  developed in this
paper, are described in Section {\ref{sec-II}}.

To understand why free radiation parametrized by the asymptotic velocities of
the charged particles must be expected to be present in \emph{all} the scattering states,
we recall a useful point of view based on \emph{classical} electrodynamics that was brought to our attention by Morchio and Strocchi.
%a mathematically precise treatment is given in Appendix \ref{app-B}
%where we refer for additional details and explanations. 

We consider a single, classical charged point-particle, e.g.,  an electron, moving along a world line
$(t,\vx(t))$ in Minkowski space, with $\vx(0)=\vec0$. We suppose that, for $t\leq0$, it moves at a constant
velocity $\vv_{in}$, and, for $t>\bar t>0$, at a constant velocity $\vv_{out}\neq\vv_{in}$, $|\vv_{out}|, |\vv_{in}|<c$ (the speed of light).
Thus, 
\begin{equation}\label{eq-introduction-1}
\vx(t)=\vv_{in}\cdot t, \quad \text{for}\,\, t\leq 0\,,
\end{equation}
and
\begin{equation}\label{eq-introduction-2}
\vx(t)=\vx_{*}+\vv_{out}\cdot t, \quad \text{for}\,\, t\geq \bar{t}\,,
\end{equation}
for some $\vx_{*}$.

\noindent
For times $t\in [0,\bar{t}]$, the particle performs an accelerated motion.  We propose to analyze the behavior of the electromagnetic field in the vicinity of the particle and the properties of the free electromagnetic radiation at very early times ($t\to -\infty$,``in") and very late times ($t\to +\infty$, ``out"). For this purpose, we  must solve Maxwell's equations for the electromagnetic field tensor, $F^{\mu\nu}(t,\vy)$, given the $4$-current density corresponding to the trajectory of the particle; (back reaction of the electromagnetic field on the motion of the charged particle is neglected):
\begin{equation}\label{eq-introduction-3}
\partial_{\mu}F^{\mu\nu}(t,\vy)=J^{\nu}(t,\vy)
\end{equation}
with
\begin{equation} 
	J^{\nu}(t,\vy) \, := \, -q \, ( \,    \, \delta^{(3)}(\vy- \vx(t)) \,,\,
	  \dot{\vx}(t) \, \delta^{(3)}(\vy- \vx (t)) \, )\,,
\end{equation}
where, in the units used in our paper, $q=2(2\pi)^3\alpha^{1/2}$ ($\alpha$ is the finestructure constant). We solve
equation (\ref{eq-introduction-3}) with prescribed spatial asymptotics
($|\vy|\to\infty$): Let $F^{\mu\nu}_{[\vv_{L.W.}]}(t,\vy)$ be a solution of
(\ref{eq-introduction-3}) that, to leading order in $|\vy|^{-1}$ ($|\vy|\to \infty$), approaches the
Li\'enard-Wiechert  field tensor for a point-particle with charge $-q$ and a \emph{constant} velocity
$\vv_{L.W.}$ at \emph{all} times.  
%Due to the uniform motion of the charge, this is a non-radiative
%solution (i.e., it does not lead to any energy dissipation for the electron) 
%which is easily obtained by computing the electrostatic field generated by
%a stationary charge, and by subsequently boosting this solution with a Poincar\'e transformation
%parametrized by $\vv_{in}$. By $F^{\mu,\nu}_{(\vx,\dot{\vx})}(t,\vy)$ we denote the Lienard-Wiechert field tensor for a particle that, at time $t$, has position $\vx(t)$ and velocity $\dot{\vx}$.
%We stress that it is always possible to choose such a solution for any
%velocity $\vv_{L.W.}$, because, by causality, the spatial asymptotics of the
%solutions is independent of the
%dynamics of the charge. 
Let us denote the Li\'enard-Wiechert  field tensor of a point-particle with charge $-q$ moving along a trajectory $(t, \vx(t))$ in Minkowski space with $\vx(0)=:\vx$ and $\dot{\vx}(t)\equiv \vv$, for \emph{all} $t$, by $F^{\mu\nu}_{\vx,\vv}(t,\vy)$. Apparently, we are looking for solutions,  $F^{\mu\nu}_{[\vv_{L.W.}]}(t,\vy)$, of (\ref{eq-introduction-3}) with the property that, for all times $t$, 
\begin{equation}\label{eq-introduction-5}
 |F^{\mu\nu}_{[\vv_{L.W.}]}(t,\vy)-F^{\mu\nu}_{\vx, \vv_{L.W.}}(t,\vy)|=o(|\vy|^{-2})\,,
\end{equation} 
as $|\vy|\to \infty$, for any $\vx$. This class of solutions of (\ref{eq-introduction-3}) is denoted by $\mathscr{C}_{\vv_{L.W.}}$. It is important to observe that, by causality, the class $\mathscr{C}_{\vv_{L.W.}}$ is \emph{non-empty}, for any $\vv_{L.W.}$, with $|\vv_{L.W.}|<c$. This can be seen by choosing Cauchy data for the solution of  (\ref{eq-introduction-3}) satisfying  (\ref{eq-introduction-5}) at some time $t_0$, e.g.,  $t_0=0$.

\noindent
Let us now consider a specific solution, $F^{\mu\nu}_{[\vv_{L.W.}]}(t,\vy)$, of Eq. (\ref{eq-introduction-3}) in the class $\mathscr{C}_{\vv_{L.W.}}$. We are interested in the behavior of this solution at very early times ($t\ll 0$). We expect that, for $|\vy-\vx(t)|=o(|t|)$,
\begin{equation}\label{eq-introduction-6}
F^{\mu\nu}_{[\vv_{L.W.}]}(t,\vy)\simeq F^{\mu\nu}_{\vec{0},\vv_{in}}(t,\vy)
\end{equation}
(here the symbol $\simeq $ means: up to a solution of the homogeneous Maxwell equation decaying at least like $\frac{1}{t^2}$). 

\noindent
However, for $|\vy-\vx(t)|\to \infty$,
\begin{equation}\label{eq-introduction-7}
F^{\mu\nu}_{[\vv_{L.W.}]}(t,\vy)\simeq F^{\mu\nu}_{\vec{0},\vv_{L.W.}}(t,\vy)\,,
\end{equation}
as quantified in (\ref{eq-introduction-5}).

\noindent
We note that, by (\ref{eq-introduction-1}), $F^{\mu\nu}_{\vec{0},\vv_{in}}(t,\vy)$ solves Eq. (\ref{eq-introduction-3}), for times $t<0$. Thus,
\begin{equation}\label{eq-introduction-8}
\phi^{\mu\nu}_{in}(t,\vy):=F^{\mu\nu}_{[\vv_{L.W.}]}(t,\vy)- F^{\mu\nu}_{\vec{0},\vv_{in}}(t,\vy)\quad\quad t<0
\end{equation}
solves the \emph{homogenous} Maxwell equation, i.e., Eq. (\ref{eq-introduction-3}) with $J^{\nu}\equiv 0$.

\noindent
For $t\gg \bar{t}$, we expect that, for $|\vy-\vx(t)|=o(t)$, 
\begin{equation}\label{eq-introduction-9}
F^{\mu\nu}_{[\vv_{L.W.}]}(t,\vy)\simeq F^{\mu\nu}_{\vec{x}_{*},\vv_{out}}(t,\vy)\,
\end{equation}
(here the symbol $\simeq $ means: up to a solution of the \emph{homogeneous} Maxwell equation decaying at least like $\frac{1}{t^2}$). 
But, for $|\vy-\vx(t)|\to \infty$, 
\begin{equation}\label{eq-introduction-10}
F^{\mu\nu}_{[\vv_{L.W.}]}(t,\vy)\simeq F^{\mu\nu}_{\vec{0},\vv_{L.W.}}(t,\vy)\,,
\end{equation}
as quantified in  (\ref{eq-introduction-5}). We note that, by  (\ref{eq-introduction-2}), $F^{\mu\nu}_{\vec{x}_{*},\vv_{out}}(t,\vy)$ solves Eq. (\ref{eq-introduction-3}), for times $t>\bar{t}$. Thus, 
\begin{equation}\label{eq-introduction-11}
\phi^{\mu\nu}_{out}(t,\vy):=F^{\mu\nu}_{[\vv_{L.W.}]}(t,\vy)- F^{\mu\nu}_{\vec{x}_{*},\vv_{out}}(t,\vy)\quad\quad t>\bar{t}
\end{equation}
solves the homogenous Maxwell equation.

\noindent
Next, we recall that $\phi^{\mu\nu}_{as}(t,\vy)$, with $as=in/out$, can be derived from an electromagnetic vector potential, $A_{as}^{\mu}$, by 
\begin{equation}
\phi^{\mu\nu}_{as}(t,\vy)=\partial ^{\mu}A^{\nu}_{as}(t,\vy)-\partial ^{\nu}A^{\mu}_{as}(t,\vy)\,.
\end{equation}
We can impose the \emph{Coulomb gauge condition} on $A_{as}^{\mu}$: $A_{as}^{\mu}=(0, \vA_{as}(t,\vy))$, with $\vnabla \cdot \vec{A}_{as}(t,\vy)\equiv 0$. It turns out (and this can be derived from formulae one finds, e.g., in \cite{Jackson}), that, to leading order in $|\vy|^{-1}$ ($|\vy|\to \infty$),  $\vA_{as}(t,\vy)$ is given by
\begin{eqnarray}\label{eq-introduction-13}
\vA_{as}(t,\vy)&:=& \; \;  \alpha^{\frac{1}{2}} \sum_\lambda \int \frac{d^3k}{\sqrt{ 
|\vk| \,}}\Big\{ \, \frac{\vv_{as}\cdot\veps^{\;*}_{\vk, \lambda}}{|\vk|^{\frac{3}{2}}
(1-\vv_{as}\cdot\hat{k})}\veps_{\vk, \lambda} e^{-i\vk \cdot \vy+i|\vk|t} \, + \,
c.c. \, \Big\}\\
&&- \, \alpha^{\frac{1}{2}} \sum_\lambda \int \frac{d^3k}{\sqrt{ 
|\vk| \,}} \Big\{ \, \frac{\vv_{L.W.}\cdot\veps^{\;*}_{\vk, \lambda}}{|\vk|^{\frac{3}{2}}
(1-\vv_{L.W.}\cdot\hat{k})} \veps_{\vk, \lambda}e^{-i\vk \cdot \vy+i|\vk|t} \, + \,
c.c. \, \Big\}\,.\nonumber
\end{eqnarray}
where we have set $c=1$, $\hat{k}$ is the unit vector in the direction of $\vk$, and $\veps_{\vk, 1}$, $\veps_{\vk, 2}$ are transverse polarization vectors with $\hat{k} \cdot \veps_{\vk, \lambda}=0$, $\lambda=1,2$, and $\veps_{\vk, \lambda}^{*}\cdot \veps_{\vk, \lambda'}=\delta_{\lambda,\lambda'}$.

\noindent
The free field 
\begin{equation}
\phi^{\mu\nu}(t,\vy)=\phi^{\mu\nu}_{out}(t,\vy)-\phi^{\mu\nu}_{in}(t,\vy)
\end{equation}
is the radiation emitted by the particle due to its accelerated motion, as $t\to \infty$. It is well known that Eqs. (\ref{eq-introduction-6})-( \ref{eq-introduction-13}) can be made precise within classical electrodynamics under some standard assumptions on the Cauchy data for the solutions in addition to  the condition in Eq. (\ref{eq-introduction-5}). We will see that analogous statements also hold in our model of quantum electrodynamics with non-relativistic matter.

In this paper, we treat the quantum theory of a system consisting of a nonrelativistic charged particle only interacting with the quantized e.m. field. The motion of the quantum particle depends on the back-reaction of the field, and the asymptotic  
$in$- and $out$-velocities of this particle are not attained at finite times. However, the infrared features of the asymptotic radiation in the classical model, described above  for a given current,  are reproduced in this interacting quantum model.

In fact, the classes $\mathscr{C}_{\vv_{L.W.}}$ correspond to superselection sectors of the quantized theory;
see  e.g. [3].
In particular, the Fock representation,
which is the usual (but not the only possible) choice for the representation of the
algebra of photon creation- and  annihilation operators,
corresponds to $\vv_{L.W.}=0$.
This implies that, in the Fock representation,
an asymptotic background radiation must be expected for all values
$\vv\neq0$ of the asymptotic velocity of the electron.
In particular, after replacing the classical velocity $\vv$ with
the spectral values of the quantum operators $\vv_{out/in}$
(where $\vv_{out/in}$ is the asymptotic velocity 
of an  outgoing or incoming asymptotic electron,
respectively), the background field with $\vv_{L.W.}=0$ (given by  (\ref{eq-introduction-13})),
corresponds to the background radiation described by the
coherent (non-Fock) representations of the asymptotic photon algebra  labeled by $\vv_{out/in}$; see also Section {\ref{ssec-scatt-ssp-1}}.
\\
\\

%\newpage

\section{Definition of the model }
\label{sec-I.2}
\resetequ

The Hilbert space of pure state vectors of the system
consisting of one non-relativistic electron interacting with the quantized
electromagnetic field is given by
\begin{equation} \label{eq-I-1}
\cH \; := \; \cH_\at \, \otimes \, \cF  \, ,
\end{equation}
where $\cH_\at = L^2(\RR^3)$ is the Hilbert space for a
single electron; (for expository convenience, we neglect
the spin of the electron).
The Fock space used to describe the states of the transverse
modes of the quantized electromagnetic field (the \emph{photons}) in
the Coulomb gauge is given by
\begin{equation} \label{eq-I-2}
\cF \ := \ \bigoplus_{N=0}^\infty \cF^{(N)} \comma \hspace{6mm}
\cF^{(0)} = \CC \, \Om \comma
\end{equation}
where $\Om$ is the vacuum vector (the state of the electromagnetic
field without any excited modes), and
\begin{equation} \label{eq-I-3}
\cF^{(N)} \ := \ \cS_N \, \bigotimes_{j=1}^N \fh \comma \hspace{6mm}
N \geq 1 \comma
\end{equation}
where the Hilbert space $\fh$ of a single photon is
\begin{equation} \label{eq-I-4}
\fh \ := \ L^2( \RR^3 \times \ZZ_2 ) \,.
\end{equation}
Here, $\RR^3$ is momentum space, and $\ZZ_2$ accounts for the
two independent transverse polarizations (or helicities) of a
photon.  In (\ref{eq-I-3}), $\cS_N$ denotes the orthogonal
projection onto the subspace of $\bigotimes_{j=1}^N \fh$ of totally
symmetric $N$-photon wave functions, to account for the fact
that photons satisfy Bose-Einstein statistics. Thus, $\cF^{(N)}$ is
the subspace of $\cF$ of state vectors for configurations of exactly
$N$ photons.  
In this paper, we use units such that Planck's constant $\hbar$,
the speed of light $c$, and the
mass of the electron are equal to unity.
The dynamics of the system is generated by the Hamiltonian
\begin{equation} \label{eq-I-6}
H \; := \; \frac{\big(-i\vnabla_{\vx} \, + \, \alpha^{1/2} \vA(\vx)
\, \big)^2}{2} \, + \, H^{f}\,.
\end{equation}
The multiplication operator $\vx\in\RR^3$ corresponds to the position of
the electron. The electron momentum operator is given by $\vp=-i\vnabla_\vx$;
$\alpha \cong 1/137$ is the finestructure constant (which, in this paper,
is treated as a small parameter), $\vA(\vx)$ denotes the
(ultraviolet regularized) vector
potential of the transverse modes of the quantized electromagnetic
field at the point $\vx$ (the electron position) in the \emph{Coulomb gauge},
\begin{equation} \label{eq-I-7}
\vnabla_\vx \cdot \vA (\vx) \ = \ 0 \, .
\end{equation}

$H^f$ is the Hamiltonian of the quantized, free
electromagnetic field, given by
\begin{equation} \label{eq-I-10}
    H^f \; := \;  \sum_{\lambda = \pm} \int d^3k \; |\vk| \,
    a^*_{\vk,\lambda} \,  a_{\vk, \lambda} \comma
\end{equation}
where $a^*_{\vk, \lambda}$ and $a_{\vk, \lambda}$ are the usual
photon creation- and annihilation operators, which satisfy the
canonical commutation relations
\begin{eqnarray}
\label{eq-I-12}
    [a_{\vk, \lambda} \, , \, a^*_{\vk', \lambda'}] & = &
    \delta_{\lambda \lambda'} \, \delta (\vk - \vk') \comma
    \\
    \label{eq-I-11}
    [a^\#_{\vk, \lambda} \, , \, a^\#_{\vk', \lambda'}] & = & 0 \, ,
\end{eqnarray}
with $a^\#=a$ or $a^*$. The vacuum vector $\Om$ obeys the condition
\begin{equation}
    a_{\vk, \lambda} \, \Om \; = \; 0 \comma  \label{eq-I-13}
\end{equation}
for all $\vk \in \RR^3$ and $\lambda\in \ZZ_2
\equiv \{+,-\}$.

The quantized electromagnetic vector potential is given by
\begin{eqnarray} \label{eq-I-14}
    \vA(\vy) \; := \;
    \sum_{\lambda = \pm} \int_{\cB_{\Lambda}} \frac{d^3k}{\sqrt{  |\vk| \,}}\,
    \big\{ \veps_{\vk, \lambda} e^{-i\vk \cdot \vy}
    a^*_{\vk, \lambda} \, + \, \veps_{\vk, \lambda}^{\;*}
    e^{i\vk \cdot \vy} a_{\vk, \lambda} \big\} \comma
\end{eqnarray}
where $\veps_{\vk, -}$, $\veps_{\vk, +}$ are photon polarization
vectors, i.e., two unit vectors in $\RR^3 \otimes\CC$ satisfying
\begin{equation} \label{eq-I-15}
    \veps_{\vk, \lambda}^{\;*} \cdot \veps_{\vk, \mu} \; = \;
    \delta_{\lambda \mu} \comma \hspace{8mm} \vk \cdot \veps_{\vk,
    \lambda} \; = \; 0 \comma
\end{equation}
for $\lambda, \mu = \pm$. The equation $\vk \cdot \veps_{\vk,
\lambda} = 0$ expresses the Coulomb gauge condition. Moreover,
$\cB_{\Lambda}$ is a ball of radius $\Lambda$ centered at the origin
in momentum space; $\Lambda$ represents an \emph{ultraviolet cutoff}
that will be kept fixed throughout our analysis. The vector
potential defined in (\ref{eq-I-14}) is thus regularized in the
ultraviolet.

Throughout this paper, it will be assumed that $\Lambda\approx 1$
(the rest energy of an electron), and that $\alpha$ is sufficiently small.
Under these assumptions, the Hamitonian $H$ is selfadjoint on $D(H_0)$,
the domain of definition of the operator
\begin{equation}
    H_0 \; := \; \frac{(-i\vnabla_{\vx})^2}{2} \, + \, H^f \;.
\end{equation}
The perturbation $H-H_0$ is small in the sense of Kato.

The operator measuring the total momentum of a state of the system consisting of the
electron and the electromagnetic field is given by
\begin{equation}
    \vP\,:=\,\vp+\vP^f \, ,
\end{equation}
where $\vp=-i\vnabla_{\vx}$ is the momentum operator for the electron, and
\begin{equation}
    \vP^f\,:=\, \sum_{\lambda = \pm} \int  d^3k \; \vk \,
    a^*_{\vk, \lambda} \, a_{\vk, \lambda}
\end{equation}
is the momentum operator for the radiation field.

The operators $H$ and $\vP$ are essentially selfadjoint on the
domain $D(H_0)$, and since the dynamics is invariant under
translations, they commute: $[H,\vP]= 0$.
The Hilbert space $\cH$ can be decomposed on the joint
spectrum, $\RR^3$,  of the component-operators of $\vP$.
Their spectral measure is
absolutely continuous with respect to Lebesgue measure,
\begin{equation}
\cH\,:=\,\int^{\oplus}\cH_{\vP} \, d^3P \, ,
\end{equation}
where each fiber space $\cH_{\vP}$ is a copy of Fock space $\cF$.
\\

\noindent {\bf{Remark:}} \emph{Throughout
this paper, the symbol $\vP$ stands for both a variable in $\RR^3$
and a vector operator in $\cH$, depending on the context.
Similarly, a double meaning is also
associated with functions  of the total
momentum operator.
(E.g.: In Eq. (\ref{eq-I-49}) $E_{\vP}$ is an operator on the Hilbert space $\cH$, while in Eq. (\ref{eq-III.3}) is a function of $\vP\in\RR^3$.)}
\\

To each fiber space $\cH_{\vP}$ there corresponds an isomorphism
\begin{equation}
    I_{\vP}\,:\,\cH_{\vP}\,\longrightarrow\,\cF^{b}\,,
\end{equation}
where $\cF^{b}$ is the Fock space corresponding to the annihilation- and
creation operators $b_{\vk,\lambda}$, $b^*_{\vk,\lambda}$,
where
$b_{\vk,\lambda}$ is given by $e^{i\vk\cdot\vx}  a_{\vk,\lambda}$, and
$b_{\vk,\lambda}^*$ by $e^{-i\vk\cdot\vx} a_{\vk,\lambda}^* $,
with vacuum $\Omega_{f}=I_{\vP}(e^{i\vP\cdot\vx} )$,
where $\vx$ is the electron position.
To define $I_{\vP}$ more precisely, we consider an (improper) vector
$\psi_{(f^{(n)};\vP)}\in \cH_{\vP}$
with a definite total momentum, which describes an electron and $n$ photons.  
Its wave function, in the variables
$(\vx;\vk_1,\dots,\vk_n;\lambda_1,\dots,\lambda_n)$, is given by
\begin{equation}
    e^{i(\vP-\vk_1-\cdots-\vk_n)\cdot\vx} 
    f^{(n)}(\vk_1,\lambda_1;\cdots\cdots;\vk_n,\lambda_n) 
\end{equation}
where $f^{(n)}$ is totally symmetric in its $n$ arguments.  
The isomorphism $I_{\vP}$ acts by way of
\begin{eqnarray}
    \lefteqn{I_{\vP}\big( e^{i(\vP-\vk_1-\cdots-\vk_n)\cdot\vx} 
    f^{(n)}(\vk_1,\lambda_1;\cdots\cdots;\vk_n,\lambda_n)=}
    \\
    &= &\frac{1}{\sqrt{n!}}\sum_{\lambda_1,\dots,\lambda_n}\int \, d^3k_1\dots d^3k_n\,
     f^{(n)}(\vk_1,\lambda_1;\cdots\cdots;\vk_n,\lambda_n)\,
    b_{\vk_1,\lambda_1}^* \cdots
    b_{\vk_n,\lambda_n}^*  \, \Omega_f \,.
\end{eqnarray}

The Hamiltonian $H$ maps each fiber space $\cH_{\vP}$ into itself, i.e.,
it can be written as
\begin{equation}
    H\,=\,\int H_{\vP}\,d^3P\,,
\end{equation}
where
\begin{equation}
    H_{\vP}\,:\,\cH_{\vP}\longrightarrow\cH_{\vP}\,.
\end{equation}
Written in terms of the operators $b_{\vk,\lambda}$, $b^*_{\vk,\lambda}$, and of the
variable $\vP$, the fiber Hamiltonian $H_{\vP}$ has the form
\begin{equation}
    H_{\vP} \; := \; \frac{\big(\vP-\vP^f +\alpha^{1/2} \vA \big)^2}{2} \;
    + \;H^{f}\,,
\end{equation}
where
\begin{eqnarray}
    \vP^f & = & \sum_{\lambda}\, \int d^3k \, \vk \, b^*_{\vk,\lambda} \, b_{\vk,\lambda} \, ,
    \\
    H^f & = & \sum_{\lambda}\, \int d^3k \, |\vk| \, b^*_{\vk,\lambda} \, b_{\vk,\lambda} \,,
\end{eqnarray}
and
\begin{equation}
    \vA \, := \,\sum_{\lambda}\, \int_{\cB_{\Lambda}}\,
    \frac{d^3k}{\sqrt{ |\vk| \,}} \, \big\{  b^*_{\vk,\lambda} \veps_{\vk,\lambda}
    \, + \, \veps^{\;*}_{\vk, \lambda}  b_{\vk,\lambda} \big\} \,.
\end{equation}
Let
\begin{equation}\label{eq-cS-def-1}
    \mathcal{S} \, := \, \lbrace\,
    \vP\in\RR^3\,:\,|\vP|<\frac{1}{3}\,\rbrace\,.
\end{equation}
In order to give precise meaning to the constructions used in this work,
we restrict the total momentum $\vP$ to the set ${\mathcal S}$, and we
introduce an infrared cut-off at an energy $\sigma>0$ in the vector potential.
{\em The removal of the infrared cutoff
in the construction of scattering states is the main problem solved in this paper.}
The restriction of  $\vP$ to ${\mathcal S}$ guarantees that the propagation speed of
a dressed electron is strictly smaller than the speed of light. However, our results can be extended to a region ${\mathcal S}$ (inside the
unit ball) of radius larger than $\frac{1}{3}$.

\noindent
We start by studying a regularized fiber Hamiltonian given by
\begin{equation}\label{eq:H-fiber}
    \HPs \, := \, \frac{\big(\vec{P}-\vec{P}^{f}
    +\alpha^{1/2} \vec{A}^{\sigma} \big)^2}{2} \,  + \, H^{f}\,
\end{equation}
acting on the
fiber space $\mathcal{H}_{\vec{P}}$, for $\vec{P}\in \mathcal{S}$,
where
\begin{eqnarray}
    \vec{A}^{\sigma}\, := \,
    \sum_{\lambda}\, \int_{\mathcal{B}_{\Lambda}\setminus \mathcal{B}_{\sigma}}\,
    \frac{d^3k}{\sqrt{  |\vk| \,}} \, \big\{  b^*_{\vk,\lambda} \veps_{\vk,\lambda}\, + \,
    \veps^{\;*}_{\vk, \lambda} b_{\vk,\lambda} \big\} \,,
\end{eqnarray}
and where $\mathcal{B}_{\sigma}$ is a ball of radius $\sigma$. 
\\

\noindent
{\bf Remark:}
In a companion paper \cite{ChFrPi2}, we construct dressed one-electron states of
fixed momentum given by the ground state vectors
$\Psi_{\vP}^{\sigma_j}$ of the Hamiltonians $H_{\vP}^{\sigma_j}$, and
we compare ground state vectors $\Psi_{\vP}^{\sigma_j}$,
$\Psi_{\vP'}^{\sigma_{j'}}$ corresponding to different fiber Hamiltonians
$H_{\vP}^{\sigma_j}$, $H_{\vP'}^{\sigma_{j'}}$ with
$\vP\not=\vP'$. We compare these ground state vectors as vectors in the Fock space $\cF^b$.
In the sequel, we use the expression
\begin{equation}
    \|\Psi_{\vP}^{\sigma_j}-\Psi_{\vP'}^{\sigma_{j'}}\|_{\cF}
\end{equation}
as an abbreviation for
\begin{equation}
    \|I_{\vP}(\Psi_{\vP}^{\sigma_j})-I_{\vP'}(\Psi_{\vP'}^{\sigma_{j'}})\|_{\cF}\,;
\end{equation}
$\| \, \cdot \, \|_{\cF}$ stands for the Fock norm.
H\"older continuity properties of $\Psi_{\vP}^{\sigma}$ in $\sigma$ and in $\vP$ 
are proven in \cite{ChFrPi2}. These properties play a crucial role in the present paper.

\subsection{Summary of contents}

In Section {\ref{sec-II}},  time-dependent vectors
$\psi_{h,\Lambdas }(t)$ approximating scattering states are constructed, and the main results of this
paper are described, along with an outline of infraparticle scattering
theory. In Sections {\ref{sec-II.2}} and {\ref{sec-II.3}},
$\psi_{h,\Lambdas }(t)$ is shown to converge to a scattering state
$\psi_{h,\Lambdas }^{out/in}$ in the Hilbert space $\cH$, as time $t$ tends to infinity.
This result is based on mathematical techniques introduced in \cite{Pizzo2005}. 
The vector $\psi_{h,\Lambdas }^{out/in}$ represents a dressed electron with a wave
function $h$ on momentum space whose support is contained
in the set $\cS$ (see details in Section {\ref{ssec:II-1}}), accompanied
by a cloud of soft photons described by a {\em Bloch-Nordsieck operator}, and with
an upper cutoff $\Lambdas $ imposed on photon frequencies.
This cutoff can be chosen arbitrarily.

In Section {\ref{sec-III}}, we construct the scattering subspaces $\cH^{out/in}$.
Vectors in these subspaces are
obtained from certain subspaces,  $\cHs$, by applying ``hard" asymptotic
photon creation operators. These spaces carry representations of the
algebras $\cA_{ph}^{out/in}$ and $\cA_{el}^{out/in}$ of asymptotic
photon creation- and annihilation operators and asymptotic electron
observables, respectively, which commute with each other.
The latter property proves asymptotic decoupling of the electron and photon dynamics.
We rigorously establish the coherent nature and the infrared properties of the
representation of $\cA_{ph}^{out/in}$ identified by Bloch and Nordsieck
in their classic paper, \cite{BlochNordsieck}.

In a companion paper \cite{ChFrPi2}, we establish the main spectral ingredients for the
construction and convergence of the vectors $\{\Psi_{\vP}^{\sigma}\}$, as $\sigma$ tends to 0.
These results are obtained with the help of a new multiscale method introduced
in \cite{Pizzo2003}, to which we refer the reader for some details of the proofs.

In the Appendix, we prove some technical results used in the proofs.

\newpage

%%%%%%%%%%%%%%%%%%%%%%%%%%%%%%%%%%%%%%%%%%%%%%%%%%%%%%%%
\section{Infraparticle Scattering States} \label{sec-II}
%%%%%%%%%%%%%%%%%%%%%%%%%%%%%%%%%%%%%%%%%%%%%%%%%%%%%%%%
\resetequ

%\subsection{Outline of infraparticle scattering theory}

\emph{Infraparticle scattering theory} is concerned with the
asymptotic dynamics in QFT models of infraparticles and massless fields.
Contrary to theories with a non-vanishing mass gap,
the picture of asymptotically freely moving particles in the Fock representation is not
valid, due to the inseparability of the dynamics of
charged massive particles and the soft modes of the massless asymptotic fields.

Our starting point is to study the
(dressed one-particle) states of a (non-relativistic) electron
when the interactions with the soft
modes of the photon field are turned off.
We then analyze their limiting behavior when this infrared cut-off is removed.
This amounts to studying vectors
$\psi^{\sigma}$, $\sigma>0$, in the
Hilbert space $\cH$ that are solutions to the equation
\begin{equation}\label{eq-I-49}
	H^{\sigma}\,\psi^{\sigma}\,=\,E^{\sigma}_{\vP}\,\psi^{\sigma}\,,
\end{equation}
where $H^{\sigma}=\int^\oplus\,\HPs\,d^3P$, and $E^{\sigma}_{\vP}$ is a
function of the vector
operator $\vP$; $E^{\sigma}_{\vP}$ is the electron energy function
defined more precisely in Section {\ref{ssec:II-1}}.
Since in our model non-relativistic matter is coupled to a
relativistic field, the form of $E^\sigma_{\vP}$ is not fixed by symmetry,
except for rotation invariance.
Furthermore, the solutions of (\ref{eq-I-49}) give rise to vectors in the physical Hilbert space describing wave packets of dressed electrons of the form
\begin{equation}\label{eq-III-2bis}
    \psi^{\sigma}(h) \, = \, \int\,h(\vP)\,\Psi_{\vP}^{\sigma}\,d^3P\,,
\end{equation}
where the support of $h$ is contained in a ball centered at $\vP=0$, chosen such that
$|\vnabla E^{\sigma}_{\vP}|<1$, as a function of $\vP$, i.e., we must impose the condition that the maximal group velocity of the electron which, a priori, is not bounded from above in our
non-relativistic model, is bounded by the speed of light. (For group velocities larger than the velocity of light, the one-electron states decay by emission of Cerenkov radiation.)

The guiding principle motivating our analysis of limiting or improper
one-particle states, $\psi^\sigma(h)$ for $\sigma\to0$,  is that refined
control of the infrared singularities, which push these vectors out of the
space $\cH$, as $\sigma\to 0$, should enable one
to characterize the soft photon cloud encountered in the scattering states.
The analysis of Bloch and Nordsieck, \cite{BlochNordsieck}, suggests that the infrared
behavior of the state describing the soft photons accompanying an electron should be singular (i.e.,  
not square-integrable at the origin in photon momentum space), 
and that it should be determined by the momentum of the asymptotic
electron.
In mathematical terms, this means that the asymptotic electron velocity is expected to
determine an asymptotic Weyl operator (creating a cloud of asymptotic photons), which
when applied to a dressed one-electron state $\psi^{\sigma=0}(h)$
yields a well defined vector in the Hilbert space
$\cH$. This vector is expected to
describe an asymptotic electron with wave function $h$
surrounded by a cloud of infinitely many asymptotic free photons,
in accordance with the observations sketched in (\ref{eq-introduction-6})-(\ref{eq-introduction-13}) .

Our goal in this paper is to translate this physical picture into rigorous
mathematics, following suggestions made in \cite{Fr73}
and methods developed in \cite{Pizzo2003,Pizzo2005,Chen,ChFr}.
%A list of spectral properties summarized below are crucial ingredients
%for the construction of the scattering states.

\subsection{Key spectral properties}
\label{ssec:II-1}
%\resetequ

In our construction of scattering states, we make extensive
use of a number of spectral properties of our model proven
in \cite{ChFrPi2}, and summarized in Theorem {\ref{thm-cfp-2}} below;
(they are analogous to those used in the analysis of Nelson's model
in \cite{Pizzo2005}).

We define the energy of a dressed one-electron state of momentum $\vP$ by
\begin{equation}\label{eq-III.3}
    E^\sigma_{\vP} \,  = \, \inf {\rm spec} H_{\vP}^\sigma 
    \quad , \quad\quad
    E_{\vP}  \,  = \, \inf {\rm spec} H_{\vP} \, = \, E^{\sigma=0}_{\vP} \,.
\end{equation}
We refer to $E_{\vP}^\sigma$ as the {\em ground state energy} of the fiber
Hamiltonian $H_{\vP}^\sigma$.
We assume that the finestructure constant $\alpha$ is so small that
\begin{equation}
    |\vnabla E_{\vP}^\sigma| \, < \, \nu_{max} \, < \, 1
\end{equation}
for all $\vP \in \mathcal{S}:=\{\vP\in \RR^3 \, : \, |\vP|<\frac13 \,\}$,
for some constant $\nu_{max}<1$, uniformly in $\sigma$.

Corresponding to $\vnabla E_{\vP}^\sigma$, we introduce a Weyl operator
\begin{equation}\label{eq-II-2}
    W_{\sigma}(\vnabla E_{\vP}^{\sigma}) \, := \, \exp\Big(\, \alpha^{\frac{1}{2}}
    \sum_{\lambda} \int_{\mathcal{B}_{\Lambda}\setminus \mathcal{B}_{\sigma}}d^3k \,
    \frac{\vnabla E_{\vP}^{\sigma}}{|\vec{k}|^{\frac{3}{2}}\delta_{\vP,\sigma}(\widehat{k})} \cdot
    (\veps_{\vk,\lambda}b_{\vk,\lambda}^{*} - h.c.)\Big) \, ,
\end{equation}
where
\begin{equation}
	\delta_{\vP,\sigma}(\widehat{k}) \, := \, 1 \, - \, \vnabla E^\sigma_{\vP}\cdot\frac{\vk}{|\vk|} \,,
\end{equation}
acting on $\cH_{\vP}$, which is unitary for $\sigma>0$. We consider the transformed
fiber Hamiltonian
\begin{equation}\label{eq-II-1}
    \HPsw\,:=\,W_{\sigma}(\vnabla
    E_{\vP}^{\sigma})\HPs W_{\sigma}^{*}(\vnabla E_{\vP}^{\sigma})\,.
\end{equation}
We note that conjugation by $W_{\sigma}(\vnabla E_{\vP}^{\sigma}) $ acts on the
creation- and annihilation operators as a linear Bogoliubov transformation (translation)
\begin{equation}\label{eq:Bog-trsf-b-1}
	W_{\sigma}(\vnabla
    E_{\vP}^{\sigma}) \, b^\#_{\vk,\lambda} \, W_{\sigma}^{*}(\vnabla E_{\vP}^{\sigma})
    \, = \, b^\#_{\vk,\lambda} \, - \, \alpha^{1/2} \,
    \frac{\one_{\sigma,\Lambda}(\vk)}{|\vec{k}|^{\frac{3}{2}}\delta_{\vP,\sigma}(\widehat{k})}
    \vnabla E_{\vP}^{\sigma}\cdot\veps_{\vk,\lambda}^{\;\#} \,,
\end{equation}
where $\one_{\sigma,\Lambda}(\vk)$ stands for the characteristic function of the set
$\mathcal{B}_{\Lambda}\setminus\mathcal{B}_{\sigma}$.
Our methods rely on proving regularity properties in $\sigma$
and $\vec{P}$ of the ground state vector, $\Phi_{\vec{P}}^{\sigma}$,
and of the ground state energy, $E^\sigma_{\vP}$, of $\HPsw$.
These regularity properties are summarized in the following theorem,
which is the main result of the companion paper \cite{ChFrPi2}.

\begin{theorem}
\label{thm-cfp-2}
For $\vP\in\mathcal{S}$ and for $\alpha>0$ sufficiently small, the following statements hold.

\begin{itemize}
\item[($\mathscr{I}1$)]
The energy $E^{\sigma}_{\vP}$ is a simple eigenvalue of the operator $\Hw_\vP^\sigma$
on $\cF^b$.
Let $\cB_{\sigma}:=\{ \vk \in \RR^3 \, | \, |\vk|\leq\sigma \}$, and
let $\cF_{\sigma}$ denote the Fock space over $L^2((\RR^3\setminus \cB_\sigma)\times\ZZ_2)$. 
%in  $b_{\vk,\lambda}$ and $b^*_{\vk,\lambda}$. Likewise,
Likewise, we define $\cF_0^{\sigma}$ to be the Fock space over $L^2(\cB_\sigma\times\ZZ_2)$;
hence $\cF^b=\cF_\sigma\otimes\cF_0^\sigma$.
On $\cF_{\sigma}$, the operator $\Hw_{\vP}^\sigma$
has a spectral gap of size $\rf\sigma$ or larger,
separating $E^{\sigma}_{\vP}$ from the rest of its spectrum,
for some constant 
$\rf$, with $0<\rf<1$.

The contour
\begin{equation}
    \gamma\,:=\lbrace z\in\CC\,|
    |z-E_{\vec{P}}^{\sigma}|=\frac{\rf\sigma}{2}\rbrace \; \;  , \; \sigma>0\,  
\end{equation}
bounds a disc which intersects the spectrum of $\Hw_{\vP}^\sigma|_{\cF_\sigma}$
in only one point, $\{E_{\vec{P}}^{\sigma}\}$. The ground state vectors
%\begin{equation}
%	\Phi_\vP^\sigma \, = \, W_\sigma(\vnabla E_\vP^\sigma) \,
%	\frac{ \Psi_\vP^\sigma }{\|\Psi_\vP^\sigma\|} \,,
%\end{equation}
of the operators $\Hw_\vP^\sigma$ are given by
\begin{equation}\label{eq-II-3}
    \Phi_{\vec{P}}^{\sigma} \, := \, \frac{\frac{1}{2\pi\,i}\int_{\gamma}
    \frac{1}{\Hw_{\vP}^\sigma-z} \,dz\,\Omega_f}
    {\|\frac{1}{2\pi\,i}\int_{\gamma}\frac{1}{\Hw_{\vP}^\sigma-z}  
    \,dz\,\Omega_f\|_{\cF} } 
    %\, \otimes \, \Omega_f
\end{equation}
and converge 
strongly   to a
non-zero vector $\Phi_{\vec{P}}\in \cF^b$, in the limit $\sigma\to 0$.
The rate of convergence
is at least of order $\sigma^{\frac{1}{2}(1-\delta)}$, for any
$0<\delta<1$. (Although it is not relevant for the purposes of this paper, we note that the results in  \cite{FrPi} imply the uniformity in $\delta$ of the range of values of $\alpha$ where the rate estimate $\sigma^{\frac{1}{2}(1-\delta)}$ holds; analogous conclusions follow for the rate estimates below.)

The dependence of the ground state energies $E_\vP^\sigma$ of the fiber Hamiltonians
$\Hw_\vP^\sigma$ on the infrared cutoff $\sigma$
is characterized by the following estimates.
%
%The ground state energies $E_{\vP}^\sigma$ are Lipschitz in $\sigma$,
\begin{equation}\label{eq-II-4}
    | \, E_{\vec{P}}^{\sigma} - E_{\vec{P}}^{\sigma'} \, | \, \leq \,  \cO(\sigma)\,,
\end{equation}
and 
%their gradients are well-defined and H\"older continuous with
\begin{equation}
	| \, \vnabla E_{\vec{P}}^{\sigma} - \vnabla E_{\vec{P}}^{\sigma'} \, | 
	\, \leq \, \cO(\sigma^{\frac{1}{2}(1-\delta)}) \,,
\end{equation}
for any $0<\delta<1$, with $\sigma>\sigma'>0$.

\item[($\mathscr{I}2$)]
The following H\"older regularity properties in  $\vec{P}\in\mathcal{S}$ hold uniformly
in $\sigma \geq 0$:
\begin{equation}\label{eq-II-5}
    \|\Phi_{\vec{P}}^{\sigma}-\Phi_{\vec{P}+\Delta\vec{P}}^{\sigma}\|_{\cF} \leq
    C_{\delta'}|\Delta \vec{P}|^{\frac{1}{4}-\delta'}
\end{equation}
and
\begin{equation}\label{eq-II-6}
    |\vnabla E_{\vec{P}}^{\sigma}-\vnabla E_{\vec{P}+\Delta\vec{P}}^{\sigma}|\leq
    C_{\delta''}|\Delta \vec{P}|^{\frac{1}{4}-\delta''} \, ,
\end{equation}
for any $0<\delta''<\delta'<\frac{1}{4}$, with $\vec{P}\,,\,\vec{P}+\Delta\vec{P} \in
\mathcal{S}$, where $C_{\delta'}$ and $C_{\delta''}$ are
finite constants depending on $\delta'$ and $\delta''$, respectively.

\item[($\mathscr{I}3$)]
Given a positive number $\nu_{min}$, there are numbers
$r = \nu_{min}/2>0$ and $\nu_{max}<1$
such that, for $\vP\in\mathcal{S}\setminus\cB_{r}$ and for $\alpha$ sufficiently small,
\begin{equation}\label{eq-II-7}
    1>\nu_{max}\geq|\vnabla E_{\vec{P}}^{\sigma}|\geq\nu_{min}>0 \, ,
\end{equation}
uniformly in $\sigma$. (We also notice that the control on the second derivative of $E_{\vec{P}}^{\sigma}$ in $\vP$ uniformly in the sharp infrared cut-off $\sigma\geq 0$ (see \cite{FrPi}) would allow us to take $\ \nu_{min}=0$ and to include velocities $\vnabla E_{\vec{P}}^{\sigma}$ arbitrarily close to $0$, but we prefer to work with an assumption self-contained in the paper.)

\item[($\mathscr{I}4$)]
For $\vP\in\mathcal{S}$ and for any $\vk\not=0$, the following inequality holds
uniformly in $\sigma$, for $\alpha$ small enough:
\begin{equation}\label{eq-II-8}
    E^{\sigma}_{\vP-\vk}>E^{\sigma}_{\vP}-C_\alpha|\vk|\,,
\end{equation}
where $E^{\sigma}_{\vP-\vk}:=\inf {\rm{spec}} \, H_{\vP-\vk}^\sigma$ and $\frac13< C_\alpha<1$,
with $C_\alpha\rightarrow\frac13$ as $\alpha\rightarrow0$.  

\item[($\mathscr{I}5$)] 
Let $\Psi_{\vP}^{\sigma}\in \cF$ denote the ground state vector of the fiber Hamiltonian $H_{\vP}^{\sigma}$, so that
\begin{equation}
\Phi^{\sigma}_{\vP}=\zeta W_{\sigma}(\vnabla E_{\vP}^{\sigma})\frac{\Psi_{\vP}^{\sigma}}{\|\Psi_{\vP}^{\sigma}\|_{\cF}}\,\quad\zeta\in \CC,\quad |\zeta|=1\,.
\end{equation}
For $\vP\in\mathcal{S}$, one has that
\begin{equation}\label{eq-II-9}
  	\| \, b_{\vk,\lambda}\frac{\Psi_{\vP}^{\sigma}}{\|\Psi_{\vP}^{\sigma}\|_{\cF}}\, \|_{\cF} \, \leq C \, \alpha^{1/2}
  	\, \frac{\one_{\sigma,\Lambda}(\vk)}{|\vk|^{3/2}}\,,
\end{equation} 
where $\Psi_\vP^\sigma$ is the ground state of $H_\vP^\sigma$;
see Lemma 6.1 of \cite{ChFr} which can be extended to $\vk\in\RR^3$ using ($\mathscr{I}4$).
\end{itemize}

\end{theorem}
\noindent
Detailed proofs of Theorem {\ref{thm-cfp-2}} based on results in \cite{Pizzo2003,ChFr}
are given in \cite{ChFrPi2}.

\subsection{Definition of the approximating vector $\Psi_{h,\Lambdas }(t)$}
\label{sec-II.2-1}
 
We construct infraparticle scattering states by using a
{\em time-dependent approach to scattering theory}. 
We define a time-dependent approximating vector
$\psi_{h,\Lambdas }(t)$ that converges to an
asymptotic vector, as $t\to\infty$.
It describes an
electron with wave function $h$ (whose momentum space support
is contained in $\cS$),
and a cloud of asymptotic free photons with an upper
photon frequency cutoff $0<\Lambdas \leq\Lambda$.
This interpretation will be justified a posteriori.

We closely follow an approach to infraparticle scattering theory developed for
Nelson's model in \cite{Pizzo2005},
(see also \cite{Fr73}). In the context of the present paper,
our task is to give a mathematically rigorous meaning to the formal expression
\begin{equation}\label{eq:wavefct-1}
	\Phi^{out}_{\Lambdas }(h) \, := \,
	\lim_{t\to\infty}\lim_{\sigma\to0}e^{iHt}\,\cW_{\Lambdas,\sigma}(\vv(t),t)\,e^{-iH^{\sigma}t}\psi^{\sigma}(h)\,,
\end{equation}
where
\begin{eqnarray}\label{eq-II-15} 
	\mathcal{W}_{\Lambdas,\sigma}(\vec{v}(t),t) 
	\, := \, \exp\Big(\alpha^{\frac{1}{2}}\sum_{\lambda}\int_{\mathcal{B}_{\Lambdas }\setminus\mathcal{B}_{\sigma}}
	\frac{d^3k}{\sqrt{|\vk|}} \frac{\vv(t)\cdot\lbrace\veps_{\vk,
	\lambda}a_{\vk,
    \lambda}^*e^{-i|\vk|t}-\veps_{\vk,
    \lambda}^{\;*}a_{\vk, \lambda}e^{i|\vk|t}\rbrace}{|\vk|(1-\widehat{k}\cdot\vv(t))}
    \Big)\nonumber\,.
\end{eqnarray}
The {\em operator} $\vv(t)$ is not known a priori; but, in the limit $t \to\infty$,
it must converge to the asymptotic velocity operator of the electron.
The latter is determined by the operator $\vnabla E_{\vP}$,
applied to the (non-Fock) vectors $\Psi_{\vP}$.
This can be seen by first considering the infrared regularized model, with $\sigma>0$,
which has dressed one-electron states $\psi^{\sigma}(h)$ in $\cH$, and by subsequently
passing to the limit $\sigma\to0$.
Formally, for $\sigma\rightarrow0$, the Weyl operator
\begin{equation}
	e^{iHt}\,\cW_{\Lambdas,\sigma}(\vv(t),t)\,e^{-iH^\sigmat}
\end{equation}
is an interpolating operator used in the L.S.Z. (Lehmann-Symanzik-Zimmermann)
approach to scattering theory for the electromagnetic field, where the photon test functions  (in the operator $\cW_{\Lambdas,\sigma}(\vv(t),t)$) are evolved backwards in time with the free evolution,
and the photon creation- and annihilation operators are evolved forward in time with
the interacting time evolution.
Moreover, the photon test functions in (\ref{eq-II-15}) coincide   with the test
functions in the Weyl operator
$W_{\sigma}(\vnabla E_{\vP}^{\sigma})$ defined in (\ref{eq-II-2}),
after replacing the operator $\vnabla E_{\vP}^{\sigma}$ by the operator $\vv(t)$.
We stress that, while the Weyl operator
$W_{\sigma}(\vnabla E_{\vP}^{\sigma})$ leaves
the fiber spaces $\cH_{\vP}$ invariant, the Weyl operator
$\cW_{\Lambdas,\sigma}(\vv(t),t)$ is expressed in terms of the operators
$\{a\,,\,a^{*}\}$, as it must be when describing real photons in
a scattering process, and hence does {\em not} preserve the fiber spaces.

Guided by the expected relation between $\vv(t)$ and $\vnabla
E^{\sigma}_{\vP}$, as $t\to\infty$ and $\sigma\to 0$,
two key ideas used to make  (\ref{eq:wavefct-1}) precise are to render
the infrared cut-off {\em time-dependent},
with $\sigma_t\to0$, as $t\rightarrow\infty$, and to discretize
the ball $\mathcal{S}=\{ \vP\in\RR^3 \, | \, |\vP|<\frac13\}$,
with a grid size decreasing in time $t$. This discretization also applies to the
velocity operator $\vv(t)$ in expression (\ref{eq:wavefct-1}).

The existence of infraparticle scattering states in $\cH$
is established by proving that the corresponding sequence of time-dependent approximating scattering
states, which depend on the cutoff $\sigma_t$ and on the discretization, defines a strongly convergent
sequence of vectors in $\cH$.
This is accomplished by appropriately tuning the convergence rates of $\sigma_t$ and of the
discretization of $\mathcal{S}$.
Our sequence of approximate infraparticle scattering states is defined as follows:
\\

\begin{itemize}

\item[i)]

We consider a wave function $h$ with support
in a region $\mathcal{R}$ contained in
$\mathcal{S}\setminus\cB_{{r}_{\alpha}}$; (see condition $(\mathscr{I}3)$
in Theorem {\ref{thm-cfp-2}}).
We introduce a time-dependent cell partition
$\mathscr{G}^{(t)}$ of $\mathcal{R}$. This partition is constructed as follows:

At time $t$, the linear dimension of each cell is $\frac{L}{2^n}$, where
$L$ is the diameter of  $\mathcal{R}$, and $n\in\NN$ is such that
\begin{equation}\label{eq-II-12}
	(2^n)^{\frac{1}{\epsilon}} \, \leq \, t < (2^{n+1})^{\frac{1}{\epsilon}} \,,
\end{equation}
for some $\epsilon>0$ to be fixed later. Thus, the total number of
cells in $\mathscr{G}^{(t)}$ is $N(t)=2^{3n}$,
where $n=\lfloor\text{log}_2 \,t^{\epsilon}\rfloor$;
($\lfloor x \rfloor$ extracts the integer part of $x$).
By $\mathscr{G}^{(t)}_j$, we denote the
$j^{th}$ cell of the partition $\mathscr{G}^{(t)}$.
\\

\item[ii)]
For each cell, we consider a one-particle state of the Hamiltonian
$H^{\sigma_t}$
\begin{equation}\label{eq-II-13}
	\psi^{(t)}_{j,\sigma_t} \, := \, \int_{\mathscr{G}^{(t)}_j}h(\vec{P})\Psi_{\vec{P}}^{\sigma_t}d^3P
\end{equation}
where
\begin{itemize}
\item[$\bullet$]
$h(\vec{P})\in C_0^1(\mathcal{S} \setminus \cB_{r_{\alpha}})$, with  ${\rm supp} \, h \subseteq \mathcal{R}$;

\item[$\bullet$]
$\sigma_t:=t^{-\beta}$, for some exponent $\beta\,(>1)$ to be fixed later;

\item[$\bullet$]
In (\ref{eq-II-13}), the ground state vector, $\Psi_{\vec{P}}^{\sigma_t}$,  of
$\HPst$ is defined by
\begin{equation}\label{eq-II-14}
	\Psi_{\vec{P}}^{\sigma_t} \, := \,
	W_{\sigma_t}^{*}(\nabla E_{\vec{P}}^{\sigma_t}) \, \Phi_{\vec{P}}^{\sigma_t}\,,
\end{equation}
where $\Phi_{\vec{P}}^{\sigma_t}$ is the ground state of $\HPstw$;
(see Theorem {\ref{thm-cfp-2}}).
$\;$
\end{itemize}

\item[iii)]
With each cell $\mathscr{G}^{(t)}_j$ we associate a soft-photon
cloud described by the following ``LSZ (Lehmann-Symanzik-Zimmermann) Weyl operator"
\begin{equation}
    	e^{iHt} \, \mathcal{W}_{\Lambdas,\sigma_t}(\vec{v}_j,t) \, e^{-iH^{\sigma_t}t}\,,
\end{equation}
where
\begin{equation}\label{eq-II-15bis}
	\mathcal{W}_{\Lambdas,\sigma_t}(\vec{v}_j,t) \, := \,
	\exp\Big(\alpha^{\frac{1}{2}}\sum_{\lambda}\int_{\mathcal{B}_{\Lambdas }\setminus\mathcal{B}_{\sigma_{t}}}
	\frac{d^3k}{\sqrt{|\vk|}} \frac{\vv_j\cdot\lbrace\veps_{\vk,\lambda}a_{\vk,\lambda}^*
	e^{-i|\vk|t}-\veps_{\vk,\lambda}^{\;*}a_{\vk, \lambda}e^{i|\vk|t}\rbrace}{|\vk|(1-\widehat{k}\cdot\vv_j)}
    	\Big)\,.
\end{equation}
\begin{itemize}
\item[$\bullet$]
Here $\Lambdas $, with $0<\Lambdas \leq\Lambda$,  is an arbitrary (but fixed) photon
energy threshold or counter threshold.
\item[$\bullet$]
$\vec{v}_j\equiv \vnabla E^{\sigma_t}(\vec{P}_j^*)$ is the c-number vector corresponding to the value of the
``velocity" $\vnabla E^{\sigma_t}(\vec{P})$ in the center, $\vec{P}_j^*$, of the
cell $\mathscr{G}^{(t)}_j$.
\\
\end{itemize}

\item[iv)]
For each cell, we consider a time-dependent phase factor
\begin{equation}\label{eq-II-17}
	e^{i\gamma_{\sigma_t}(\vv_j,\vnabla E_{\vec{P}}^{\sigma_t},t)} \,,
\end{equation}
with
\begin{equation}\label{eq-II-18} 
	\gamma_{\sigma_t}(\vv_j,\vnabla E^{\sigma_t}_{\vP},t) 
	\, := \, - \, \alpha\,\int_{1}^{t}
	\vnabla E^{\sigma_t}_{\vP}\cdot\int_{\mathcal{B}_{\sigma_\tau^{S}}\setminus\mathcal{B}_{\sigma_t}}
	\vec{\Sigma}_{\vv_j}(\vk) \,
	\cos(\vk\cdot\vnabla E^{\sigma_t}_{\vP}\tau-|\vk|\tau) \, d^3k \, d\tau\,, 
\end{equation}
and
\begin{eqnarray}
	\Sigma^{l}_{\vv_j}(\vk) \, := \, 2 \, \sum_{l'}(\delta_{l,l'}-\frac{k^lk^{l'}}{|\vk|^2})
	\, v^{l'}_{j} \, \frac{1}{|\vk|^2(1-\widehat{k}\cdot\vv_j)}\label{eq-II-19} \, .
\end{eqnarray}
Here, $\sigma_\tau^{S}:=\tau^{-\theta}$, and
the exponent $0<\theta<1$ will be chosen later.
Note that, in (\ref{eq-II-17}), (\ref{eq-II-18}), $\vnabla E^{\sigma_t}_{\vP}$ is interpreted as an {\em operator}.
\\

\item[v)]
The approximate scattering state at time $t$ is given by the expression
\begin{equation}\label{eq-II-20}
	\psi_{h,\Lambdas }(t) \, := \,
	e^{iHt}\sum_{j=1}^{N(t)}\cW_{\Lambdas,\sigma_t}(\vv_j,t) \,
	e^{i\gamma_{\sigma_t}(\vv_j,\vnabla E_{\vec{P}}^{\sigma_t},t)}\,
	e^{-iE_{\vec{P}}^{\sigma_t}t}\psi_{j,\sigma_t}^{(t)}\,,
\end{equation}
where $N(t)$ is the number of cells in $\mathscr{G}^{(t)}$.
\end{itemize}

The role played by the phase factor $e^{i\gamma_{\sigma_t}(\vv_j,\vnabla
E_{\vec{P}}^{\sigma_t},t)}$ is similar to that of the Coulomb phase in Coulomb
scattering. However, in the present case, the phase has a {\em limit}, as $t\to\infty$, and
is introduced to control an oscillatory term in the Cook
argument which is not absolutely convergent; (see Section {\ref{ssec-state-res-1}}).

%\centerline{\epsffile{cfp-fig1.pdf} }
\centerline{\epsffile{cfp-fig1.epsf} }

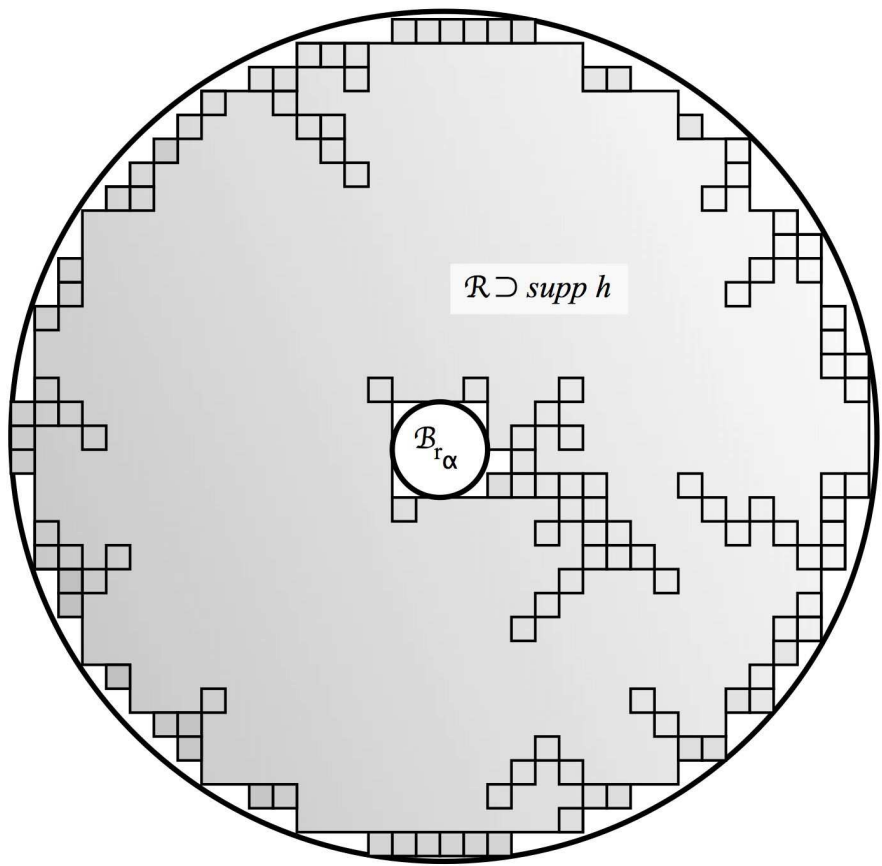
\noindent Figure 1. $\mathcal{R}$ can be described as a union of cubes with sides of length
$\frac{L}{2^{n_0}}$, for some $n_0<\infty$.
\\

\subsection{Statement of the main result}
\label{ssec-state-res-1}
The main result of this paper is Theorem {\ref{theo-II.2}}, below, from which
the asymptotic picture described in Section {\ref{ssec-scatt-ssp-1}}, below, emerges. It relies on
the assumptions summarized in the following hypothesis.

\begin{hypothesis}
\label{hyp-main}
The following assumptions hold throughout this paper:
\begin{enumerate}
\item
The conserved momentum $\vP$ takes values in $\mathcal{S}$; see (\ref{eq-cS-def-1}).
\item
The finestructure constant $\alpha$ satisfies $\alpha<\alpha_c$, for
some small constant $\alpha_c\ll1$ independent of the infrared cutoff.
\item
The wave function $h$ is supported in a set $\mathcal{R}$ and is of class $C^1$, where
$\mathcal{R}$ is contained in $\mathcal{S}\setminus\cB_{r_\alpha}$, as indicated in
Fig. 1, and $r_\alpha$ is introduced in ($\mathscr{I}3$).  
\end{enumerate}
\end{hypothesis}

\begin{theorem}\label{theo-II.2}
Given the Main Assumption {\ref{hyp-main}},
the following holds:
There exist positive real numbers $\beta>1$,
$\theta<1$ and $\epsilon>0$ such that the limit
\begin{equation}\label{eq-II-73}
	s-\lim_{t\to+\infty}\psi_{h,\Lambdas }(t) \, =: \, \psi_{h,\Lambdas }^{(out)}
\end{equation}
(where $\psi_{h,\Lambdas }(t)$ is defined in Eq. (\ref{eq-II-20}) and $\Lambdas$, see (\ref{eq-II-15bis}), is the threshold frequency) exists as a vector in $\cH$,
and $\|\psi_{h,\Lambdas }^{(out)}\|^2 = \int |h(\vec{P})|^2d^3P$. Furthermore, the rate of
convergence is at least of order $t^{-\rho'}$, for some $\rho'>0$.
\end{theorem}

We note that this result corresponds to Theorem~3.1 of \cite{Pizzo2005}
for Nelson's model.

The limiting state is the
desired infraparticle scattering state without infrared cut-offs.
We shall verify that $\{\psi_{h,\Lambdas }(t)\}$
is a {\em Cauchy sequence} in $\cH$, as $t\to\infty$;
(or $t\to-\infty$).

In   Section {\ref{sec:convscheme}},
we outline the key mechanisms responsible for the convergence
of the approximating vectors
$\psi_{h,\Lambdas }(t)$, as $t\rightarrow\infty$.
We note that, in (\ref{eq-II-73}),
{\em three different convergence rates} are involved:
\begin{itemize}
\item
The rate $t^{-\beta}$ related to the fast infrared cut-off $\sigma_t$;

\item
the rate $t^{-\theta}$, related to the slow infrared cut-off $\sigma_t^{S}$ (see (\ref{eq-II-18}));

\item
the rate $t^{-\epsilon}$ of the grid size of the cell partition.
\end{itemize}

\noindent
We anticipate that, in order to control the interaction,
\begin{itemize}
\item
$\beta$ has to be larger than $1$, due to
the time-energy uncertainty principle.

\item
The exponent $\theta$ has to be smaller than $1$, in order to ensure
the cancelation of some ``infrared tails'' discussed
in Section {\ref{sec-II.2}}.
\item
The exponent $\epsilon$, which controls the rate of refinement of
the cell decomposition, will have to be chosen small enough
to be able to prove certain decay estimates.
\\
\end{itemize}

\newpage

\subsection{Strategy of convergence proof}
\label{sec:convscheme}
Here we outline the key mechanisms used to prove that
the approximating vectors $\psi_{h,\Lambdas }(t)$ converge to
a nonzero vector in $\cH$, as $t\to \pm\infty$.

Among other things, we will prove that
\begin{equation}\label{eq-I-75}
	\lim_{t\to\infty}\|\psi_{h,\Lambdas }(t)\| \, = \, \|h\|_{2} \, := \,
	( \, \int |h(\vP)|^2\,d^3P \, )^{\frac{1}{2}}\,.
\end{equation}
>From its definition, see (\ref{eq-II-20}), one sees that the square of the norm of the vector
$\psi_{h,\Lambdas }(t)$ involves a double sum over cells of the partitions $\mathscr{G}^{(t)}$,
i.e.,
\begin{equation}\label{eq-I-78}
	\|\psi_{h,\Lambdas }(t)\|^2 \, = \,
	\sum_{l,j=1}^{N(t)}\,\Bra
	\, e^{i\gamma_{\sigma_t}(\vv_l,\vnabla E_{\vec{P}}^{\sigma_t},t)}
	\, e^{-iE_{\vec{P}}^{\sigma_t}t}\psi_{l,\sigma_t}^{(t)}\,,\,
  	\, \cW_{\Lambdas,\sigma_t}^{*}(\vv_l,t) \, \cW_{\Lambdas,\sigma_t}(\vv_j,t)
  	\, e^{i\gamma_{\sigma_t}(\vv_j,\vnabla E_{\vec{P}}^{\sigma_t},t)}
  	\, e^{-iE_{\vec{P}}^{\sigma_t}t}\psi_{j,\sigma_t}^{(t)} \, \Ket \,,
\end{equation}
where the individual terms, labeled by $(l,j)$, are inner products between vectors
labeled by cells $\mathscr{G}^{(t)}_l$ and $\mathscr{G}^{(t)}_j$ of $\mathscr{G}^{(t)}$.

A heuristic argument to see where (\ref{eq-I-75}) comes from is as follows.
Assuming that

\begin{itemize}
\item
the vectors $\psi_{h,\Lambdas }(t)$ converge to an asymptotic
vector of the form
\begin{equation}
	\lim_{t\to\pm\infty}\lim_{\sigma\to0}e^{iHt} \, \cW_{\Lambdas,\sigma }(\vv(t),t)
	\,e^{-iH^\sigmat}\psi^{\sigma}(h) \, = \, \cW^{out/in}_{\Lambdas,\sigma=0}(\vv(\pm\infty)) \, \psi^{\sigma=0}(h) \,,
\end{equation}
where
\begin{eqnarray}\label{eq-II-15bisbis} 
	\cW^{out/in}_{\Lambdas,\sigma=0}(\vv(\pm\infty)) 
	\, := \, \exp\Big( \, \alpha^{\frac{1}{2}}\sum_{\lambda}\int_{\mathcal{B}_{\Lambdas }}
	\frac{d^3k}{\sqrt{|\vk|}} \frac{\vv(\pm\infty)\cdot\lbrace
	\veps_{\vk, \lambda}a_{\vk,\lambda}^{out/in\,*} -
	\veps_{\vk,\lambda}^{\;*}a_{\vk, \lambda}^{out/in}\rbrace}{|\vk|(1-\widehat{k}\cdot\vv(\pm\infty))}
    	\, \Big)\nonumber\,,
\end{eqnarray}
and $a_{\vk,\lambda}^{out/in\,*}\,,\,a_{\vk, \lambda}^{out/in}$ are the creation- and
annihilation operators of the asymptotic photons;

\item
the operators $\vv(\pm\infty)$ commute with the algebra of
asymptotic creation- and annihilation operators
$\lbrace a_{\vk,\lambda}^{out/in\,*}\,,\,a_{\vk, \lambda}^{out/in}\rbrace$;
(this can be expected to be a consequence of asymptotic decoupling
of the photon dynamics from the dynamics of the electron);

\item
the restriction of the asymptotic velocity operators, $\vec v(\pm\infty)$,
to the improper dressed one-electron state is given by the operator $\vnabla E_{\vP}$, i.e.,
\begin{equation}
	\vv(\pm\infty)\Psi_{\vP}\equiv\vnabla E_{\vP}\Psi_{\vP}\,;
\end{equation}
\end{itemize}
then, the two vectors
\begin{equation}\label{eq-I-80}
	\cW_{\Lambdas,\sigma_t}(\vv_j,t) \,
	e^{i\gamma_{\sigma_t}(\vv_j,\nabla E_{\vec{P}}^{\sigma_t},t)}
	\, e^{-iE_{\vec{P}}^{\sigma_t}t}
	\, \psi_{j,\sigma_t}^{(t)}\quad\text{and}
	\quad \cW_{\Lambdas,\sigma_t}(\vv_l,t)
	\, e^{i\gamma_{\sigma_t}(\vv_l,\nabla E_{\vec{P}}^{\sigma_t},t)}
	\, e^{-iE_{\vec{P}}^{\sigma_t}t}\psi_{l,\sigma_t}^{(t)}
\end{equation}
corresponding to two different cells of $\mathscr{G}^{(t)}$ (i.e., $j\neq l$)
turn out to be orthogonal in the limit $t\to\pm\infty$.
One can then show that the diagonal terms in the sum (\ref{eq-I-78}) 
are the only ones that survive in the limit $t\to\infty$. 
The fact that their sum converges to $\|h\|_{2}^2$ is comparatively easy to prove.

A mathematically precise formulation of this mechanism is presented
in Section {\ref{sec-II.2}}.
In Section {\ref{sec-II.2.1}}, part \underline{\em A.},
the analysis of the scalar products between the cell vectors
in (\ref{eq-I-80}) is reduced to the study of
an ODE. To prove (\ref{eq-I-75}), we invoke the following properties of
the one-particle states $\psi_{j,\sigma_t}^{(t)}$ and
$\psi_{l,\sigma_t}^{(t)}$ located in the $j$-th and $l$-th cell:
\begin{itemize}
\item
Their spectral supports
with respect to  the momentum operator  $\vP$
are disjoint up to sets of measure zero.
\item
They are vacua for asymptotic annihilation operators,  as long as an
infrared cut-off $\sigma_t$ for a fixed time $t$
is imposed: For Schwartz test functions $g^\lambda$,
we define
\begin{equation}
	a^{out/in}_{\sigma_t}(g) \, := \, \lim_{s\to\pm\infty}\, e^{iH^{\sigma_t} s}\,
	\sum_{\lambda}\,\int a_{\vk,\lambda} \, \overline{g^{\lambda}(\vk)}
	\, e^{i|\vk|s} \, e^{-iH^{\sigma_t} s} \, d^3k \,,
\end{equation}
on the domain of $H^{\sigma_t}$.
\end{itemize}

An important step in the proof of (\ref{eq-I-75}) is to control the decay
in time of the off-diagonal terms. After completion of this step, one can
choose the rate, $t^{-\epsilon}$, by which the diameter of the cells
of the partition $\mathscr{G}^{(t)}$ tends to 0 in such a way that the sum of the off-diagonal
terms vanishes, as $t\to\infty$. Precise control is achieved in Section
{\ref{sec-II.2.1}}, part \underline{\em B.}, where we invoke Cook's argument
and analyze the decay in time $s$ of
\begin{eqnarray}\label{eq-I-82}
	\lefteqn{\frac{d}{ds}
	\Big( \,e^{iH^{\sigma_t}s} \, \cW_{\Lambdas,\sigma_t}(\vv_j,s)
	\, e^{i\gamma_{\sigma_t}(\vv_j,\vnabla E_{\vec{P}}^{\sigma_t},s)}
	\, e^{-iE_{\vec{P}}^{\sigma_t}s}\psi_{j,\sigma_t}^{(t)} \, \Big)\,}
	\\
	&= &i \, e^{iH^{\sigma_t}s}[H^{\sigma_t}_I \, , \,
	\cW_{\Lambdas,\sigma_t}(\vv_j,s)]
	\, e^{i\gamma_{\sigma_t}(\vv_j,\vnabla E_{\vec{P}}^{\sigma_t},s)}
	\, e^{-iE_{\vec{P}}^{\sigma_t}s}
	\, \psi_{j,\sigma_t}^{(t)}
	\label{eq-I-83} \\
	& &+ \, i \, e^{iH^{\sigma_t}s}
	\, \cW_{\Lambdas,\sigma_t}(\vv_j,s)
	\, \frac{d\gamma_{\sigma_t}(\vv_j,\vnabla E_{\vec{P}}^{\sigma_t},s)}{ds}
	\, e^{i\gamma_{\sigma_t}(\vv_j,\vnabla E_{\vec{P}}^{\sigma_t},s)}
	\, e^{-iE_{\vec{P}}^{\sigma_t}s} \, \psi_{j,\sigma_t}^{(t)} \,, \quad
  	\quad\label{eq-I-84}
\end{eqnarray}
for a \emph{fixed} infrared cut-off $\sigma_t$, and a \emph{fixed} partition.
As we will show, the term in (\ref{eq-I-83}) can be
written (up to a unitary operator) as
\begin{equation}\label{eq-I-86}
	\alpha^{\frac{1}{2}} \, \int\,  d^3y \, \Big\lbrace \vec{J}_{\sigma_t}(t,\vy)\,
	\int_{\mathcal{B}_{\Lambdas } \setminus\mathcal{B}_{\sigma_t}}
	\vec{\Sigma}_{\vv_j}(\vq)\cos(\vq\cdot\vy-|\vq|s) \, d^3q\,\Big\rbrace
	\,e^{i\gamma_{\sigma_t}(\vv_j,\vnabla E_{\vec{P}}^{\sigma_t},s)}\, \psi_{j,\sigma_t}^{(t)}
\end{equation}
plus subleading terms, where $\vec{J}_{\sigma_t}(t,\vy)$ is 
essentially the electron current at time $t$, which is proportional to the velocity operator
\begin{equation}
	\label{eq:velop-def}
	 i \, [ H^{\sigma_t} \, , \, \vec x ] \, = \, \vec p \, + \, \alpha^{\frac12} \, 
	 \vec A_{\sigma_t}(\vec x) \,.
\end{equation}
In (\ref{eq-I-86}), the electron current is
smeared out with the vector function
\begin{equation}
	\vec{g}_t(s,\vy) \, := \, \int_{\mathcal{B}_{\Lambdas } \setminus\mathcal{B}_{\sigma_t}}
	\vec{\Sigma}_{\vv_j}(\vq) \, \cos(\vq\cdot\vy-|\vq|s) \, d^3q\,,
\end{equation}
which solves the wave equation
\begin{equation}\label{eq-I-89}
	\Box_{s,\vy}\,\vec{g}_t(s,\vy)=0\,,
\end{equation}
and is then applied to the one-particle state $e^{i\gamma_{\sigma_t}(\vv_j,\vnabla E_{\vec{P}}^{\sigma_t},s)}\psi_{j,\sigma_t}^{(t)}$.
Because of the dispersive properties of the dynamics of the
system, the resulting vector is expected to
converge to 0 in norm at an integrable rate, as $s\to\infty$.
An intuitive explanation proceeds as follows:
\begin{itemize}
\item[i)]
A vector function $\vec{g}_t(s,\vy)$ that solves (\ref{eq-I-89})
propagates along the light cone, and $\sup_{\vy\in\RR^3}|\vec{g}_t(s,\vy)|$ decays
in time like $s^{-1}$, while a much faster decay is observed when $\vy$ is restricted
to the interior of the light cone (i.e., $|\frac{|\vec y|}{s}|<1$).
\item[ii)]
Because of the support in $\vP$ of the vector $\psi_{j,\sigma_t}^{(t)}$,
the propagation of  the electron current in (\ref{eq-I-83}) is
limited to the interior of the
light cone, up to subleading tails.
\end{itemize}
Combination of i) and ii)
is expected to suffice to exhibit decay of the vector norm of
(\ref{eq-I-83}) and to complete our argument. An
important refinement of this reasoning process, 
involving the term (\ref{eq-I-84}), is, however, necessary:

A mathematically precise version of statement ii) is as follows:
Let $\chi_{h}$ be a smooth, approximate characteristic function of the
support of $h$. We will prove a {\em propagation estimate}  
\begin{eqnarray}\label{eq-II-49}
	\lefteqn{
	\Big\| \, \chi_h(\frac{\vx}{s})
	\, e^{i\gamma_{\sigma_t}(\vv_j,\vnabla E_{\vec{P}}^{\sigma_t},s)}
	\, e^{-iE_{\vec{P}}^{\sigma_t}s} \, \psi_{j,\sigma_t}^{(t)} 
	}
  	\nonumber\\
  	&&\quad\quad\quad\quad
  	\quad\quad-\,\chi_h(\vnabla E^{\sigma_t}_{\vP})
	\, e^{i\gamma_{\sigma_t}(\vv_j,\vnabla E_{\vec{P}}^{\sigma_t},s)}
	\, e^{-iE_{\vec{P}}^{\sigma_t}s} \, \psi_{j,\sigma_t}^{(t)}\, \Big\|
  	\nonumber\\ 
	\label{eq-II-50}
	&\leq&\frac{1}{s^{\nu}}\,\frac{1}{t^{\frac{3\epsilon}{2}}}\,|\ln(\sigma_t)|\,, 
\end{eqnarray}
as $ s\rightarrow\infty$,
where $\nu>0$ is independent of $\epsilon$. Using result ($\mathscr{I}3$) of
Theorem {\ref{thm-cfp-2}},
and our assumption on the support of $h$ formulated in point ii) of Section {\ref{sec-II.2-1}}, this
estimate provides sufficient control of the asymptotic dynamics of the electron.

An important modification of the argument above is necessary because of
the dependence of
\begin{equation}
	\vec{g}_t(s,\vy) \, := \, \int_{\mathcal{B}_{\Lambdas } \setminus\mathcal{B}_{\sigma_t}}
	\vec{\Sigma}_{\vv_j}(\vq) \, \cos(\vq\cdot\vy-|\vq|s) \, d^3q\,,
\end{equation}
on $t$,
which cannot be neglected even if $\vy$ is in the interior of the light cone.
In order to exhibit the desired decay, it is
necessary to split $\vec{g}_t(s,\vy)$ into two pieces,
\begin{eqnarray}
	\int_{\mathcal{B}_{\Lambdas } \setminus\mathcal{B}_{\sigma_s^S}}
	\vec{\Sigma}_{\vv_j}(\vq)\cos(\vq\cdot\vy-|\vq|s)d^3q \label{eq-I-93}
\end{eqnarray}
and
\begin{eqnarray}
	\int_{\mathcal{B}_{\sigma_s^S}\setminus\mathcal{B}_{\sigma_t}}
	\vec{\Sigma}_{\vv_j}(\vq)\cos(\vq\cdot\vy-|\vq|s) \, d^3q 
	\label{eq-I-94}
\end{eqnarray}
for $s$ such that $\sigma_s^S>\sigma_t$,
where $\sigma_s^S=s^{-\theta}$, with $0<\theta<1$.
(The same procedure will also be used in  (\ref{eq-II-70bis}), below.)
The function (\ref{eq-I-93}) has good
decay properties inside the light cone. Expression (\ref{eq-I-86}),
with $\vec{g}_t(s,\vy)$ replaced  by (\ref{eq-I-93}), can be controlled by standard dispersive
estimates. The other contribution, proportional to (\ref{eq-I-94}), is in principle singular
in the infrared region, but is canceled by (\ref{eq-I-84}). This can be seen by using
a propagation estimate similar to (\ref{eq-II-49}).
This strategy has been designed in \cite{Pizzo2005}. However, because
of the vector nature of the interaction in non-relativistic QED, the cancellation
in our proof
is technically more subtle than the one in \cite{Pizzo2005}.

After having proven  the uniform boundedness of the norms of the
approximating vectors $\psi_{h,\Lambdas }(t)$,
one must prove that they define a Cauchy sequence in $\cH$. To this end,
we compare these vectors at two different
times, $t_2>t_1$ (for the limit $t\rightarrow\infty$), and split their difference into
\begin{eqnarray}
 	\psi_{h,\Lambdas }(t_1) - \psi_{h,\Lambdas }(t_2) &=&
	\Delta \psi (t_2,\sigma_{t_2},\mathscr{G}^{(t_2)}\rightarrow \mathscr{G}^{(t_1)})
	\, + \,
	\Delta \psi (t_2\rightarrow t_1,\sigma_{t_2},\mathscr{G}^{(t_1)})
	\nonumber\\
	&&
	\, + \,
	\Delta \psi (t_1,\sigma_{t_2}\rightarrow\sigma_{t_1}, \mathscr{G}^{(t_1)}) \, ,
\end{eqnarray}
where the three terms on the r.h.s. correspond to
\begin{itemize}
\item[I)] changing the partition $\mathscr{G}^{(t_2)}\rightarrow \mathscr{G}^{(t_1)}$ in $\psi_{h,\Lambdas }(t_2)$:
\begin{eqnarray}\label{eq-II-69bis}
	\lefteqn{
	\Delta \psi (t_2,\sigma_{t_2},\mathscr{G}^{(t_2)}\rightarrow \mathscr{G}^{(t_1)})
	}
	\nonumber\\
	&&:= \; \; 
	e^{iHt_2}\sum_{j=1}^{N(t_1)}\cW_{\Lambdas,\sigma_{t_2}}(\vv_j,t_2)
	\, e^{i\gamma_{\sigma_{t_2}}(\vv_j,\vnabla E_{\vec{P}}^{\sigma_{t_2}},t_2)}
	e^{-iE_{\vec{P}}^{\sigma_{t_2}}t_2}
	\, \psi_{j,\sigma_{t_2}}^{(t_1)}
	\\
	&& \; \; \, - \; e^{iHt_2}\sum_{j=1}^{N(t_1)} \sum_{l(j)}\cW_{\Lambdas,\sigma_{t_2}}(\vv_{l(j)},t_2)
	e^{i\gamma_{\sigma_{t_2}}(\vv_{l(j)},	
	\vnabla E_{\vec{P}}^{\sigma_{t_2}},t_2)}
	\, e^{-iE_{\vec{P}}^{\sigma_{t_2}}t_2} \, \psi_{l(j),\sigma_{t_2}}^{(t_2)}
	\, ,\nonumber
\end{eqnarray}
where $l(j)$ labels all cells of
$\mathscr{G}^{(t_2)}$ contained in the $j$-th cell $\mathscr{G}^{(t_1)}_j$
of $\mathscr{G}^{(t_1)}$. Moreover,
$$
	\vv_{l(j)}\equiv\vnabla
  	E_{\vec{P}_{l(j)}^*}^{\sigma_{t_2}}
  	\; \; \; \; {\rm and} \; \; \; \;
  	\vv_{j}\equiv\vnabla
  	E_{\vec{P}_{j}^*}^{\sigma_{t_1}} \, ;
$$
\item[II)] subsequently changing the time, $t_2\rightarrow t_1$, for the fixed partition $\mathscr{G}^{(t_1)}$,
and  the fixed infrared cut-off $\sigma_{t_2}$:
\begin{eqnarray}\label{eq-II-70bis}
	\lefteqn{\Delta \psi (t_2\rightarrow t_1,\sigma_{t_2},\mathscr{G}^{(t_1)})
	}
	\nonumber\\
	&&:= \; \;
	e^{iHt_1}\sum_{j=1}^{N(t_1)}\cW_{\Lambdas,\sigma_{t_2}}(\vv_j,t_1)
	\, e^{i\gamma_{\sigma_{t_2}}(\vv_j,\vnabla E_{\vec{P}}^{\sigma_{t_2}},t_1)}
  	\, e^{-iE_{\vec{P}}^{\sigma_{t_2}}t_1} \, \psi_{j,\sigma_{t_2}}^{(t_1)}\\
	&& \; \; \,  - \; e^{iHt_2}\sum_{j=1}^{N(t_1)}\cW_{\Lambdas,\sigma_{t_2}}(\vv_j,t_2)
	\, e^{i\gamma_{\sigma_{t_2}}(\vv_j,\vnabla E_{\vec{P}}^{\sigma_{t_2}},t_2)}
  	\,e^{-iE_{\vec{P}}^{\sigma_{t_2}}t_2} \, \psi_{j,\sigma_{t_2}}^{(t_1)} \nonumber\,;
\end{eqnarray}
and, finally,
\item[III)] shifting the infrared cut-off, $\sigma_{t_2}\rightarrow\sigma_{t_1}$:
\begin{eqnarray}\label{eq-II-71bis}
	\lefteqn{\Delta \psi (t_1,\sigma_{t_2}\rightarrow\sigma_{t_1}, \mathscr{G}^{(t_1)})
	}
	\nonumber\\
	&&:= \; \;
	e^{iHt_1}\sum_{j=1}^{N(t_1)}\cW_{\Lambdas,\sigma_{t_1}}(\vv_j,t_1)e^{i\gamma_{\sigma_{t_1}}(\vv_j,\vnabla
  	E_{\vec{P}}^{\sigma_{t_1}},t_1)}e^{-iE_{\vec{P}}^{\sigma_{t_1}}t_1}\psi_{j,\sigma_{t_1}}^{(t_1)}
	\\
	&& \; \; \, - \; e^{iHt_1}\sum_{j=1}^{N(t_1)}\cW_{\Lambdas,\sigma_{t_2}}(\vv_j,t_1)e^{i\gamma_{\sigma_{t_2}}(\vv_j,\vnabla
  	E_{\vec{P}}^{\sigma_{t_2}},t_1)}e^{-iE_{\vec{P}}^{\sigma_{t_2}}t_1}\psi_{j,\sigma_{t_2}}^{(t_1)}
	\nonumber\,.
\end{eqnarray}
\end{itemize}
It is important to take these three steps in the order indicated above.

In step I), the size of
$\|\Delta \psi (t_2,\sigma_{t_2},\mathscr{G}^{(t_2)}\rightarrow \mathscr{G}^{(t_1)})\|$
in (\ref{eq-II-69bis}) is controlled as follows: The sum of   off-diagonal
terms yields a subleading contribution. The diagonal terms are shown to tend to 0 by controlling
the differences
$$\vv_{l(j)}-\vv_{j} \,.$$

In step II), Cook's argument, combined with the cancelation of an infrared tail
(as in the mechanism described above), yields the desired decay in $t_1$.

Step III) is  more involved.
But the basic idea is quite simple to grasp: It consists in rewriting
\begin{equation}\label{eq-II-83bis}
	e^{iHt_1}\sum_{j=1}^{N(t_1)} \, \cW_{\Lambdas,\sigma_{t_2}}(\vv_j,t_1)
	\, e^{i\gamma_{\sigma_{t_2}}(\vv_j,\vnabla E_{\vec{P}}^{\sigma_{t_2}},t_1)}
  	\, e^{-iE_{\vec{P}}^{\sigma_{t_2}}t_1}
	\, \psi_{j,\sigma_{t_2}}^{(t_1)}
\end{equation}
as
\begin{equation}\label{eq-II-84bis}
 	e^{iHt_1}\sum_{j=1}^{N(t_1)} 
	\, \cW_{\Lambdas,\sigma_{t_2}}(\vv_j,t_1)
	\, W_{\sigma_{t_2}}^{*}(\vnabla E^{\sigma_{t_2}}_{\vP})
	\, W_{\sigma_{t_2}}(\vnabla E^{\sigma_{t_2}}_{\vP})
	\, e^{i\gamma_{\sigma_{t_2}}(\vv_j,\vnabla E_{\vec{P}}^{\sigma_{t_2}},t_1)}
	\, e^{-iE_{\vec{P}}^{\sigma_{t_2}}t_1}
	\, \psi_{j,\sigma_{t_2}}^{(t_1)}\,.
\end{equation}
The term in (\ref{eq-II-84bis}) corresponding to the cell $\mathscr{G}^{(t_1)}_j$
of $\mathscr{G}^{(t_1)}$ can then be obtained by acting with the  ``{\em dressing operator}"
\begin{equation}\label{eq-I-100}
	e^{iHt_1}\cW_{\Lambdas,\sigma_{t_2}}(\vv_j,t_1)
	\, W_{\sigma_{t_2}}^{*}(\vnabla E^{\sigma_{t_2}}_{\vP})
	\, e^{-iE_{\vec{P}}^{\sigma_{t_2}}t_1}\,,
\end{equation}
on the ``{\em infrared-regular}'' vector
\begin{equation}\label{eq-II-86bis}
e^{i\gamma_{\sigma_{t_2}}(\vv_j,\vnabla E_{\vec{P}}^{\sigma_{t_2}},t_1)}\Phi^{(t_1)}_{j,\sigma_{t_2}}:=e^{i\gamma_{\sigma_{t_2}}(\vv_j,\vnabla E_{\vec{P}}^{\sigma_{t_2}},t_1)}\int_{\mathscr{G}^{(t_1)}_j}h(\vec{P})\Phi_{\vec{P}}^{\sigma_{t_2}}d^3P
\end{equation}
corresponding to the vectors $\Phi_{\vec{P}}^{\sigma}=W_{\sigma}(\vnabla
E^{\sigma}_{\vP})\Psi_{\vec{P}}^{\sigma}$ (see (\ref{eq-II-14})), for all $j$.
The advantage of (\ref{eq-II-84bis}) over (\ref{eq-II-83bis}) is that the vector
$\Phi_{j,\sigma_{t_2}}^{(t_1)}$ inherits  the regularity properties
of $\Phi_{\vec{P}}^{\sigma}$ described in Theorem {\ref{thm-cfp-2}}.
In particular, the vectors $\Phi_{j,\sigma_{t_2}}^{(t_1)}$ converge strongly, as
$\sigma_{t_2}\to0$, and the vector
\begin{equation}
    e^{-i\vq\cdot\vx}
    \Phi^{(t_1)}_{j,\sigma_{t_2}}
\end{equation}
depends on $\vq$ in a strongly H\"older continuous manner, uniformly in $\sigma_{t_2}$. This last
property entails enough decay to offset various logarithmic divergences
appearing in
the removal of the infrared cut-off in the dressing operator
(\ref{eq-I-100}).

Our analysis of the strong convergence of the sequence of approximating vectors
culminates in the estimate
\begin{equation}\label{eq-II-72bis}
	\|\psi_{h,\Lambdas }(t_2)-\psi_{h,\Lambdas }(t_1)\|
	\, \leq \, \cO\big((\ln(t_2))^2 /t_1^{\rho}\big) \, ,
\end{equation}
for some $\rho>0$.
By telescoping, this bound suffices to prove Theorem {\ref{theo-II.2}}. Indeed, to estimate the difference
between the two vectors at times $t_2$ and $t_1$, respectively, where $ t_2>t_1>1$, 
we may consider a sequence of times
$\{t_1^2,...,t_1^n\}$, such that $t_1^n<t_2<t_1^{n+1}$, and use Estimate (\ref{eq-II-72bis}) for
each difference 
\begin{eqnarray}
& &\psi_{h,\Lambdas }(t_2)-\psi_{h,\Lambdas }(t_1^{n})\,, \\
& &\psi_{h,\Lambdas }(t_1^m)-\psi_{h,\Lambdas }(t_1^{m-1})\,,\quad 2\leq m\leq n \,.
\end{eqnarray}
Then, one can show that there exists a constant  $\rho'>0$ such that the rate
of convergence of the time-dependent vector is at least of order
$t^{-\rho'}$, as stated in Theorem \ref{theo-II.2}.

\subsection{A space of scattering states}
\label{ssec-scatt-ssp-1}

We use the asymptotic states
$\psi_{h,\Lambdas }^{(out/in)}$ to construct a
subspace, $\cHs_{\Lambdas }$, of scattering states invariant under space-time
translations, and with a photon energy threshold $\Lambdas $,
\begin{equation}
	\cHs_{\Lambdas }:=\overline{\Big\lbrace
  	\bigvee\,\psi_{h,\Lambdas }^{out/in}(\tau,\vec{a})\,: \, h(\vP)\in\, C_{0}^{1}
  	(\cS\setminus\mathcal{B}_{r_{\alpha}})\,,\,\tau\in\RR,\,\vec{a}\in\RR^3\Big\rbrace}\,,
	\label{eq-III-8}
\end{equation}
where
\begin{equation}
\psi_{h,\Lambdas }^{out/in}(\tau,\vec{a})\equiv e^{-i\vec{a}\cdot\vec{P}}e^{-iH\tau}\psi_{h,\Lambdas }^{out}\,.
\end{equation}
This space contains
states describing an asymptotically freely moving electron,
accompanied by asymptotic free photons with energy smaller than $\Lambdas $.

Spaces of scattering states are obtained from the space
$\cHs_{\Lambdas }$ by adding ``hard'' photons, i.e.,
\begin{equation}\label{eq-III-17}
\cH^{out/in}:=\overline{
  \Big\{\bigvee \,\psi_{h,\vec{F}}^{out/in}\,: \, h(\vP)\in\, C_{0}^{1}(\cS\setminus\mathcal{B}_{r_{\alpha}})\,,\,
  \hat{\vF}\in\, C_{0}^{\infty}(\RR^3\setminus0\,;\,\CC^3) \, \Big\} } \,,
\end{equation}
where
\begin{equation}\label{eq-III-14}
\psi_{h,\vec{F}}^{out/in}:=s-\lim_{t\to
  +/-\infty}e^{i\big(\vA[\vF_{t},t]-\vA[\overline{\vF_t},t]\big)}\psi_{h,\Lambdas }(t)\,,
\end{equation}
and
\begin{equation}
\vA[\vF_{t},t]\,:= \, i\int\big(\vA(t,\vy)\cdot\frac{\partial\vF_{t}(\vy)}{\partial t}-
	\frac{\partial\vA(t,\vy)}{\partial t}\cdot\vF_{t}(\vy)\big)d^3y
\end{equation}
is the L.S.Z. photon field smeared out with the vector test function
\begin{equation}\label{eq-III-11}
\vF_{t}(\vy):=\sum_{\lambda = \pm}\int \frac{d^3k}{(2\pi)^{3}2\sqrt{|k|}} \,
\veps^{\;*}_{\vk, \lambda}\hat{F}^{\lambda}(\vk) \, e^{-i|k|t+i\vk \cdot \vy}
\end{equation}
 with
\begin{equation}
\hat{\vF}(\vk):=\sum_{\lambda}\veps^{\;*}_{\vk, \lambda}\hat{F}^{\lambda}(\vk) \; \; \in \; \;
C_0^{\infty}(\RR^3\backslash\lbrace0\rbrace\, ; \, \CC^3)\,.
\end{equation}

An \emph{a posteriori} physical interpretation of the scattering states constructed here
emerges by studying how certain
algebras of asymptotic operators are represented on the
spaces of scattering states:
\begin{itemize}
\item
The Weyl algebra, $\cA^{out/in}_{ph}$, associated with the asymptotic
electromagnetic field.
\item
The algebra $\cA^{out/in}_{el}$ generated by smooth functions of compact support
of the asymptotic velocity of the electron.
\end{itemize}
These algebras will be defined in terms of
the limits
(\ref{eq-III-19bis}) and (\ref{eq-III-21bis}), below, whose existence is established in Section
{\ref{ssec-III.2}}.
\begin{theorem}
Functions $f\in C_{0}^{\infty}(\RR^3)$, of the variable $e^{iHt}\,\frac{\vec{x}}{t}\,e^{-iHt}$, have
strong limits, as $t\to\pm\infty$, as operators acting on $\cH^{out/in}$,
\begin{equation}\label{eq-III-19bis}
	s-\lim_{t\to +/-\infty}e^{iHt}\,f(\frac{\vx}{t})\,e^{-iHt}\psi_{h,\vec{F}}^{out/in}
	\, =: \, \psi_{f_{\vnabla E}   \, h\, , \, \vec{F}}^{out/in}
\end{equation}
where $f_{\vnabla E}(\vP):=\lim_{\sigma\to 0}f(\vnabla E_{\vP}^{\sigma})$.
\end{theorem}

\begin{theorem}
The LSZ Weyl operators
\begin{equation}\label{eq-III-20bis}
	\big\lbrace \, e^{i\big(\vA[\vG_{t},t]-\vA[\overline{\vG_t},t]\big)}\,:\,
	\hat{G}^{\lambda}(\vk)\in L^{2}(\RR^3,(1+|\vk|^{-1})d^3k)\,,\,\lambda \, = \, \pm \, \big\rbrace \, ,
\end{equation}
have strong limits in $\cH^{out/in}$; i.e.,
\begin{equation}\label{eq-III-21bis}
	\cW^{out/in}(\vec{G}) \, := \, s-\lim_{t\to+/-\infty} e^{i\big(\vA[\vG_{t},t]-\vA[\overline{\vG_t},t]\big)}\,.
\end{equation}
exists.

The limiting operators are unitary and satisfy the following properties:
\begin{itemize}
\item[i)]
\begin{equation}\label{eq-III-22bis}
	\cW^{out/in}(\vec{G}) \, \cW^{out/in}(\vec{G}') \, = \,
	\cW^{out/in}(\vec{G}+\vec{G}') \, e^{-\frac{\rho(\vec{G},\vec{G}')}{2}} \, ,
\end{equation}
where
\begin{equation}
	\rho(\vec{G},\vec{G}') \, = \, 2i \,
	Im \, \big( \, \sum_{\lambda}\int\hat{G}^{\lambda}(\vk)\overline{\hat{G'}^{\lambda}(\vk)}
	\, d^3k \, \big) \, .
\end{equation}
\item[ii)]
The mapping $\RR\ni s\longrightarrow\cW^{out/in}(s\,\vec{G})$ defines a strongly
continuous one-parameter group of unitary operators.
\item[iii)]
\begin{equation}\label{eq-III-23bis}
	e^{iH\tau} \, \cW^{out/in}(\vec{G}) \, e^{-iH\tau} \, = \, \cW^{out/in}(\vec{G}_{-\tau})
\end{equation}
where $\vec{G}_{-\tau}$ is the freely time-evolved (vector) test function
at time $-\tau$.
\end{itemize}
\end{theorem}
The two algebras, $\cA^{out/in}_{ph}$ and $\cA^{out/in}_{el}$,  commute.
This is the precise mathematical expression of the
asymptotic decoupling of the dynamics of photons from the one of the electron.
The proof is non-trivial, because non-Fock representations of the 
asymptotic photon creation- and annihilation operators appear. (The 
representation of $\cA^{out/in}_{ph}$, which is {\em non-Fock} but {\em locally} Fock; see
Section {\ref{ssec-III.2}}). We will show that
\begin{equation}\label{eq-III-27bis}
	\bra \, \psi_{h,\Lambdas }^{out/in}\,,\,\cW^{out}(\vec{G})\psi_{h,\Lambdas }^{out/in} \, \ket \, = \,
	\int e^{-\frac{\|\vec{G}\|_2^2}{2}}e^{\varrho_{\vnabla E_{\vec{P}}}(\vec{G})}
	\, | \, h(\vP) \, |^2 \, d^3P\,,
\end{equation}
where
\begin{equation}\label{eq-III-28bis}
 \| \, \vec{G} \|_2 \,=\, \big( \, \int|\vec{G}(\vk)|^2d^3k \, \big)^{1/2}\,,
\end{equation}
and
\begin{equation}\label{eq-III-29bis}
	\varrho_{\vec{u}}(\vec{G})\,:=\,2iRe\big(\alpha^{\frac{1}{2}}\sum_{\lambda}
	\int_{\mathcal{B}_{\Lambdas }}\hat{G}^{\lambda}(\vk)\frac{\vec{u}\cdot
	\veps_{\vk,\lambda}^{\;*}}{|\vec{k}|^{\frac{3}{2}}(1-\vec{u}\cdot\hat{k})}d^3k\big)\,.
\end{equation}
More precisely, the representation of $\cA^{out/in}_{ph}$ on the space of scattering states
can be decomposed in a direct integral of inequivalent irreducible representations labelled by
the asymptotic velocity of the electron. For different values
of the asymptotic velocity,  these representations turn
out to be inequivalent. Only for a vanishing electron velocity, the representation is
Fock; for non-zero velocity, it is a coherent non-Fock representation.
The coherent photon cloud,
labeled by the asymptotic velocity, is the one anticipated by Bloch and Nordsieck.

These results can be interpreted as follows: 
{\em In every scattering state, an asymptotically
freely moving electron is observed} 
(with an asymptotic velocity whose size is
strictly smaller than the speed of light, by construction)
{\em accompanied by a cloud of asymptotic photons
propagating along the light cone}.
\\

\noindent
\emph{\bf{Remark:}}
We point out that, in our definition of scattering states, we can directly accommodate an
arbitrarily large number of ``hard'' photons without energy
restriction, i.e., we can construct the limiting vector
\begin{equation}\label{eq-III-18}
\vA^{out}[\vF^{(m)}]\dots\vA^{out}[\vF^{(1)}]\psi_{h,\Lambdas }^{out} \,:= \,
s-\lim_{t\to +\infty}\vA[\vF_{t}^{(m)},t]\dots\vA[\vF_{t}^{(1)},t]\psi_{h,\Lambdas }(t)
\end{equation}
which represents the  state $\psi_{h,\Lambdas }^{out}$ plus $m$ asymptotic
photons with wave functions $\vF^{(m)},\dots,\vF^{(1)}$, respectively.
Analogously, we define
\begin{equation}\label{eq-III-18in}
\vA^{in}[\vG^{(m')}]\dots\vA^{in}[\vG^{(1)}]\psi_{h,\Lambdas }^{in} \,:= \,
s-\lim_{t\to -\infty}\vA[\vG_{t}^{(m')},t]
\dots\vA[\vG_{t}^{(1)},t]\psi_{h,\Lambdas }(t)
\end{equation}
 This is possible because, apart from some \emph{higher order estimates} to
 control the commutator $i[H\,,\,\vx]$, and the photon creation operators
 in (\ref{eq-III-18}) (see for example \cite{FGS}), we use the propagation estimate
(\ref{eq-II-49}), which only limits the asymptotic
velocity of the electron. This fact is very important for estimating
scattering amplitudes involving an arbitrary number of ``hard'' photons.

In particular, for any $m,m'\in\NN$, we can define the S-matrix element
\begin{equation}
	S_\alpha^{m,m'} (\, \{\vF_i\} \,, \, \{ \vG_j\} \,)
	\, = \, 
	\Big( \, \vA^{out}[\vF^{(m)}]\dots\vA^{out}[\vF^{(1)}]\psi_{h,\Lambdas^{out} }^{out} \, , \, 
	\vA^{in}[\vG^{(m')}]\dots\vA^{in}[\vG^{(1)}]\psi_{h,\Lambdas^{in} }^{in}	\, \Big)
\end{equation}
which corresponds to the transition amplitude between two states describing an
incoming electron with wave function $h^{in}$, accompanied by a soft photon cloud of
free photons of energy smaller than $\Lambdas ^{in}$, plus $m'$ hard photons (with wave functions $\vG^{(1)}, \dots ,\vG^{(m')}$), and
an outgoing electron with wave function $h^{out}$ and soft photon energy threshold
$\Lambdas ^{out}$, plus $m$ hard photons, respectively.

The expansion of $S_\alpha^{m,m'} (\, \{\vF_i\} \,, \, \{ \vG_j\} \,)$ in
the finestructure constant $\alpha$ can be carried out, at least to leading order,
along the lines of \cite{BFP}. 
This yields a rigorous proof of the transition amplitudes for Compton scattering 
in leading order, and in the non-relativistic approximation, that one can find in textbooks.
\\

Moreover, as expected from classical electromagnetism,
``close" to the electron a Li\'enard-Wiechert electromagnetic field is observed.
The precise mathematical statement is
\begin{eqnarray}\label{eq-III-72}
    &\lim_{|\vec{d}|\to\infty} \, \lim_{t\to\pm\infty} \, |\vec{d}|^2&
    \,\Big\{\Big\langle \, \psi_{h,\vF}^{out/in}
    \, , \, e^{iHt}\int d^3y \, F_{\mu\nu}(0,\vy) \,
    \tilde{\delta}_\Lambda(\vy-\vx-\vec{d}) \, e^{-iHt}\psi_{h,\vF}^{out/in} \, \Big\rangle
    \nonumber\\
    & &\quad \quad - \int F_{\mu\nu}^{\vnabla E_{\vP}}(0,\vec{d})
    \langle \, \psi_{\vP}^{\sigma_t} \, , \, \psi_{\vP}^{\sigma_t} \, \rangle \,
    |h(\vP)|^2d^3P\Big\} \, = \, 0 \, ,
\end{eqnarray}
where $\tilde{\delta}_\Lambda$ is a smooth, $\Lambda$-dependent delta function
which has the property that its Fourier transform is supported in $B_\Lambda$,
$\vx$ is the electron position,
\begin{equation}\label{eq-III-72bis}
	F_{\mu\nu} \, = \, \partial_\mu A_\nu \, - \, \partial_\nu A_\mu \,
\end{equation}
with
\begin{eqnarray}
	A_0(0,\vy) & := &- \, \frac{ (2\pi)^{2}\alpha^{1/2}}{|\vy - \vx|}
	\\
	A_i(0,\vy) & := & - \, \sum_\lambda \int \frac{d^3k}{\sqrt{ |\vk|}}\Big(
	\, (\vec\epsilon_{\vk,\lambda})^i \, e^{-i\vk\cdot\vy} a_{\vk,\lambda}^*
	\, + \, (\vec\epsilon_{\vk,\lambda}^{\;*})^i \, e^{i\vk\cdot\vy} a_{\vk,\lambda} \, \Big) \, ,
\end{eqnarray}
and $F_{\mu\nu}^{\vnabla E_{\vP}} $ is the electromagnetic field tensor corresponding to a
Li\'enard-Wiechert solution for the current
\begin{equation}
    J_{\mu}(t,\vy) \, := \, \big( \, -2(2\pi)^{3}\alpha^{\frac{1}{2}}\delta^{(3)}(\vy-\vnabla E_{\vP} t)\,,\,
    2(2\pi)^{3}\alpha^{\frac{1}{2}} \, \vnabla E_{\vP} \, \delta^{(3)}(\vy-\vnabla E_{\vP}  t) \, \big)
    \; \; \; , \; \; \;
    |\vnabla E_{\vP}| \, < \, 1 \,;
\end{equation}
see the discussion in Section {\ref{sec-intro}}.

\newpage

\section{Uniform boundedness of the limiting norm}
\label{sec-II.2}
\resetequ

Our first aim is to prove the {\em uniform boundedness} of
$\|\psi_{h,\Lambdas }(t)\|$, as $t\to\infty$; more precisely, that
\begin{equation}\label{eq-II-21}
	\lim_{t\to\infty}\bra \, \psi_{h,\Lambdas }(t)\,,\,\psi_{h,\Lambdas }(t) \, \ket \, = \,
	\int |h(\vec{P})|^2d^3P \, .
\end{equation}
The sum of the diagonal terms -- with
respect to the partition $\mathscr{G}^{(t)}$ introduced above -- is easily seen to yield
$\int |h(\vec{P})|^2d^3P$ in the limit $t\rightarrow\infty$, as one expects.
Thus, our main task is to show that the sum of the off-diagonal
terms vanishes in this limit.

In Section {\ref{sec-II.3}}, we prove that the norm-bounded sequence $\{\psi_{h,\Lambdas }(t)\}$
is, in fact, Cauchy.
\\
 
We recall that the definition of the vector
$\psi_{h,\Lambdas }(t)$ involves three different rates:
\begin{itemize}
\item
The rate $t^{-\beta}$ related to the fast infrared cut-off
$\sigma_t$;
\item
the rate $t^{-\theta}$ of the slow infrared cut-off
$\sigma_t^{S}$ (see (\ref{eq-II-18}));
\item
the rate $t^{-\epsilon}$ of refinement of the cell partitions $\mathscr{G}^{(t)}$.
\\
\end{itemize}

\subsection{Control of the off-diagonal terms}
\label{sec-II.2.1}

We denote the off-diagonal term labeled by the pair $(l,j)$ of cell indices $l\neq j$
contributing to the l.h.s. of (\ref{eq-II-21}) by
\begin{equation}\label{eq-II-22}
    	M_{l,j}(t) \, := \,
    	\Bra \, e^{i\gamma_{\sigma_t}(\vv_l,\nabla E_{\vec{P}}^{\sigma_t},t)}
	\, e^{-iE_{\vec{P}}^{\sigma_t}t}\psi_{l,\sigma_t}^{(t)}
	\, , \,
    	\cW_{\Lambdas,\sigma_t,l,j}(t)e^{i\gamma_{\sigma_t}(\vv_j,\nabla E_{\vec{P}}^{\sigma_t},t)}
    	\, e^{-iE_{\vec{P}}^{\sigma_t}t}
	\, \psi_{j,\sigma_t}^{(t)} \, \Ket
\end{equation}
where we use the notation
\begin{equation}\label{eq-II-23}
    	\cW_{\Lambdas,\sigma_t,l,j}(t) \, := \,
    	\exp\Big( \, \alpha^{\frac{1}{2}}\sum_{\lambda}
	\, \int_{\mathcal{B}_{\Lambdas }\setminus\mathcal{B}_{\sigma_t}}
	\vec{\eta}_{l,j}(\vk)\cdot
	\big\lbrace \, \veps_{\vk, \lambda}
	\, a_{\vk,\lambda}^*e^{-i|\vk|t}-\veps_{\vk,\lambda}^{\;*} \, a_{\vk, \lambda}	
	\, e^{i|\vk|t} \, \big\rbrace \, d^3k \, \Big)\,,
\end{equation}
where $\vv_j\equiv \vnabla E^{\sigma_t}(\vP_j^*)$, $\vP_j^*$ being the center of the cell $\mathscr{G}_j^{(t)}$, and
\begin{equation}\label{eq-II-24}
    \vec{\eta}_{l,j}(\vk) \, := \, \alpha^{\frac{1}{2}}\frac{\vv_j}{|\vk|^{\frac{3}{2}}
    (1-\widehat{k}\cdot\vv_j)}-\alpha^{\frac{1}{2}}\frac{\vv_l}{|\vk|^{\frac{3}{2}}(1-\widehat{k}\cdot\vv_l)}\,.
\end{equation}
We study the limit $t\to +\infty$;
the case $t\rightarrow-\infty$ is analogous.
\\

\noindent
{\underline{\emph{A. Asymptotic orthogonality}}}
\\
\\
In order to prove that the off-diagonal terms (IV.2) vanish in the  
limit $ t \to ë\infty$, we
separate the role played by the time variable $t$  as the parameter  
determining the
dynamical cell decomposition and infrared cutoffs, from its usual role  
as the conjugate
variable to the energy. For the latter, we introduce an auxiliary  
variable $s\geq t$.
Then, for fixed  $t $ (such that the cell decomposition and the cutoffs  
are constant),
we interpret the terms (IV.2) as special values $M_{l,j}(t)=M_{l,j}^1(t,t)$
of families $M_{l,j}^\mu(t,s)$ introduced
below, which depend on $t, s$, and an additional auxiliary parameter $\mu 
\in\RR$.
Our strategy will be based on proving that the dispersive properties of
$M_{l,j}^\mu(t,s)$ as a function of $s\geq t$ alone, for fixed $t$ and $\mu$,  
imply that $M_{l,j}(t)$
has a sufficiently fast decay in $t$ such that our desired result of
asymptotic orthogonality follows. 

\noindent
More precisely, we introduce  a family of operators
\begin{equation}\label{eq-II-26}
    \cW_{\Lambdas,\sigma_t,l,j}^{\mu}(s) \, := \,
    \exp \Big(\mu\,\alpha^{\frac{1}{2}}\sum_{\lambda}\int_{\mathcal{B}_{\Lambdas }\setminus\mathcal{B}_{\sigma_t}}
    \vec{\eta}_{l,j}(\vk)\cdot\lbrace\veps_{\vk, \lambda}a_{\vk,
    \lambda}^*e^{-i|\vk|s}-\veps_{\vk,\lambda}^{\;*}a_{\vk, \lambda}e^{i|\vk|s}\rbrace d^3k\Big)
\end{equation}
depending on a parameter $\mu\in\RR$, and define
\begin{equation}\label{eq-II-25}
    \widehat{M}_{l,j}^{\mu}(t,s) \, := \,
    \Bra \, e^{i\gamma_{\sigma_t}(\vv_l,\nabla
    E_{\vec{P}}^{\sigma_t},s)}e^{-iE_{\vec{P}}^{\sigma_t}s}\psi_{l,\sigma_t}^{(t)}\,,\,
    \cW_{\Lambdas,\sigma_t,l,j}^{\mu}(s)e^{i\gamma_{\sigma_t}(\vv_j,\nabla
    E_{\vec{P}}^{\sigma_t},s)}e^{-iE_{\vec{P}}^{\sigma_t}s}\psi_{j,\sigma_t}^{(t)} \, \Ket
\end{equation}
for $s\geq t(\gg  1)$.
Obviously, $\widehat M_{l,j}^{\mu=1}(t,t)=M_{l,j}(t)$.

The phase factor $\gamma_{\sigma_t}(\vv_j,\nabla E_{\vec{P}}^{\sigma_t},s)$ 
is chosen as follows:
\begin{eqnarray}\label{eq-II-27} 
	\gamma_{\sigma_t}(\vv_j,\nabla
    	E_{\vec{P}}^{\sigma_t},s) 
	\, := \, -\alpha\,\int_{1}^{s}\vnabla
    	E_{\vec{P}}^{\sigma_t}\cdot\int_{\mathcal{B}_{\sigma_\tau^{S}}
    	\setminus\mathcal{B}_{\sigma_t}}\vec{\Sigma}_{\vv_j}(\vk)\cos(\vk\cdot\vnabla
    	E_{\vec{P}}^{\sigma_t}\tau-|\vk|\tau)d^3kd\tau 
\end{eqnarray}
for $s^{-\theta}\geq\sigma_t$, and
\begin{eqnarray}\label{eq-II-28} 
	\gamma_{\sigma_t}(\vv_j,\nabla
    	E_{\vec{P}}^{\sigma_t},s) 
	\, := \, - \, \alpha\,\int_{1}^{\sigma_t^{-\frac{1}{\theta}}}\vnabla E_{\vec{P}}^{\sigma_t}
    	\cdot\int_{\mathcal{B}_{\sigma_\tau^{S}}\setminus\mathcal{B}_{\sigma_t}}\vec{\Sigma}_{\vv_j}(\vk)
    	\cos(\vk\cdot\vnabla E_{\vec{P}}^{\sigma_t}\tau-|\vk|\tau)d^3kd\tau\,, 
\end{eqnarray}
for $s^{-\theta}<\sigma_t$.
As a function of $\mu$, the scalar product in  (\ref{eq-II-25}) satisfies the ordinary
differential equation
\begin{equation}\label{eq-II-29}
    \frac{d\widehat{M}_{l,j}^{\mu}(t,s)}{d\mu} \, = \, - \, \mu \,C_{l,j,\sigma_t}\,\widehat{M}_{l,j}^{\mu}(t,s)+
    r^{\mu}_{\sigma_t}(t,s)\,,
\end{equation}
where
\begin{equation}\label{eq-II-30}
    	C_{l,j,\sigma_t} \, := \,
	\int_{\mathcal{B}_{\Lambdas }\setminus\mathcal{B}_{\sigma_t}}|\vec{\eta}_{l,j}(\vk)|^2d^3k\,,
\end{equation}
and
\begin{eqnarray}\label{eq-II-31}
    r^{\mu}_{\sigma_t}(t,s)
    &:=&- \, \Bra \, e^{i\gamma_{\sigma_t}(\vv_l,\vnabla
    E_{\vec{P}}^{\sigma_t},s)}e^{-iE_{\vec{P}}^{\sigma_t}s}\psi_{l,\sigma_t}^{(t)}\,,\,
    \nonumber\\
    &&\quad\quad\quad\quad\quad\quad
    \cW_{\Lambdas,\sigma_t,l,j}^{\mu}(s)\,\vec{a}_{\sigma_t}(\overline{\vec{\eta}_{l,j}})(s)\,
    e^{i\gamma_{\sigma_t}(\vv_j,\vnabla E_{\vec{P}}^{\sigma_t},s)}
    e^{-iE_{\vec{P}}^{\sigma_t}s}\psi_{j,\sigma_t}^{(t)} \, \Ket
    \nonumber\\
    & &+ \, \Bra \, \vec{a}_{\sigma_t}(\overline{\vec{\eta}_{l,j}})(s)e^{i\gamma_{\sigma_t}
    (\vv_l,\vnabla E_{\vec{P}}^{\sigma_t},s)}e^{-iE_{\vec{P}}^{\sigma_t}s}\psi_{l,\sigma_t}^{(t)}\,,\,
    \nonumber\\
    &&\quad\quad\quad\quad\quad\quad \cW_{\Lambdas,\sigma_t,l,j}^{\mu}(s)e^{i\gamma_{\sigma_t}(\vv_j,\vnabla
    E_{\vec{P}}^{\sigma_t},s)}e^{-iE_{\vec{P}}^{\sigma_t}s}\psi_{j,\sigma_t}^{(t)} \, \Ket
    \,,\nonumber
\end{eqnarray}
with
\begin{equation}\label{eq-II-32}
    \vec{a}_{\sigma_t}(\overline{\vec{\eta}_{l,j}})(s)
	\, := \, \sum_{\lambda}
    \int_{\mathcal{B}_{\Lambdas }\setminus\mathcal{B}_{\sigma_t}}
    \vec{\eta}_{l,j}(\vk)\cdot\veps_{\vk, \lambda}^{\;*}a_{\vk,
    \lambda} \, e^{i|\vk|s} \, d^3k\,.
\end{equation}
The solution of the ODE (\ref{eq-II-29})  is given by
\begin{equation}\label{eq-II-33}
	\widehat{M}_{l,j}^{\mu}(t,s) \, = \,
	e^{-\frac{C_{l,j,\sigma_t}}{2}\mu^2}
	\, \widehat{M}_{l,j}^{0}(t,s)
	\, + \,
	\int_{0}^{\mu}r^{\mu'}_{\sigma_t}(t,s)
	\, e^{-\frac{C_{l,j,\sigma_t}}{2}(\mu^2-\mu'^2)} \, d\mu'\,,
\end{equation}
where the initial condition at $\mu=0$ is given by
\begin{equation}\label{eq-II-34}
	\widehat{M}_{l,j}^{0}(t,s) \, = \, 0\,,
\end{equation}
since the supports in $\vec{P}$ of the two vectors $\psi_{l,\sigma_t}^{(t)}$,
$\psi_{j,\sigma_t}^{(t)}$ are disjoint (up to sets of measure $0$), for arbitrary $t$ and $s$.

Furthermore, condition $(\mathscr{I}4)$ in Theorem {\ref{thm-cfp-2}} implies that the vectors $\psi_{l,\sigma_t}^{(t)}$,
$\psi_{j,\sigma_t}^{(t)}$ are vacua for the annihilation part of the
asymptotic photon field
\footnote{The existence of the asymptotic field operator for a fixed cut-off dynamics is
derived as explained in part \underline{\em B.} of this section.}
under the dynamics generated by the Hamiltonian $H^{\sigma_t}$.  As a consequence, we find that
\begin{equation}\label{eq-II-35}
\lim_{s\to +\infty}\,r^{\mu}_{\sigma_t}(t,s)=0\,,
\end{equation}
for fixed $\mu$ and $t$.
To arrive at this conclusion, the following is used: The one-particle state, multiplied by 
the phase $e^{i\gamma_{\sigma_t}(\vv_j,\vnabla
    E_{\vec{P}}^{\sigma_t},s)}$ continues to be a one-particle 
state for the Hamiltonian $H_{\sigma_t}$; for large $s$ (see (\ref{eq-II-28})) the phase is $s$-independent; the operator $E_{\vP}^{\sigma_t}$ coincides with the operator $H^{\sigma_t}$ when applied to one-particle states of the Hamiltonian $H^{\sigma_t}$.

Therefore, by dominated convergence, it follows that
\begin{equation}\label{eq-II-36}
	\lim_{s\to +\infty}\,\widehat{M}_{l,j}^{1}(t,s)=0\,.
\end{equation}
Since $\widehat{M}_{l,j}^{1}(t,t)\equiv M_{l,j}(t)$,  we have
\begin{equation}\label{eq-II-37}
	|M_{l,j}(t)| \, \leq \, \int_{t}^{+\infty}
	\, \big|\frac{d}{ds}\widehat{M}_{l,j}^{1}(t,s)\big| \, ds \,.
\end{equation}
To estimate $|\frac{d}{ds}\widehat{M}_{l,j}^{1}(t,s)|$ (see (\ref{eq-II-25})),
we proceed as follows.

Since we are interested in the limit $t\rightarrow+\infty$,
and the integration domain on the r.h.s. of (\ref{eq-II-37}) is $[t,+\infty)$,
our aim is to show that
\begin{equation}\label{eq-II-38}
    \Big| \, \frac{d}{ds}\Big(e^{iH^{\sigma_t}s}\cW_{\Lambdas,\sigma_t}(\vv_j,s)e^{i\gamma_{\sigma_t}(\vv_j,\nabla
    E_{\vec{P}}^{\sigma_t},s)}e^{-iE_{\vec{P}}^{\sigma_t}s}\psi_{j,\sigma_t}^{(t)}\Big) \, \Big|
\end{equation}
is integrable in $s$, and that the rate at which the time integral in (\ref{eq-II-37})
converges to zero offsets the growth of the number of cells in
the partition. This allows us to conclude that
\begin{equation}\label{eq-II-39}
	\sum_{l,j\,(l\not=j)} M_{l,j}(t) \, \longrightarrow \, 0
\end{equation}
in the limit $t\to+\infty$, and,  as a corollary,
\begin{equation}\label{eq-II-68}
	\lim_{t\to+\infty}\sum_{l,j}^{N(t)}M_{l,j}(t) \, = \, \int |h(\vec{P})|^2d^3P \, ,
\end{equation}
as asserted in Theorem {\ref{theo-II.2}}.
The convergence (\ref{eq-II-39}) follows from the following theorem.

\begin{theorem}
\label{thm:II-1}
The off-diagonal terms $M_{l,j}(t)$, $l\not=j$, satisfy
\begin{equation}\label{eq-II-67}
    | \, M_{l,j}(t) \, | \, \leq \, C \, \frac{1}{t^{\eta}}\,|\ln\,\sigma_t|^{2}\,t^{-3\epsilon}   \,,
\end{equation}
for some constants $C<\infty$ and $\eta>0$, both independent of $l$, $j$, and $\epsilon>0$. In particular, $M_{l,j}(t)\rightarrow0$, as $t\to+\infty$.
\end{theorem}
As a corollary, we find
\begin{equation}\label{eq-Mljsum-est-1}
    \sum_{1\leq l\neq j\leq N(t)} | \, M_{l,j}(t) \, | \, \leq \,
    C \, N^2(t) \, \frac{1}{t^{\eta}}\,|\ln\,\sigma_t|^{2}\,t^{-3\epsilon}
    \, \leq \, C' \,  \frac{1}{t^{\eta}}\,|\ln\,\sigma_t|^{2}\,t^{3\epsilon}
\end{equation}
since $N(t)\approx t^{3\epsilon}$.
We conclude that, for $\epsilon<\frac{\eta}{4}$,
(\ref{eq-II-39}) follows.
\\
\\

\noindent
%%%%%%%%%%%%%%%%%%%%%%%%%%%%%%%%%%%%%%%%%%%%%%%%%%%%%%%%%%%%%%%%%%%%%%%%%%%%%%
{\underline{\emph{B. Time derivative and infrared tail}}}
%%%%%%%%%%%%%%%%%%%%%%%%%%%%%%%%%%%%%%%%%%%%%%%%%%%%%%%%%%%%%%%%%%%%%%%%%%%%%%
\\
\\
We now proceed to prove Theorem {\ref{thm:II-1}}.
The arguments developed here will
also be relevant for the proof of the (strong) convergence of the vectors
$\psi_{h,\Lambdas }(t)$, as $t\to+\infty$, which we discuss in Section {\ref{sec-II.3}}.

To control (\ref{eq-II-37}), we focus on the derivative
\begin{eqnarray}\label{eq-II-40}
	\lefteqn{\frac{d}{ds}\Big(\,e^{iH^{\sigma_t}s}\cW_{\Lambdas,\sigma_t}(\vv_j,s)
	\, e^{i\gamma_{\sigma_t}(\vv_j,\vnabla E_{\vec{P}}^{\sigma_t},s)}
	\, e^{-iE_{\vec{P}}^{\sigma_t}s}
	\, \psi_{j,\sigma_t}^{(t)}\,\Big)\, }\\
	&= &i \, e^{iH^{\sigma_t}s}[H^{\sigma_t}_I,\cW_{\Lambdas,\sigma_t}(\vv_j,s)]
	\, e^{i\gamma_{\sigma_t}(\vv_j,\vnabla E_{\vec{P}}^{\sigma_t},s)}
	\, e^{-iE_{\vec{P}}^{\sigma_t}s} 
	\, \psi_{j,\sigma_t}^{(t)}
	\label{eq-II-41} \\
    & &+ \, i \, e^{iH^{\sigma_t}s} \, \cW_{\Lambdas,\sigma_t}(\vv_j,s) 
    \,\frac{d\gamma_{\sigma_t}(\vv_j,\vnabla E_{\vec{P}}^{\sigma_t},s)}{ds}
    \, e^{i\gamma_{\sigma_t}(\vv_j,\vnabla E_{\vec{P}}^{\sigma_t},s)}
    \, e^{-iE_{\vec{P}}^{\sigma_t}s}\psi_{j,\sigma_t}^{(t)} \,, \quad
    \quad\label{eq-II-42}
\end{eqnarray}
where
\begin{equation}\label{eq-II-43}
    H^{\sigma_t}_I:=\alpha^{\frac{1}{2}}\,\vp\cdot\vA^{\sigma_t}(\vx)+
    \alpha\,\frac{\vA^{\sigma_t}(\vx)\cdot\vA^{\sigma_t}(\vx)}{2}\,.
\end{equation}
We have used that $\cW_{\Lambdas,\sigma_t}(\vv_j,s)=e^{-isH_0}\cW_{\Lambdas,\sigma_t}(\vv_j,0)e^{isH_0}$,
where $H_0:=H^{\sigma_t}-H^{\sigma_t}_I$ is the free Hamiltonian,
to obtain the commutator in (\ref{eq-II-41}).
We rewrite the latter in the form
\begin{equation}
	[ \, H^{\sigma_t}_I \, , \, \cW_{\Lambdas,\sigma_t}(\vv_j,s) \, ] \, = \, \cW_{\Lambdas,\sigma_t}(\vv_j,s)
	\, \Big( \, \cW_{\Lambdas,\sigma_t}(\vv_j,s)^* H^{\sigma_t}_I\cW_{\Lambdas,\sigma_t}(\vv_j,s) 
	\, - \, H^{\sigma_t}_I \, \Big)  \, ,
\end{equation}
and use that
\begin{equation}
	\cW_{\Lambdas,\sigma_t}(\vv_j,s)^* \, \vA^{\sigma_t}(\vx) \, \cW_{\Lambdas,\sigma_t}(\vv_j,s)
	\, = \, \vA^{\sigma_t}(\vx) 
	\, + \, \alpha^{\frac12} \int_{\mathcal{B}_{\Lambdas }\setminus\mathcal{B}_{\sigma_t}} 
	\, \vec{\Sigma}_{\vv_j}(\vk)
	\, \cos(\vk\cdot\vx-|\vk|s) \, d^3k 
\end{equation}
(see (\ref{eq:Bog-trsf-b-1})), where $\Sigma^{l}_{\vv_j}(\vk)$ is defined in (\ref{eq-II-19}).
We can then write the term (\ref{eq-II-41}) as
\begin{eqnarray} 
    (\ref{eq-II-41}) 
    &=& \; \; \; 
    i \, e^{iH^{\sigma_t}s}\cW_{\Lambdas,\sigma_t}(\vv_j,s)\alpha\,i[H^{\sigma_t},\vx]\cdot
    \int_{\mathcal{B}_{\Lambdas }\setminus\mathcal{B}_{\sigma_t}} d^3k \, \vec{\Sigma}_{\vv_j}(\vk)
    \times\label{eq-II-44} \\
    & &\quad\quad \quad \times \cos(\vk\cdot\vx-|\vk|s)
    \; e^{-iE_{\vec{P}}^{\sigma_t}s}e^{i\gamma_{\sigma_t}(\vv_j,\vnabla
    E_{\vec{P}}^{\sigma_t},s)}\psi_{j,\sigma_t}^{(t)}\nonumber\\
    & &+ \; 
    i \, e^{iH^{\sigma_t}s}\cW_{\Lambdas,\sigma_t}(\vv_j,s)\frac{\alpha^{2}}{2}
    \Big( \int_{\mathcal{B}_{\Lambdas }
    \setminus\mathcal{B}_{\sigma_t}}\vec{\Sigma}_{\vv_j}(\vk)\cos(\vk\cdot\vx-|\vk|s)d^3k \, \Big)^2 
    \label{eq-II-45} \\
    & &\quad\quad \quad  
    \times e^{i\gamma_{\sigma_t}(\vv_j,\vnabla
    E_{\vec{P}}^{\sigma_t},s)}e^{-iE_{\vec{P}}^{\sigma_t}s}\psi_{j,\sigma_t}^{(t)}\,,\nonumber
\end{eqnarray}
where we recall that $i[H^{\sigma_t},\vx]=\vp+\alpha^{\frac12}\vA^{\sigma_t}(\vx)$; see (\ref{eq:velop-def}).

>From the decay estimates provided by Lemma {\ref{lemm-A2}} in the Appendix
one concludes that the norm of (\ref{eq-II-45})  is integrable in
$s$, and that
\begin{equation}\label{eq-II-45bis}
	\int_t^\infty ds \, \| \, (\ref{eq-II-45}) \, \| \, \leq \,
    \frac{1}{t^{\eta}}\, | \, \ln\,\sigma_t \, |^{2} \, t^{-\frac{3\epsilon}{2}}\,
\end{equation}
for some $\eta>0$ independent of $\epsilon$.

The analysis of (\ref{eq-II-44}) is more involved.
Our argument will eventually
involve the derivative of the phase factor in  (\ref{eq-II-42}).
To begin with, we write (\ref{eq-II-44}) as
\begin{eqnarray}
    \lefteqn{
	(\ref{eq-II-44}) \, = \,
    i \, e^{iH^{\sigma_{t}s}}\cW_{\Lambdas,\sigma_t}(\vv_j,s)\alpha\,i[H^{\sigma_t},\vx]\cdot
    \frac{1}{H^{\sigma_t}+i}\times\label{eq-II-46} \quad
    }
    \\
    & & \times(H^{\sigma_t}+i)\int_{\mathcal{B}_{\Lambdas }
    \setminus\mathcal{B}_{\sigma_t}} d^3k \, \vec{\Sigma}_{\vv_j}(\vk)
    \cos(\vk\cdot\vx-|\vk|s) \; e^{-iE_{\vec{P}}^{\sigma_t}s}
    e^{i\gamma_{\sigma_t}(\vv_j,\vnabla E_{\vec{P}}^{\sigma_t},s)}
    \psi_{j,\sigma_t}^{(t)}\,.\nonumber
\end{eqnarray}
Pulling the operator $(H^{\sigma_t}+i)$ through to the right, the vector (\ref{eq-II-46})
splits into the sum of a term involving the commutator $[H^{\sigma_t},\vx]$, 
\begin{eqnarray}
    \lefteqn{
    i \, e^{iH^{\sigma_{t}s}}\cW_{\Lambdas,\sigma_t}(\vv_j,s)\alpha\,i[H^{\sigma_t},\vx]\cdot
    \frac{1}{H^{\sigma_t}+i}\times\quad
    }\label{eq-II-47} \\
    & &\quad \times\int_{\mathcal{B}_{\Lambdas }\setminus\mathcal{B}_{\sigma_t}}
    \,d^3k \, \vec{\Sigma}_{\vv_j}(\vk)\,[ \, H^{\sigma_t} \, , \, \cos(\vk\cdot\vx-|\vk|s) \, ] \;
    e^{i\gamma_{\sigma_t}(\vv_j,\vnabla
    E_{\vec{P}}^{\sigma_t},s)} e^{-iE_{\vec{P}}^{\sigma_t}s}\psi_{j,\sigma_t}^{(t)}\,,\nonumber
\end{eqnarray}
and
\begin{eqnarray}\label{eq-II-48}
    \lefteqn{
    i \, e^{iH^{\sigma_t}s}\cW_{\Lambdas,\sigma_t}(\vv_j,s)\alpha\,i[H^{\sigma_t},\vx]\cdot
    \int_{\mathcal{B}_{\Lambdas }\setminus\mathcal{B}_{\sigma_t}}
    d^3k \, \vec{\Sigma}_{\vv_j}(\vk)\times
    }
    \\
    & &\quad\times\frac{1}{H^{\sigma_t}+i}\cos(\vk\cdot\vx-|\vk|s)
    e^{i\gamma_{\sigma_t}(\vv_j,\vnabla
    E_{\vec{P}}^{\sigma_t},s)}e^{-iE_{\vec{P}}^{\sigma_{t}}s}(E_{\vec{P}}^{\sigma_t}+i)\psi_{j,\sigma_t}^{(t)}\,.
    \nonumber
\end{eqnarray}
Note that $[H^{\sigma_t},\vec{x}]\frac{1}{H^{\sigma_t}+i}$ is bounded in the operator norm,
uniformly in $\sigma_t$.
To control (\ref{eq-II-47}) and (\ref{eq-II-48}),
we invoke a propagation estimate for the electron position operator as follows.
Due to condition $(\mathscr{I}3)$ in Theorem {\ref{thm-cfp-2}}, we can introduce a $C^{\infty}-$function
$\chi_h(\vec{y})$, $\vec{y}\in\RR^3$, such that
\begin{itemize}
\item%[i)]
$\chi_h(\vnabla E^{\sigma_t}_{\vP})\equiv 1$ for
$\vP\in\text{supp}\,h$,  uniformly in $\sigma_t$.\\
\item%[ii)]
$\chi_h(\vec{y})=0$ for
$|\vec{y}|<\nu_{min}$ and $|\vec{y}|>\nu_{max}\,$.
\end{itemize}
It is shown in Theorem {\ref{theo-A'2}} of the Appendix
that, for $\theta<1$ sufficiently close to 1 and $s$ large, the propagation estimate
\begin{eqnarray}
\lefteqn{
    \Big\| \, \chi_h(\frac{\vx}{s})e^{i\gamma_{\sigma_t}(\vv_j,\vnabla
  E_{\vec{P}}^{\sigma_t},s)}e^{-iE_{\vec{P}}^{\sigma_t}s}\psi_{j,\sigma_t}^{(t)}
  }
  \nonumber\\
  &&\quad\quad\quad\quad 
  \quad\quad - \, \chi_h(\vnabla E^{\sigma_t}_{\vP})e^{i\gamma_{\sigma_t}(\vv_j,\vnabla
  E_{\vec{P}}^{\sigma_t},s)}e^{-iE_{\vec{P}}^{\sigma_t}s}\psi_{j,\sigma_t}^{(t)}\, \Big\|
  \label{eq-II-49bis}\\ 
 &&\quad\quad\quad\quad\quad\quad
    \leq \, \frac{1}{s^{\nu}}\,\frac{1}{t^{\frac{3\epsilon}{2}}}\,|\ln(\sigma_t)|\,,
    \nonumber
\end{eqnarray}
holds, where $\nu>0$ is independent of $\epsilon$.
The argument uses 
the H\"older regularity of $\vnabla E^\sigma_{\vP}$
and of $\Phi_{\vP}^\sigma$
listed under properties $(\mathscr{I}2)$ in Theorem {\ref{thm-cfp-2}}, 
differentiability of $h(\vP)$, 
and (\ref{eq-II-9}).

We continue with the discussion of the expressions (\ref{eq-II-47}) and (\ref{eq-II-48}).
We split (\ref{eq-II-48}) into two parts 
\begin{eqnarray}
& &
    i \, e^{iH^{\sigma_t}s} \, \cW_{\Lambdas,\sigma_t}(\vv_j,s) \, \cJ|_{\sigma^{S}_s}^{\Lambdas }(s)\,
    e^{i\gamma_{\sigma_t}(\vv_j,\vnabla E_{\vec{P}}^{\sigma_t},s)}
    \, e^{-iE_{\vec{P}}^{\sigma_t}s}(E_{\vec{P}}^{\sigma_{t}}+i)
    \, \psi_{j,\sigma_t}^{(t)} \\
    &+&i \, e^{iH^{\sigma_{t}}s}\cW_{\Lambdas,\sigma_t}(\vv_j,s)\,\cJ|_{\sigma_{t}}^{\sigma^{S}_s}(s)\,
    \, e^{i\gamma_{\sigma_t}(\vv_j,\vnabla E_{\vec{P}}^{\sigma_t},s)}
    \, e^{-iE_{\vec{P}}^{\sigma_{t}}s}(E_{\vec{P}}^{\sigma_t}+i) \, \psi_{j,\sigma_t}^{(t)}
\end{eqnarray}
using the definitions
\begin{eqnarray}
	\cJ|_{\sigma^{S}_s}^{\Lambdas }(s) & := &  
	\alpha\,i[H^{\sigma_t},\vx]\cdot
	\frac{1}{H^{\sigma_t}+i}\int_{\mathcal{B}_{\Lambdas }\setminus\mathcal{B}_{\sigma_s^{S}}}
	\vec{\Sigma}_{\vv_j}(\vk) \,\cos(\vk\cdot\vx-|\vk|s)d^3k 
	\; \; \; {\rm if \;} s^{-\theta} \, \geq \, \sigma_{t}
	\nonumber \\
	& := & \alpha\,i[H^{\sigma_t},\vx]\cdot
    \int_{\mathcal{B}_{\Lambdas }\setminus\mathcal{B}_{\sigma_t}}
    d^3k \, \vec{\Sigma}_{\vv_j}(\vk)\,\frac{1}{H^{\sigma_t}+i}\cos(\vk\cdot\vx-|\vk|s)
	\; \; \; {\rm if \; } s^{-\theta} \, < \, \sigma_{t} \,, \quad \quad\quad
	\label{eq-II-51}
\end{eqnarray}
and
\begin{eqnarray}\label{eq-II-52}
	\cJ|_{\sigma_{t}}^{\sigma^{S}_s}(s) & := &  
	\alpha\,i[H^{\sigma_t},\vx]\cdot
	\frac{1}{H^{\sigma_t}+i}\int_{\mathcal{B}_{\sigma_s^{S}}\setminus\mathcal{B}_{\sigma_t}}
	\vec{\Sigma}_{\vv_j}(\vk) \,\cos(\vk\cdot\vx-|\vk|s)d^3k 
	\; \; \; {\rm if \; } s^{-\theta} \, \geq \, \sigma_t 
	\nonumber\\
	& := & 0 
	\; \; \; {\rm if \; } s^{-\theta} \, < \, \sigma_{t} \,, 
\end{eqnarray}
where we refer to $\sigma_{s}^S:=s^{-\theta}$ as the {\em slow infrared cut-off}.

To control $\cJ|_{\sigma_{t}}^{\sigma^{S}_s}(s)$ in (\ref{eq-II-52}), we define the {\em ``infrared tail''}
\begin{eqnarray} \label{eq-II-52bis} 
    \frac{d\widehat{\gamma}_{\sigma_t}(\vv_j,\frac{\vx}{s},s)}{ds} & := &
	\alpha \, e^{-iH^{\sigma_t}s} \, \frac{1}{H^{\sigma_t}+i}
    \, \frac{d( e^{iH^{\sigma_t}s} \vec{x}_{h}(s) e^{- iH^{\sigma_t}s} )}{ds} \, e^{iH^{\sigma_t}s} \, \cdot
   \\
    && \quad \quad \cdot \, \int_{\mathcal{B}_{\sigma_s^{S}}\setminus\mathcal{B}_{\sigma_t}}
    \vec{\Sigma}_{\vv_j}(\vk) \, \cos(\vk\cdot\vnabla E_\vP^{\sigma_t}s-|\vk|s) \, d^3k
    \; \; \; {\rm if \; } s^{-\theta} \, \geq \, \sigma_t \,,
    \nonumber\\
     & := & 0 \; \; \; {\rm if \; } s^{-\theta} \, < \, \sigma_t
     \nonumber
\end{eqnarray}
where $\vec{x}_{h}(s):= \vx\chi_{h}(\frac{\vx}{s})$.
Summarizing, we can write (\ref{eq-II-40}) as
\begin{eqnarray}\label{eq-II-53}
	\lefteqn{
	 (\ref{eq-II-40}) \, = \,  (\ref{eq-II-45}) \, +
	}
	\nonumber\\
    & + &
    i \, e^{iH^{\sigma_t}s} \, \cW_{\Lambdas,\sigma_t}(\vv_j,s) \, \cJ|_{\sigma^{S}_s}^{\Lambdas }(s)\,
    e^{i\gamma_{\sigma_t}(\vv_j,\vnabla E_{\vec{P}}^{\sigma_t},s)}
    \, e^{-iE_{\vec{P}}^{\sigma_t}s}(E_{\vec{P}}^{\sigma_{t}}+i)
    \, \psi_{j,\sigma_t}^{(t)}\label{eq-II-54} \\
    &+&i \, e^{iH^{\sigma_{t}}s}\cW_{\Lambdas,\sigma_t}(\vv_j,s)\,\cJ|_{\sigma_{t}}^{\sigma^{S}_s}(s)\,
    \, e^{i\gamma_{\sigma_t}(\vv_j,\vnabla E_{\vec{P}}^{\sigma_t},s)}
    \, e^{-iE_{\vec{P}}^{\sigma_{t}}s}(E_{\vec{P}}^{\sigma_t}+i) \, \psi_{j,\sigma_t}^{(t)}
    \label{eq-II-55} \\
    & +&i \, e^{iH^{\sigma_t}s}\cW_{\Lambdas,\sigma_t}(\vv_j,s)\,\frac{d\gamma_{\sigma_t}(\vv_j,\vnabla
    E_{\vec{P}}^{\sigma_t},s)}{ds} \, e^{i\gamma_{\sigma_t}(\vv_j,\nabla
    E_{\vec{P}}^{\sigma_t},s)}
    \, e^{-iE_{\vec{P}}^{\sigma_t}s} \, \psi_{j,\sigma_t}^{(t)}\quad\quad\quad
    \label{eq-II-56} \\
    &+&i \, e^{iH^{\sigma_{t}}s} \, \cW_{\Lambdas,\sigma_t}(\vv_j,s) \, \alpha \, i
	\, [H^{\sigma_t},\vx]\cdot\frac{1}{H^{\sigma_t}+i}
    \times\quad\label{eq-II-57} \\
    & &\quad \times\int_{\mathcal{B}_{\Lambdas }\setminus\mathcal{B}_{\sigma_t}}
    d^3k \, \vec{\Sigma}_{\vv_j}(\vk) \, [H^{\sigma_t},\cos(\vk\cdot\vx-|\vk|s)] \,
    e^{i\gamma_{\sigma_t}(\vv_j,\vnabla E_{\vec{P}}^{\sigma_t},s)} \,
    e^{-iE_{\vec{P}}^{\sigma_t}s} \, \psi_{j,\sigma_t}^{(t)}\,,\nonumber
\end{eqnarray}
where we recall that (\ref{eq-II-45}) satisfies (\ref{eq-II-45bis}). 
We claim that
\begin{equation}\label{eq-II-58}
	\Big\| \, \int_t^{\infty}[(\ref{eq-II-54})+(\ref{eq-II-55})+(\ref{eq-II-56})]\,ds \, \Big\|
	\, \leq \, \frac{1}{t^{\eta}}\,|\ln\,\sigma_t|^{2}\,t^{-3\epsilon}\,,
\end{equation}
for some $\eta>0$ depending on $\theta$, but independent of $\epsilon$.
This is obtained from
\begin{eqnarray}
	\lefteqn{
	\Big\| \, \int_t^{+\infty}[(\ref{eq-II-54})+(\ref{eq-II-55})+(\ref{eq-II-56})]\,ds \, \Big\|
	}
	\label{eq-II-59} \\
	&&\leq \; \;
	\int_t^{+\infty} ds
    \, \Big\| \, \alpha \, i[H^{\sigma_t},\vx] \cdot \frac{1}{H^{\sigma_t}+i}
	\int_{\mathcal{B}_{\Lambdas }\setminus\mathcal{B}_{\sigma_t}}
	d^3k \, \vec{\Sigma}_{\vv_j}(\vk) \, \cos(\vk\cdot\vx-|\vk|s)\times\quad\quad\nonumber \\
	& &\quad \quad \quad \quad \times
    \, \Big[ \, \chi_h(\vnabla E^{\sigma_t}_{\vP})-\chi_h(\frac{\vx}{s}) \, \Big]
	\, e^{i\gamma_{\sigma_t}(\vv_j,\vnabla E_{\vec{P}}^{\sigma_t},s)}
    \, e^{-iE_{\vec{P}}^{\sigma_t}s}
    \, (E_{\vec{P}}^{\sigma_{t}}+i)
    \, \psi_{j,\sigma_t}^{(t)} \, \Big\| \,
    \label{eq-II-60}
	\\
	&  & \; \; + \; \int_t^{+\infty} ds \, \Big\| \,
	\cJ|_{\sigma^{S}_s}^{\Lambdas }(s)\chi_h(\frac{\vx}{s})e^{i\gamma_{\sigma_t}(\vv_j,\vnabla
  	E_{\vec{P}}^{\sigma_t},s)}e^{-iE_{\vec{P}}^{\sigma_t}s}(E_{\vec{P}}^{\sigma_{t}}+i)
	\, \psi_{j,\sigma_t}^{(t)}
  	\, \Big\| \,
  	\label{eq-II-61}
	\\
	&  & \; \; + \; \Big\|\int_t^{+\infty}ds
    \, e^{i H^{\sigma_t} s} \, \mathcal{W}_{\Lambdas,\sigma_t}(\vec v_j, s)
    \, \Big[ \, \cJ|_{\sigma_{t}}^{\sigma^{S}_s}(s)\chi_h(\frac{\vx}{s}) -
	\frac{d\widehat{\gamma}_{\sigma_t}(\vv_j,\frac{\vx}{s},s)}{ds} \, \Big] \, \times
	\label{eq-II-62}
	\\
	& &\quad\quad\quad\quad\quad\quad\quad\quad\quad\times
	\, e^{i\gamma_{\sigma_t}(\vv_j,\vnabla E_{\vec{P}}^{\sigma_t},s)}
	\, e^{-iE_{\vec{P}}^{\sigma_t}s}
  	\, (E_{\vec{P}}^{\sigma_{t}}+i)
	\, \psi_{j,\sigma_t}^{(t)} \, \Big\|\nonumber\\
	&  & \; \; + \; \Big\| \, \int_t^{+\infty} ds
    \, e^{i H^{\sigma_t} s} \, \mathcal{W}_{\Lambdas,\sigma_t}(\vec v_j, s)
    \, \Big[\frac{d\gamma_{\sigma_t}(\vv_j,\vnabla E_{\vec{P}}^{\sigma_t},s)}{ds}-
	\frac{d\widehat{\gamma}_{\sigma_t}(\vv_j,\frac{\vx}{s},s)}{ds}\Big]
	\times\quad\quad\quad
	\label{eq-II-63} \\
	& &\quad\quad\quad\quad\quad\quad\quad\quad\quad\times
	\, e^{i\gamma_{\sigma_t}(\vv_j,\vnabla E_{\vec{P}}^{\sigma_t},s)}
	\, e^{-iE_{\vec{P}}^{\sigma_t}s}
  	\, (E_{\vec{P}}^{\sigma_{t}}+i) \, \psi_{j,\sigma_t}^{(t)} \, \Big\|
    \nonumber 
\end{eqnarray}
using the following arguments:

\begin{itemize}
\item
The term  (\ref{eq-II-60}) can be bounded from above by 
$\frac{1}{t^{\eta}}\,|\ln\,\sigma_t|^{2}\,t^{-3\epsilon}$,
for some $\eta>0$ independent of $\epsilon$,
due to the propagation
estimate for (\ref{eq-II-49}) and Lemma {\ref{lemm-A2}}, 
which show that the integrand has a sufficiently strong decay in $s$.

\item
In (\ref{eq-II-61}), the slow cut-off $\sigma_{s}^{S}$ and the function
$\chi_h(\frac{\vx}{s})$ make the norm integrable in $s$ with the desired rate (i.e., to get a bound as in (\ref{eq-II-45bis})),
for a suitable choice of $\theta<1$. In particular, we can
exploit that
\begin{equation}\label{eq-II-65}
	\sup_{\vx\in\RR^3}\, \Big| \, \int_{\mathcal{B}_{\Lambdas }\setminus\mathcal{B}_{\sigma_{s}^{S}}} d^3k \,
	\vec{\Sigma}(\vk,\vv_j)\cos(\vk\cdot\vx-|\vk|s)\,\chi_h(\frac{\vx}{s}) \, \Big|
	\, \leq \, \cO(\frac{s^{\theta}}{s^2})\,,
\end{equation}
see Lemma~\ref{lemm-A2} in the Appendix.

\item
In (\ref{eq-II-62}), only terms integrable in $s$ and decaying fast enough to satisfy
the bound (\ref{eq-II-45bis}) are left after subtracting
\begin{equation}\label{eq-II-66}
\frac{d\widehat{\gamma}_{\sigma_t}(\vv_j,\frac{\vx}{s},s)}{ds}\,
\end{equation}
from $\cJ|_{\sigma_{t}}^{\sigma^{S}_s}(s)$.
This is explained in detail in the proof of Theorem~\ref{theo-A4} in the Appendix.

\item
To bound (\ref{eq-II-63}), we use the electron propagation estimate, 
combined with an integration by parts, to show that the
derivative of the phase factor tends to the ``infrared tail'' for large $s$, at an integrable rate that provides a bound as in (\ref{eq-II-45bis}).
We note that due to the vector interaction in non-relativistic QED,
this argument is more complicated here than in the Nelson model treated
in \cite{Pizzo2005} where the interaction term is scalar.
Here, we have to show (see Theorem~\ref{theo-A4})
that, in the integral with respect to $s$,
the pointwise velocity $e^{iH^{\sigma_t}s}i[\vx,H^{\sigma_t}]e^{-iH^{\sigma_t}s}$ can be  replaced by the
mean velocity $\vnabla E_{\vP}^{\sigma_t}$ at asymptotic times.
 
\end{itemize}

Finally, to control (\ref{eq-II-57}), we observe that the commutator introduces additional
decay in $s$ into the integrand when $\frac{\vx}{s}$ is restricted to
the support of $\chi_h$. It then follows that the propagation estimate suffices (without infrared
tail) to control the norm, by the same arguments that were applied to $\cJ|_{\sigma^{S}_s}^{\Lambdas }(s)$
in (\ref{eq-II-60}), (\ref{eq-II-61}).

Combining the above arguments, the proof of Theorem {\ref{thm:II-1}} is completed.
\\

\newpage
%%%%%%%%%%%%%%%%%%%%%%%%%%%%%%%%%%%%%%%%%%%%%%%%%%%%%%%%%%%%%%%%%%%%%%%%%%%%%%%%%%%%%%%%%%%%%%%%%%%%%
%%%%%%%%%%%%%%%%%%%%%%%%%%%%%%%%%%%%%%%%%%%%%%%%%%%%%%%%%%%%%%%%%%%%%%%%% C O N V E R G E N C E %%%%%
%%%%%%%%%%%%%%%%%%%%%%%%%%%%%%%%%%%%%%%%%%%%%%%%%%%%%%%%%%%%%%%%%%%%%%%%%%%%%%%%%%%%%%%%%%%%%%%%%%%%%
%%%%%%%%%%%%%%%%%%%%%%%%%%%%%%%%%%%%%%%%%%%%%%%%%%%%%%%%%%%%%%%%%%%%%%%%%%%%%%%%%%%%%%%%%%%%%%%%

\section{Proof of convergence of $\psi_{h,\Lambdas }(t)$}
\label{sec-II.3}
\resetequ

In this section, we prove that $\psi_{h,\Lambdas }(t)$
defines a bounded Cauchy sequence in $\cH$, as $t\to\infty$.
To this end, it is necessary
to control the norm difference between vectors $\psi_{h,\Lambdas }(t_i)$, $i=1,2$,
at times $t_2>t_1$.

\subsection{Three key steps}

As anticipated in Section {\ref{sec:convscheme}}, we decompose the difference
of $\psi_{h,\Lambdas }(t_1)$ and $\psi_{h,\Lambdas }(t_2)$
into three terms
\begin{eqnarray}
 	\psi_{h,\Lambdas }(t_1) - \psi_{h,\Lambdas }(t_2) &=&
	\Delta \psi (t_2,\sigma_{t_2},\mathscr{G}^{(t_2)}\rightarrow \mathscr{G}^{(t_1)})
	\, + \,
	\Delta \psi (t_2\rightarrow t_1,\sigma_{t_2},\mathscr{G}^{(t_1)})
	\nonumber\\
	&&
	\, + \,
	\Delta \psi (t_1,\sigma_{t_2}\rightarrow\sigma_{t_1}, \mathscr{G}^{(t_1)}) \, ,
\end{eqnarray}
where we recall from (\ref{eq-II-69bis}) -- (\ref{eq-II-71bis}):

\begin{itemize}
\item[I)] The term
\begin{eqnarray}\label{eq-II-69}
    \lefteqn{
	\Delta \psi (t_2,\sigma_{t_2},\mathscr{G}^{(t_2)}\rightarrow \mathscr{G}^{(t_1)})
	}
	\nonumber\\
	&& = \; \; 
	e^{iHt_2}\sum_{j=1}^{N(t_1)}\cW_{\Lambdas,\sigma_{t_2}}(\vv_j,t_2)
	\, e^{i\gamma_{\sigma_{t_2}}(\vv_j,\vnabla E_{\vec{P}}^{\sigma_{t_2}},t_2)}
	e^{-iE_{\vec{P}}^{\sigma_{t_2}}t_2}
	\, \psi_{j,\sigma_{t_2}}^{(t_1)}
	\\
	&& \; \; - \; e^{iHt_2}\sum_{j=1}^{N(t_1)} \sum_{l(j)}\cW_{\Lambdas,\sigma_{t_2}}(\vv_{l(j)},t_2)
	e^{i\gamma_{\sigma_{t_2}}(\vv_{l(j)},	
	\vnabla E_{\vec{P}}^{\sigma_{t_2}},t_2)}
	\, e^{-iE_{\vec{P}}^{\sigma_{t_2}}t_2} \, \psi_{l(j),\sigma_{t_2}}^{(t_2)}
	\, ,\nonumber
\end{eqnarray}
accounts for the change of the partition  $\mathscr{G}^{(t_2)}\rightarrow \mathscr{G}^{(t_1)}$
in $\psi_{h,\Lambdas }(t_2)$, where $l(j)$ labels the sub-cells  belonging to the sub-partition
$\mathscr{G}^{(t_2)}\cap\mathscr{G}^{(t_1)}_j$ of $\mathscr{G}^{(t_1)}_j$, and where we define
$$
    \vv_{l(j)}\equiv\vnabla
    E_{\vec{P}_{l(j)}^*}^{\sigma_{t_2}}
    \; \; \; \; {\rm and} \; \; \; \;
    \vv_{j}\equiv\vnabla
    E_{\vec{P}_{j}^*}^{\sigma_{t_1}} \, ;
$$
\item[II)] the term
\begin{eqnarray}\label{eq-II-70}
	\lefteqn{\Delta \psi (t_2\rightarrow t_1,\sigma_{t_2},\mathscr{G}^{(t_1)})
	}
	\nonumber\\
	&& = \; \;
	e^{iHt_1}\sum_{j=1}^{N(t_1)}\cW_{\Lambdas,\sigma_{t_2}}(\vv_j,t_1)
	\, e^{i\gamma_{\sigma_{t_2}}(\vv_j,\vnabla E_{\vec{P}}^{\sigma_{t_2}},t_1)}
  	\, e^{-iE_{\vec{P}}^{\sigma_{t_2}}t_1} \, \psi_{j,\sigma_{t_2}}^{(t_1)}\\
	& & \; \; - \; e^{iHt_2}\sum_{j=1}^{N(t_1)}\cW_{\Lambdas,\sigma_{t_2}}(\vv_j,t_2)
	\, e^{i\gamma_{\sigma_{t_2}}(\vv_j,\vnabla E_{\vec{P}}^{\sigma_{t_2}},t_2)}
  	\,e^{-iE_{\vec{P}}^{\sigma_{t_2}}t_2} \, \psi_{j,\sigma_{t_2}}^{(t_1)} \nonumber \,,
\end{eqnarray}
accounts for the subsequent change of the time variable, $t_2\rightarrow t_1$, for the fixed
partition $\mathscr{G}^{(t_1)}$, and the fixed infrared cut-off $\sigma_{t_2}$; and finally,
\item[III)] the term
\begin{eqnarray}\label{eq-II-71}
	\lefteqn{\Delta \psi (t_1,\sigma_{t_2}\rightarrow\sigma_{t_1}, \mathscr{G}^{(t_1)})
	}
	\nonumber\\
	&& = \; \;
	e^{iHt_1}\sum_{j=1}^{N(t_1)}\cW_{\Lambdas,\sigma_{t_1}}(\vv_j,t_1)e^{i\gamma_{\sigma_{t_1}}(\vv_j,\vnabla
  	E_{\vec{P}}^{\sigma_{t_1}},t_1)}e^{-iE_{\vec{P}}^{\sigma_{t_1}}t_1}\psi_{j,\sigma_{t_1}}^{(t_1)}
	\\
	& & \; \; - \; 
	e^{iHt_1}\sum_{j=1}^{N(t_1)}\cW_{\Lambdas,\sigma_{t_2}}(\vv_j,t_1)e^{i\gamma_{\sigma_{t_2}}(\vv_j,\vnabla
  	E_{\vec{P}}^{\sigma_{t_2}},t_1)}e^{-iE_{\vec{P}}^{\sigma_{t_2}}t_1}\psi_{j,\sigma_{t_2}}^{(t_1)}
	\nonumber \,
\end{eqnarray}
accounts for the change of the infrared cut-off, $\sigma_{t_2}\rightarrow\sigma_{t_1}$.
\end{itemize}
Our goal is to prove
\begin{equation}\label{eq-II-72}
    \| \, \psi_{h,\Lambdas }(t_2)-\psi_{h,\Lambdas }(t_1) \, \| \, \leq \, \cO\big((\ln(t_2))^2 /t_1^{\rho}\big) \, ,
\end{equation}
for some $\rho>0$.
To this end, it is necessary to perform the three steps in the order displayed above.
As a corollary of the bound (\ref{eq-II-72}), we obtain Theorem {\ref{theo-II.2}}
by telescoping (see the comment after Eq. (\ref{eq-II-72bis})).

The arguments in our proof are very similar to those in
\cite{Pizzo2005}, but a number of modifications are necessary because of the vector nature of
the QED interaction. For these modifications, we provide detailed explanations.

%%%%%%%%%%%%%%%%%%%%%%%%%%%%%%%%%%%%%%%%%%%%%%%%%%%%%%%%%%%%%%%%%%%%%%%%%%%%%%%%%%%%%%%%%%%%%%%%
%%%%%%%%%%%%%%%%%%%%%%%%%%%%%%%%%%%%%%%%%%%%%%%%%%%%%%%%%%%%%%%
\noindent
\subsection{Refining the cell partition}
%%%%%%%%%%%%%%%%%%%%%%%%%%%%%%%%%%%%%%%%%%%%%%%%%%%%%%%%%%%%%%%%%%%%%%%%%%%%%%%%%%%%%%%%%%%%%%%%
%%%%%%%%%%%%%%%%%%%%%%%%%%%%%%%%%%%%%%%%%%%%%%%%%%%%%%%%%

In this section, we discuss step (\ref{eq-II-69}) in which the
momentum space cell partition is refined. It is possible to apply the
methods developed in \cite{Pizzo2005}, up to some minor modifications. 

We will prove that
\begin{eqnarray} 
	\big\| \, \Delta \psi (t_2,\sigma_{t_2},\mathscr{G}^{(t_2)}\rightarrow \mathscr{G}^{(t_1)})\, \big\|
	\, \leq \, \cO\big((\ln(t_2))^2 /t_1^{\rho}\big)  \, 
    \label{eq-II-74} 
\end{eqnarray}
for some $\rho>0$. The contributions from the off-diagonal terms with
respect to the sub-partition $\mathscr{G}^{(t_2)}$ of $\mathscr{G}^{(t_1)}$ can be estimated by the same
arguments that have culminated in the proof of Theorem {\ref{thm:II-1}}.
That is, we first express $\psi_{j,\sigma_{t_2}}^{(t_1)}$ as
\begin{equation}
    	\psi_{j,\sigma_{t_2}}^{(t_1)}
	\, = \, \int_{\mathscr{G}^{(t_1)}_j}h(\vec{P})\Psi_{\vec{P}}^{\sigma_{t_2}}d^3P
    	\, = \, \sum_{l(j)}\,\int_{\mathscr{G}^{(t_2)}_{l(j)}}h(\vec{P})\Psi_{\vec{P}}^{\sigma_{t_2}}d^3P
    	\, = \, \sum_{l(j)}\,\psi_{l(j),\sigma_{t_2}}^{(t_2)}\,.
\end{equation}
Then,
\begin{eqnarray}
    \lefteqn{(\ref{eq-II-74})
    \, = \, \sum_{j,j'=1}^{N(t_1)} \sum_{l(j),l'(j')}\,
    \Bra \, [\widehat{\cW}_{\Lambdas,\sigma_{t_2}}(\vv_{l(j)},t_2)
    -\widehat{\cW}_{\Lambdas,\sigma_{t_2}}(\vv_{j},t_2)]
    e^{-iE_{\vec{P}}^{\sigma_{t_2}}t_2}\psi_{l(j),\sigma_{t_2}}^{(t_2)}\,,
    }
	\nonumber\\
    \nonumber\\
    & &\quad\quad\quad\quad\quad\quad\quad\quad\quad\quad\quad
	\, \,[\widehat{\cW}_{\Lambdas,\sigma_{t_2}}(\vv_{l'(j')},t_2)
    -\widehat{\cW}_{\Lambdas,\sigma_{t_2}}(\vv_{j'},t_2)]
    e^{-iE_{\vec{P}}^{\sigma_{t_2}}t_2}\psi_{l'(j'),\sigma_{t_2}}^{(t_2)}\,\Ket\,,
    \quad\quad
    \label{eq-II-74-1}
\end{eqnarray}
where we define
\begin{eqnarray}
	\widehat{\cW}_{\Lambdas,\sigma_{t_2}}(\vv_{j},t_2)&:=&
	\cW_{\Lambdas,\sigma_{t_2}}(\vv_{j},t_2)\,
	e^{i\gamma_{\sigma_{t_2}}(\vv_{j},\vnabla E_{\vec{P}}^{\sigma_{t_2}},t_2)}\,,
	\\
	\widehat{\cW}_{\Lambdas,\sigma_{t_2}}(\vv_{l(j)},t_2)&:=&
	\cW_{\Lambdas,\sigma_{t_2}}(\vv_{l(j)},t_2)\,
	e^{i\gamma_{\sigma_{t_2}}(\vv_{l(j)},\vnabla E_{\vec{P}}^{\sigma_{t_2}},t_2)} \,.
\end{eqnarray}
Following the analysis in Section~\ref{sec-II.2.1}, one 
finds that the sum over pairs
$(l'(j')\,,\,l(j))$ with either $l\not= l'$ or $j\not= j'$ can be bounded
by $\cO(t_2^{-\epsilon})$, provided that $\epsilon<\frac{\eta}{4}$, as in (\ref{eq-Mljsum-est-1}).

Let $\langle \quad \rangle_{\widetilde\Psi}$ stand for the expectation value
with respect to the vector $\widetilde\Psi$.
Then, we are left with the diagonal terms
\begin{equation}  \label{eq-II-74-2}
	\sum_{j=1}^{N(t_1)} \sum_{l(j)}
	\,\Big\langle \,\big[ \, 2-\widehat{\cW}_{\Lambdas,\sigma_{t_2}}^{*}(\vv_{l(j)},t_2) \,
	\widehat{\cW}_{\Lambdas,\sigma_{t_2}}(\vv_{j},t_2)-\widehat{\cW}_{\Lambdas,\sigma_{t_2}}^{*}(\vv_{j},t_2)
	\widehat{\cW}_{\Lambdas,\sigma_{t_2}}(\vv_{l(j)},t_2) \, \big] \, \Big\rangle_{\widetilde\Psi}\,,
\end{equation}
labeled by pairs $(l(j)\,,\,l(j))$,
where $\widetilde\Psi\equiv e^{-iE_{\vec{P}}^{\sigma_{t_2}}t_2}\psi_{l(j),
\sigma_{t_2}}^{(t_2)}$ in the case considered here.
For each term
\begin{equation}
	\big\langle \,\widehat{\cW}_{\Lambdas,\sigma_{t_2}}^{*}(\vv_{l(j)},t_2)
	\widehat{\cW}_{\Lambdas,\sigma_{t_2}}(\vv_{j},t_2)\,\big\rangle_{\widetilde\Psi}\,,
\end{equation}
we can again invoke the arguments developed for off-diagonal elements
indexed by $(l,j)$ (where $l\not=j$) from Section~\ref{sec-II.2.1}.
\\

In particular, we define for $s>t_2$
\begin{equation} 
	\widehat{M}_{[l(j),\vv_{j}],[l(j),\vv_{l(j)}]}^{\mu}(t_2,s) \, 
	:=  \, \Bra \, e^{-iE_{\vec{P}}^{\sigma_{t}}s}\psi_{l(j),\sigma_{t_2}}^{(t_2)}\,,\,
    	\widehat{\cW}_{\Lambdas,\sigma_{t_2}}^{*\,\mu}(\vv_{j},s) \,
	\widehat{\cW}_{\Lambdas,\sigma_{t_2}}^{\mu}(\vv_{l(j)},s) \,
	e^{-iE_{\vec{P}}^{\sigma_t}s}\psi_{l(j),\sigma_{t_2}}^{(t_2)} \, \Ket\,, 
\end{equation}
where
\begin{equation} 
	\widehat{\cW}_{\Lambdas,\sigma_{t_2}}^{*\,\mu}(\vv_{j},s) \,
	\widehat{\cW}_{\Lambdas,\sigma_{t_2}}^{\mu}(\vv_{l(j)},s) \, 
	= 
	\, e^{-i\gamma_{\sigma_{t_2}}(\vv_{j},\vnabla E_{\vec{P}}^{\sigma_{t_2}},s)}
	\, \cW_{\Lambdas,\sigma_{t_2}}^{\mu\,*}(\vv_{j},s)
	\, \cW_{\Lambdas,\sigma_{t_2}}^{\mu}(\vv_{l(j)},s) \;
	\, e^{i\gamma_{\sigma_{t_2}}(\vv_{l(j)},\vnabla E_{\vec{P}}^{\sigma_{t_2}},s)} \,, 
\end{equation}
and
\begin{equation} 
	\cW_{\Lambdas,\sigma_{t_2}}^{\mu\,*}(\vv_{j},s)
	\, \cW_{\Lambdas,\sigma_{t_2}}^{\mu}(\vv_{l(j)},s) \, 
	= \, \exp \Big( \, \mu\,\alpha^{\frac{1}{2}}\sum_{\lambda}\int_{\mathcal{B}_{\Lambdas }\setminus\mathcal{B}_{\sigma_{t_2}}}
    	\vec{\eta}_{j,l(j)}(\vk)\cdot
	\big\lbrace \, \veps_{\vk, \lambda}a_{\vk,
    	\lambda}^*e^{-i|\vk|s}-\veps_{\vk,\lambda}^{\;*}a_{\vk, \lambda}e^{i|\vk|s}
	\, \big\rbrace d^3k \, \Big) \,, 
\end{equation}
with $\mu$ a real parameter.
\\

Proceeding similarly as in (\ref{eq-II-33}),
the solution of the ODE analogous to (\ref{eq-II-29}) for
$\widehat{M}_{[l(j),\vv_{j}],[l(j),\vv_{l(j)}]}^{\mu}(t_2,s)$ at $\mu=1$
consists of a contribution at $\mu=0$, which remains non-zero as $s\to\infty$,
and a remainder term that vanishes in the limit $s\to\infty$. In fact,
\begin{eqnarray}
    \lefteqn{
    \lim_{s\to+\infty}\widehat{M}_{[l(j),\vv_{j}],[l(j),\vv_{l(j)}]}^{1}(t_2,s)
    \;
    }
    \\
    &=&
    e^{-\frac{C_{j,l(j)}\,\sigma_{t_2}}{2}}\,\Bra \, e^{i\gamma_{\sigma_{t_2}}(\vv_{j},\vnabla
    E_{\vec{P}}^{\sigma_{t_2}},\sigma_{t_2}^{-\frac{1}{\theta}})}\psi_{l(j),\sigma_{t_2}}^{(t_2)}\,,
    \,e^{i\gamma_{\sigma_{t_2}}(\vv_{l(j)},\vnabla
    E_{\vec{P}}^{\sigma_{t_2}},\sigma_{t_2}^{-\frac{1}{\theta}})}\psi_{l(j),\sigma_{t_2}}^{(t_2)} \, \Ket\,,
    \nonumber
\end{eqnarray}
where
\begin{equation}\label{eq-II-77}
	C_{j,l(j),\sigma_{t_2}} \, := \, \int_{\mathcal{B}_{\Lambdas }\setminus\mathcal{B}_{\sigma_{t_2}}}
	|\vec{\eta}_{j,l(j)}(\vk)|^2 d^3k\,,
\end{equation}
as in (\ref{eq-II-30}), with $\vec{\eta}_{j,l(j)}(\vk)$ defined in (\ref{eq-II-24}).
Hence, (\ref{eq-II-74-2}) is given by the sum of
\begin{eqnarray}\label{eq-II-75bis}
    & - & \sum_{j=1}^{N(t_2)} \sum_{\ell(j)}
    \int_{t_2}^{+\infty} 
    \frac{d}{ds} \widehat M^1_{ [\ell(j),\vec v_j] , [\ell(j) , \vec v_{\ell(j)}] } (t_2,s) \, ds
    	\nonumber\\
	& - & \sum_{j=1}^{N(t_2)} \sum_{\ell(j)}
    \int_{t_2}^{+\infty} \frac{d}{ds} \widehat M^1_{[\ell(j) , \vec v_{\ell(j)}] ,  [\ell(j),\vec v_j] } (t_2,s) \, ds
\end{eqnarray}
and
\begin{equation}\label{eq-II-75}
    \sum_{j=1}^{N(t_2)} \sum_{\ell(j)}
	\Bra \, \psi_{l(j),\sigma_{t_2}}^{(t_2)} \, ,
    \,\big[2-2\cos\big(\Delta\gamma_{\sigma_{t_2}}(\vv_j-\vv_{l(j)},\vnabla
  	E_{\vec{P}}^{\sigma_{t_2}},t_2)\big)
    e^{-\frac{C_{l(j),j,\sigma_{t_2}}}{2}} \big]\,
  	\psi_{l(j),\sigma_{t_2}}^{(t_2)} \, \Ket \,,
\end{equation}
where
\begin{equation}\label{eq-II-76}
	\Delta\gamma_{\sigma_{t_2}}(\vv_j-\vv_{l(j)},\vnabla E_{\vec{P}}^{\sigma_{t_2}},t_2)
	\, := \,
	\gamma_{\sigma_{t_2}}(\vv_{l(j)},\vnabla E_{\vec{P}}^{\sigma_{t_2}},\sigma_{t_2}^{-\frac{1}{\theta}})
	\, - \,
	\gamma_{\sigma_{t_2}}(\vv_j,\vnabla E_{\vec{P}}^{\sigma_{t_2}},\sigma_{t_2}^{-\frac{1}{\theta}}) \,.
\end{equation}
The arguments that have culminated in Theorem {\ref{thm:II-1}} also imply that
the sum (\ref{eq-II-75bis}) can be bounded by $\cO(t_2^{-4\epsilon})$, for $\eta>4\epsilon$.
The leading contribution in (\ref{eq-II-74}) is represented by the sum (\ref{eq-II-75}) of diagonal
terms (with respect to  $\mathscr{G}^{(t_2)}$), which can now be bounded from above.
It suffices to show that
\begin{equation}\label{eq-II-78}
	\sup_{\vP\in\mathcal{S}}\Big| \, 2-2\cos\big(\Delta\gamma_{\sigma_{t_2}}(\vv_j-\vv_{l(j)},\vnabla
  	E_{\vec{P}}^{\sigma_{t_2}},t_2)\big)e^{-\frac{C_{l(j),j,\sigma_{t_2}}}{2}} \, \Big|
	\, \leq \, \frac{1}{t_1^{\eta'}}\ln t_2
\end{equation}
for some $\eta'>0$ that depends on $\epsilon$.
To see this, we note that
the lower integration
bound in the integral (\ref{eq-II-77}) contributes a factor to (\ref{eq-II-78}) proportional to $\ln t_2$.
In Lemma~\ref{lemm-A1}, it is proven that
\begin{equation}
	\big| \, \gamma_{\sigma_{t_2}}(\vv_j,\vnabla
  	E_{\vec{P}}^{\sigma_{t_2}},(\sigma_{t_2})^{-\frac{1}{\theta}})-\gamma_{\sigma_{t_2}}(\vv_{l(j)},\vnabla
  	E_{\vec{P}}^{\sigma_{t_2}},(\sigma_{t_2})^{-\frac{1}{\theta}}) \, \big |
  	\, \leq \, \cO(|\vv_j-\vv_{l(j)}|)\, .
\end{equation}
We can estimate the difference $\vv_j-\vv_{l(j)}=\vnabla E_{\vP_j^*}^{\sigma_{t_1}}
-\vnabla E_{\vP_{l(j)}^*}^{\sigma_{t_2}}$, which also appears in $\vec{\eta}_{l(j),j}(\vk)$,
using condition $(\mathscr{I}2)$ of Theorem {\ref{thm-cfp-2}}.
This yields the $\epsilon$-dependent negative power of $t_1$ in (\ref{eq-II-78}).
\\

%%%%%%%%%%%%%%%%%%%%%%%%%%%%%%%%%%%%%%%%%%%%%%%%%%%%%%%%%%%%%%%%%%%%%%%%%%%%%%%%%%%%%%%%%%%%
%%%%%%%%%%%%%%%%%%%%%%%%%%%%%%%%%%%%%%%%%%%%%%%%%%%%%%%%%%%%%%%%%%%%
\noindent
\subsection{Shifting the time variable for a fixed cell partition and infrared cut-off}
%%%%%%%%%%%%%%%%%%%%%%%%%%%%%%%%%%%%%%%%%%%%%%%%%%%%%%%%%%%%%%%%%%%%%%%%%%%%%%%%%%%%%%%%%%%%
%%%%%%%%%%%%%%%%%%%%%%%%%%%%%%%%%%%%%%%%%%%%%%%%%%%%%%%%%%%%%%%%%%%

%%% New March 20

In this subsection, we prove that
\begin{equation}
 	\big\| \, \Delta \psi (t_2\rightarrow t_1,\sigma_{t_2},\mathscr{G}^{(t_1)}) \, \big\| 
	\, \leq \, \cO\big((\ln(t_2))^2 /t_1^{\rho}\big) \,
\end{equation}
for some $\rho>0$; see (\ref{eq-II-70}).
This accounts for the change of the time variable, while both the cell partition
and the infrared cutoff are
kept fixed. It can be controlled by a standard Cook argument, and methods similar to
those used in the discussion of (\ref{eq-II-40}).

For $t_1\leq s\leq t_2$, we define
\begin{equation}\label{eq-II-79}
	\gamma_{\sigma_{t_2}}(\vv_j,\vnabla
  	E_{\vec{P}}^{\sigma_{t_2}},s)
 	\, := \,-\int_{1}^{s}\vnabla
	E_{\vec{P}}^{\sigma_{t_2}}\cdot\int_{\mathcal{B}_{\sigma_\tau^{S}}\setminus\mathcal{B}_{\sigma_{t_2}}}
	\vec{\Sigma}_{\vv_j}(\vk)\cos(\vk\cdot\vnabla
	E_{\vec{P}}^{\sigma_{t_2}}\tau-|\vk|\tau) \, d^3k \, d\tau \,.
\end{equation}
Then, we estimate
\begin{equation}\label{eq-II-80}
    \int_{t_1}^{t_2}\,ds\,\frac{d}{ds}\Big(\,e^{iH^{\sigma_{t_2}}s}\cW_{\Lambdas,\sigma_{t_2}}(\vv_j,s)
    e^{i\gamma_{\sigma_{t_2}}(\vv_j,\vnabla
    E_{\vec{P}}^{\sigma_{t_2}},s)}e^{-iE_{\vec{P}}^{\sigma_{t_2}}s}\psi_{j,\sigma_{t_2}}^{(t_1)}\,
    \Big)\,
\end{equation}
cell by cell. To this end, we can essentially
apply the same arguments that entered the
treatment of the time derivative in
(\ref{eq-II-40}), see also the remark after Theorem {\ref{theo-A'2}}, 
by defining a tail in a similar fashion. 
The only modification to be added is that,
apart from two terms analogous to (\ref{eq-II-41}), 
(\ref{eq-II-42}), we now also have to consider
\begin{equation}\label{eq-II-81}
    i\,e^{iHs}(H-H^{\sigma_{t_2}})\,\cW_{\Lambdas,\sigma_{t_2}}(\vv_j,s)e^{i\gamma_{\sigma_{t_2}}(\vv_j,\vnabla
    E_{\vec{P}}^{\sigma_{t_2}},s)}e^{-iE_{\vec{P}}^{\sigma_{t_2}}s}\psi_{j,\sigma_{t_2}}^{(t_1)}\,,
\end{equation}
which enters from the derivative in $s$ of the operator underlined in
\begin{equation}\label{eq-II-81bis}
    \underline{e^{iHs}e^{-iH^{\sigma_{t_2}}s}}e^{iH^{\sigma_{t_2}}s}\,
    \cW_{\Lambdas,\sigma_{t_2}}(\vv_j,s)e^{i\gamma_{\sigma_{t_2}}(\vv_j,\vnabla
    E_{\vec{P}}^{\sigma_{t_2}},s)}e^{-iE_{\vec{P}}^{\sigma_{t_2}}s}\psi_{j,\sigma_{t_2}}^{(t_1)}\,.
\end{equation}
To control the norm of (\ref{eq-II-81}), we observe that
\begin{equation}
    H-H^{\sigma_{t_2}} \, =  \, \alpha^{\frac{1}{2}}\,i[H,\vx]\,\cdot\,\vA_{<\sigma_{t_2}}
    \, - \, \alpha\,\frac{ \vec{A}_{<\sigma_{t_2}}\cdot \vec{A}_{<\sigma_{t_2}}}{2}\, ,
\end{equation}
where
\begin{eqnarray}
    \vec{A}_{<\sigma_{t_2}} \, := \,
    \sum_{\lambda = \pm} \int_{\mathcal{B}_{\sigma_{t_2}}}\frac{d^3k}{\sqrt{|\vk| \,}} \,
    \big\{ \veps_{\vk, \lambda} b^*_{\vk, \lambda} \, + \,
    \veps_{\vk, \lambda}^{\;*} b_{\vk, \lambda} \big\} \,,
\end{eqnarray}
and we note that
$$
    [H,\vx]\,\cdot\,\vA_{<\sigma_{t_2}} \, = \, \vA_{<\sigma_{t_2}}\,\cdot\,[H,\vx]\,,
$$
because of the Coulomb gauge condition.
Moreover,
\begin{eqnarray} 
    \cW_{\Lambdas,\sigma_{t_2}}^{*}(\vv_j,s)\,i[H,\vx]\,\cW_{\sigma_{t_2}}(\vv_j,s) 
    \, = \, \,i[H,\vx]\,+\vec{h}_s(\vx)
\end{eqnarray}
with $\|\vec{h}_s(\vx)\|< \cO(1)$ and
$$
    [b_{\vk, \lambda} \,,\,\vec{h}_s(\vx)] \, = \,
    [\vec{h}_s(\vx)\,,\,\vec{A}_{<\sigma_{t_2}}] \, = \, 0 \,.
$$
Furthermore, we have
$$
    b_{\vk, \lambda}\psi_{j,\sigma_{t_2}}^{(t_1)} \, = \, 0
$$
for $\vk\in\mathcal{B}_{\sigma_{t_2}}$, and
\begin{equation}
    \| \vec{A}_{<\sigma_{t_2}}\,\psi_{j,\sigma_{t_2}}^{(t_1)}\|
    \, , \, \| \vec{A}_{<\sigma_{t_2}} \cdot \vec{A}_{<\sigma_{t_2}} \, \psi_{j,\sigma_{t_2}}^{(t_1)}\|
    \, \leq \, \cO(\sigma_{t_2}) \,.
\end{equation}
The estimate
\begin{equation}
    \|\vec{A}_{<\sigma_{t_2}}\cdot[H,\vx] \, \psi_{j,\sigma_{t_2}}^{(t_1)}\|
    \, \leq \, \cO( \, \sigma_{t_2} \,
    \big\{ \, \|[H,\vx] \, \psi_{j,\sigma_{t_2}}^{(t_1)}\|
    \, + \, \| \psi_{j,\sigma_{t_2}}^{(t_1)} \|
    \,  \big\}\, )\,,
\end{equation}
holds, where
\begin{equation}
    \|[H,\vx]\psi_{j,\sigma_{t_2}}^{(t_1)}\| \,\leq \, \cO(t_{1}^{-\frac{3\epsilon}{2}})\,,
\end{equation}
because
$$
    \|[H,\vx]\psi_{j,\sigma_{t_2}}^{(t_1)}\| \, \leq \,
    c_1\|(H^{\sigma_{t_2}}+i)\psi_{j,\sigma_{t_2}}^{(t_1)}\|
$$
for some constant $c_1$, and
$$
    \|\psi_{j,\sigma_{t_2}}^{(t_1)}\| \, = \, \cO(t_{1}^{-\frac{3\epsilon}{2}}) \, .
$$
Consequently, we obtain that
\begin{eqnarray}
    \lefteqn{
    \Big\| \, (H-H^{\sigma_{t_2}})\,\cW_{\Lambdas,\sigma_{t_2}}(\vv_j,s)e^{i\gamma_{\sigma_{t_2}}(\vv_j,\vnabla
    E_{\vec{P}}^{\sigma_{t_2}},s)}e^{-iE_{\vec{P}}^{\sigma_{t_2}}s}\psi_{j,\sigma_{t_2}}^{(t_1)} \, \Big\|
    }
    \nonumber\\
    &&\leq \; \cO( \, \sigma_{t_2} \,
    \big\{ \, \|[H,\vx] \, \psi_{j,\sigma_{t_2}}^{(t_1)}\|
    \, + \, \| \psi_{j,\sigma_{t_2}}^{(t_1)} \|
    \,  \big\}\, )
    \; \leq\; \cO ( \, \sigma_{t_2}\, t^{-3/2\epsilon} \,  ) \,.
    \quad\quad\quad
    \label{eq-II-82}
\end{eqnarray}
Following the procedure in Section II.2.1, \underline{{\em B.}}, one can also check that
\begin{eqnarray}
    \Big\| \, \int_{t_1}^{t_2}\,e^{iHs}e^{-iH^{\sigma_{t_2}}s}\frac{d}{ds}\Big(\,e^{iH^{\sigma_{t_2}}s}\,
    \cW_{\Lambdas,\sigma_{t_2}}(\vv_j,s)e^{i\gamma_{\sigma_{t_2}}(\vv_j,\vnabla
    E_{\vec{P}}^{\sigma_{t_2}},s)}e^{-iE_{\vec{P}}^{\sigma_{t_2}}s}\psi_{j,\sigma_{t_2}}^{(t_1)}\,
    \Big)ds\, \Big\|
    \nonumber\\
    \leq \; \cO(\,\frac{1}{t_1^{\eta}}|\ln \sigma_{t_2}|^2\,t_1^{-\frac{3\epsilon}{2}}\,) \, ,
    \quad\quad\quad
    \label{eq-II-81bisbis}
\end{eqnarray}
for some $\eta>0$ independent of $\epsilon$. Similarly as in (\ref{eq-Mljsum-est-1}), we
choose $\epsilon$ small enough that $\frac{\eta}{4}>\epsilon$.

The number of cells in the partition $\mathscr{G}^{(t_1)}$ is $N(t_1)\approx t_1^{3\epsilon}$.
Therefore, summing over all cells, we get
\begin{equation}
	\cO\Big(\,N(t_1)\,\,t_1^{-\frac{3\epsilon}{2}}\,\sigma_{t_2}t_2 \, \Big) \,+\,
	\cO\Big( \, N(t_1)\,\frac{1}{t_1^{\eta}}|\ln \sigma_{t_2}|^2\,t_1^{-\frac{3\epsilon}{2}}\Big)\,,
\end{equation}
as an upper bound on the norm of the term in (\ref{eq-II-70}).

The parameter $\beta$ in the definition of $\sigma_{t_2}=t_2^{-\beta}$
can be chosen arbitrarily large, independently of $\epsilon$.
Hereby, we arrive at the upper bound claimed in (\ref{eq-II-72}).
\\

%%%%%%%%%%%%%%%%%%%%%%%%%%%%%%%%%%%%%%%%%%%%%%%%%%%%%%%%%%%%%%%%%%%%%%%%%%%%%%%%%%%%%%%%%%%%%%%%%%%%%%%%%%%%%%%%%%%%%%%%%%%%%%%%%%%%%%%%%%%%%%%%%%%%%%%%%%%%%%%
\subsection{Shifting the infrared cut-off}
%%%%%%%%%%%%%%%%%%%%%%%%%%%%%%%%%%%%%%%%%%%%%%%%%%%%%%%%%%%%%%%%%%%%%%%%%%%%%%%%%%%%%%%%%%%%%%%%%%%%%%%%%%%%%%%%%%%%%%%%%%%%%%%%%%%%%%%%%%%%%%%%%%%%%%%%%%%%%%%

In this section, we prove that
\begin{equation}\label{eq:IRchange-ineq-1}
	\| \, \Delta \psi (t_1,\sigma_{t_2}\rightarrow\sigma_{t_1}, \mathscr{G}^{(t_1)}) \, \|
	\, \leq \, \cO\big((\ln(t_2))^2 /t_1^{\rho}\big)  \,
\end{equation}
for some $\rho>0$; see (\ref{eq-II-71}).
The analysis of this last step is the most involved one, and
will require extensive use of our previous results.

The starting idea is to rewrite the last term in (\ref{eq-II-71}),
\begin{equation}\label{eq-II-83}
	e^{iHt_1}\sum_{j=1}^{N(t_1)}\cW_{\Lambdas,\sigma_{t_2}}(\vv_j,t_1)
	\, e^{i\gamma_{\sigma_{t_2}}(\vv_j,\vnabla E_{\vec{P}}^{\sigma_{t_2}},t_1)}
	\, e^{-iE_{\vec{P}}^{\sigma_{t_2}}t_1} \, \psi_{j,\sigma_{t_2}}^{(t_1)} \, ,
\end{equation}
as
\begin{equation}\label{eq-II-84}
	e^{iHt_1}\sum_{j=1}^{N(t_1)}\cW_{\Lambdas,\sigma_{t_2}}(\vv_j,t_1)
	\, W_{\sigma_{t_2}}^{*}(\vnabla E^{\sigma_{t_2}}_{\vP})
	\, W_{\sigma_{t_2}}(\vnabla E^{\sigma_{t_2}}_{\vP})
	\, e^{i\gamma_{\sigma_{t_2}}(\vv_j,\vnabla E_{\vec{P}}^{\sigma_{t_2}},t_1)}
	\, e^{-iE_{\vec{P}}^{\sigma_{t_2}}t_1}
	\, \psi_{j,\sigma_{t_2}}^{(t_1)}\,,
\end{equation}
and to group the terms appearing in (\ref{eq-II-84}) in such a way that, cell by cell,
we consider the new {\em dressing operator}
\begin{equation}\label{eq-II-85}
    	e^{iHt_1} \, \cW_{\Lambdas,\sigma_{t_2}}(\vv_j,t_1)
	\, W_{\sigma_{t_2}}^{*}(\vnabla E^{\sigma_{t_2}}_{\vP})
	\, e^{-iE_{\vec{P}}^{\sigma_{t_2}}t_1}\,,
\end{equation}
which acts on
\begin{equation}\label{eq-II-86}
    	\Phi^{(t_1)}_{j,\sigma_{t_2}} \, := \,
	\int_{\mathscr{G}^{(t_1)}_j}h(\vec{P}) \, \Phi_{\vec{P}}^{\sigma_{t_2}} \, d^3P \, ,
\end{equation}
where $\Phi_{\vec{P}}^{\sigma} =W_{\sigma}(\vnabla
E^{\sigma}_{\vP})\Psi_{\vec{P}}^{\sigma}$, see (\ref{eq-II-14}).
The key advantage is that the vector
$\Phi_{j,\sigma_{t_2}}^{(t_1)}$ inherits  the H\"older regularity
of $\Phi_{\vec{P}}^{\sigma}$; see (\ref{eq-II-5}) in condition ($\mathscr{I}2$) of Theorem {\ref{thm-cfp-2}}.
We will refer to (\ref{eq-II-86}) as an {\em infrared-regular vector}.

Accordingly, (\ref{eq-II-84}) now reads
\begin{equation}\label{eq-II-87}
    	e^{iHt_1}\sum_{j=1}^{N(t_1)}\cW_{\Lambdas,\sigma_{t_2}}(\vv_j,t_1)
	\, W_{\sigma_{t_2}}^{*}(\vnabla E^{\sigma_{t_2}}_{\vP})
	\, e^{i\gamma_{\sigma_{t_2}}(\vv_j,\vnabla E_{\vec{P}}^{\sigma_{t_2}},t_1)}
	\, e^{-iE_{\vec{P}}^{\sigma_{t_2}}t_1}
	\, \Phi_{j,\sigma_{t_2}}^{(t_1)}\,,
\end{equation}
and we proceed as follows.
\\

\noindent
%%%%%%%%%%%%%%%%%%%%%%%%%%%%%%%%%%%%%%%%%%%%%%%%%%
\emph{\underline{A. Shifting the IR cutoff in the infrared-regular vector}}
%%%%%%%%%%%%%%%%%%%%%%%%%%%%%%%%%%%%%%%%%%%%%%%%%%
\\
\\
First, we substitute
\begin{eqnarray}\label{eq-II-88}
    	\lefteqn{e^{iHt_1}\sum_{j=1}^{N(t_1)}\cW_{\Lambdas,\sigma_{t_2}}(\vv_j,t_1)W_{\sigma_{t_2}}^{*}(\vnabla
    	E^{\sigma_{t_2}}_{\vP}) \,
	\underline{e^{i\gamma_{\sigma_{t_2}}(\vv_j,\vnabla E_{\vec{P}}^{\sigma_{t_2}},t_1)}
	e^{-iE_{\vec{P}}^{\sigma_{t_2}}t_1}\Phi_{j,\sigma_{t_2}}^{(t_1)}}}\\
    	&\longrightarrow & e^{iHt_1}\sum_{j=1}^{N(t_1)}\cW_{\Lambdas,\sigma_{t_2}}(\vv_j,t_1)
	W_{\sigma_{t_2}}^{*}(\vnabla E^{\sigma_{t_2}}_{\vP}) \,
	\underline{e^{i\gamma_{\sigma_{t_1}}(\vv_j,\vnabla E_{\vec{P}}^{\sigma_{t_1}},t_1)}
	e^{-iE_{\vec{P}}^{\sigma_{t_1}}t_1}\Phi_{j,\sigma_{t_1}}^{(t_1)}} \, ,
	\nonumber\,
\end{eqnarray}
where $\sigma_{t_2}$ is replaced by $\sigma_{t_1}$ in the underlined terms.
We prove that the norm difference of these two vectors is bounded by the r.h.s. of (\ref{eq:IRchange-ineq-1}).
The necessary ingredients are:
\begin{itemize}
\item[1)]
Condition $(\mathscr{I}1)$  in Theorem {\ref{thm-cfp-2}}.
\item[2)]
The estimate
\begin{eqnarray} 
    |\gamma_{\sigma_{t_2}}(\vv_j,\vnabla
    E_{\vec{P}}^{\sigma_{t_2}},t_1)-\gamma_{\sigma_{t_1}}(\vv_j,\vnabla
    E_{\vec{P}}^{\sigma_{t_1}},t_1)| 
	\, \leq \,
    \cO\big(\, \sigma_{t_1}^{{\frac{1}{2}(1-\delta)}}\,t_1^{1-\theta} \, \big)
	\, + \, \cO\big( \, t_1 \,\sigma_{t_1} \,\big)  \, ,
    \nonumber
\end{eqnarray}
for $t_2>t_1\gg1$, proven in Lemma \ref{lemm-A1}.
The parameter $0<\theta<1$ is the same as the one in (\ref{eq-II-17}).
\item[3)]
The cell partition $\mathscr{G}^{(t_1)}$ depends on $t_1<t_2$.
\item[4)]
The parameter $\beta$ can be chosen arbitrarily large, independently of $\epsilon$,
so that the infrared cutoff $\sigma_{t_1}=t_1^{-\beta}$ can be made as small as one wishes.
\end{itemize}
First of all, it is clear that the norm
difference of the two vectors in (\ref{eq-II-88}) is bounded by
the norm difference of the two underlined vectors, summed over all $N(t_1)$ cells.
Using 1) and 2), one straightforwardly derives that the norm difference between
the two underlined vectors in (\ref{eq-II-88}) is bounded from above by
\begin{equation}
    \cO(\,t_1\,\sigma_{t_1}^{\frac{1}{2}(1-\delta)}\,t_1^{-\frac{3}{2}\epsilon}\,) \, ,
\end{equation}
where the last factor, $t_1^{-\frac{3}{2}\epsilon}$, accounts for the volume of an individual cell
in $\mathscr{G}^{(t_1)}$, by 3).
The sum over all cells in $\mathscr{G}^{(t_1)}$ yields a bound
\begin{equation}  \label{eq-II-89}
	\cO( \, N(t_1)\,\sigma_{t_1}^{\frac{1}{2}(1-\delta)}\,t_1^{1-\frac{3}{2}\epsilon} \, )
\end{equation}
where $N(t_1)\approx t_1^{3\epsilon}$, by 3).
Picking $\beta$ sufficiently large, by 4),  we find that the norm difference of the two
vectors in (\ref{eq-II-88})
is bounded by $t_1^{-\eta}$, for some $\eta>0$. This agrees with the bound stated in (\ref{eq:IRchange-ineq-1}).
\\

\noindent
%%%%%%%%%%%%%%%%%%%%%%%%%%%%%%%%%%%%%%%%%%%%%%%%%%%
\emph{\underline{B. Shifting the IR cutoff in the dressing operator}}
%%%%%%%%%%%%%%%%%%%%%%%%%%%%%%%%%%%%%%%%%%%%%%%%%%%
\\
\\
Subsequently to (\ref{eq-II-88}), we substitute
\begin{eqnarray}\label{eq-II-89bis}
    \lefteqn{
    e^{iHt_1}\sum_{j=1}^{N(t_1)}\underline{\cW_{\Lambdas,\sigma_{t_2}}(\vv_j,t_1)W_{\sigma_{t_2}}^{*}(\vnabla
    E^{\sigma_{t_2}}_{\vP})} e^{i\gamma_{\sigma_{t_1}}(\vv_j,\vnabla
    E_{\vec{P}}^{\sigma_{t_1}},t_1)}e^{-iE_{\vec{P}}^{\sigma_{t_1}}t_1}\Phi_{j,\sigma_{t_1}}^{(t_1)}
    }
    \\
    &\longrightarrow & e^{iHt_1}\sum_{j=1}^{N(t_1)}
    \underline{\cW_{\Lambdas,\sigma_{t_1}}(\vv_j,t_1)W_{\sigma_{t_1}}^{*}(\vnabla
    E^{\sigma_{t_1}}_{\vP})} e^{i\gamma_{\sigma_{t_1}}(\vv_j,\vnabla
    E_{\vec{P}}^{\sigma_{t_1}},t_1)}e^{-iE_{\vec{P}}^{\sigma_{t_1}}t_1}\Phi_{j,\sigma_{t_1}}^{(t_1)}
    \nonumber\, ,
\end{eqnarray}
where $\sigma_{t_2}\rightarrow\sigma_{t_1}$ in the underlined operators.
A crucial point in our argument is that when $\sigma_{t_1}(>\sigma_{t_2})$
tends to $0$, the H\"older continuity of
$\Phi_{\vec{P}}^{\sigma_{t_1}}$ in $\vP$ offsets the (logarithmic)
divergence in $t_2$ which arises from the dressing operator.

We subdivide the shift $\sigma_{t_2}\rightarrow\sigma_{t_1}$ in
\begin{equation}\label{eq-II-90}
    \cW_{\Lambdas,\sigma_{t_2}}(\vv_j,t_1)W_{\sigma_{t_2}}^{*}(\vnabla E^{\sigma_{t_2}}_{\vP})
    \, \longrightarrow \, \cW_{\Lambdas,\sigma_{t_1}}(\vv_j,t_1)W_{\sigma_{t_1}}^{*}(\vnabla
    E^{\sigma_{t_1}}_{\vP})\,
\end{equation}
into the following three intermediate steps,
where the operators modified in each step are underlined:
\begin{itemize}
\item[]\underline{\em Step a)}
\begin{eqnarray}\label{eq-II-91}
    & &\underline{\cW_{\Lambdas,\sigma_{t_2}}(\vv_j,t_1)W_{\sigma_{t_2}}^{*}(\vv_j)}
    W_{\sigma_{t_2}}(\vv_j)W_{\sigma_{t_2}}^{*}(\vnabla E^{\sigma_{t_2}}_{\vP})\\
    & &\longrightarrow \underline{\cW_{\Lambdas,\sigma_{t_1}}(\vv_j,t_1)W_{\sigma_{t_1}}^{*}(\vv_j)}
    W_{\sigma_{t_2}}(\vv_j)
    W_{\sigma_{t_2}}^{*}(\vnabla E^{\sigma_{t_2}}_{\vP}) \nonumber\,,
\end{eqnarray}
\item[]\underline{\em Step b)}
\begin{eqnarray}\label{eq-II-92}
    & &\cW_{\Lambdas,\sigma_{t_1}}(\vv_j,t_1)W_{\sigma_{t_1}}^{*}(\vv_j)W_{\sigma_{t_2}}(\vv_j)
    \underline{W_{\sigma_{t_2}}^{*}(\vnabla E^{\sigma_{t_2}}_{\vP})} \\
    & &\longrightarrow \cW_{\Lambdas,\sigma_{t_1}}(\vv_j,t_1)
    W_{\sigma_{t_1}}^{*}(\vv_j)W_{\sigma_{t_2}}(\vv_j)\underline{W_{\sigma_{t_2}}^{*}
    (\vnabla E^{\sigma_{t_1}}_{\vP})}\nonumber\,,
\end{eqnarray}
\item[]\underline{\em Step c)}
\begin{eqnarray}\label{eq-II-93}
    & &\cW_{\Lambdas,\sigma_{t_1}}(\vv_j,t_1)W_{\sigma_{t_1}}^{*}(\vv_j)
    \underline{W_{\sigma_{t_2}}(\vv_j)W_{\sigma_{t_2}}^{*}(\vnabla
    E^{\sigma_{t_1}}_{\vP})} \\
    & &\longrightarrow \cW_{\Lambdas,\sigma_{t_1}}(\vv_j,t_1)W_{\sigma_{t_1}}^{*}(\vv_j)
    \underline{W_{\sigma_{t_1}}(\vv_j)W_{\sigma_{t_1}}^{*}
    (\vnabla E^{\sigma_{t_1}}_{\vP})}\nonumber\,.
\end{eqnarray}
\end{itemize}

\noindent
%%%%%%%%%%%%%%%%%%%%%%%%%%%%%%%%%%%%%%%%%%%%%%%%%%%
\emph{\underline{Analysis of Step a)}}
%%%%%%%%%%%%%%%%%%%%%%%%%%%%%%%%%%%%%%%%%%%%%%%%%%%
$\;$\\

\noindent
In step {\em a)}, we analyze the difference between the vectors
\begin{equation}\label{eq-II-93bis}
    e^{iHt_1}\underline{\cW_{\Lambdas,\sigma_{t_2}}(\vv_j,t_1)
    W_{\sigma_{t_2}}^{*}(\vv_j)}W_{\sigma_{t_2}}(\vv_j)W_{\sigma_{t_2}}^{*}(\vnabla
    E^{\sigma_{t_2}}_{\vP})e^{i\gamma_{\sigma_{t_1}}(\vv_j,\vnabla E_{\vec{P}}^{\sigma_{t_1}},t_1)}
    e^{-iE_{\vec{P}}^{\sigma_{t_1}}t_1}\Phi_{j,\sigma_{t_1}}^{(t_1)}
\end{equation}
and
\begin{equation}\label{eq-II-93bisbis}
    e^{iHt_1}\underline{\cW_{\Lambdas,\sigma_{t_1}}(\vv_j,t_1)W_{\sigma_{t_1}}^{*}(\vv_j)}
    W_{\sigma_{t_2}}(\vv_j)W_{\sigma_{t_2}}^{*}(\vnabla E^{\sigma_{t_2}}_{\vP})
    e^{i\gamma_{\sigma_{t_1}}(\vv_j,\vnabla E_{\vec{P}}^{\sigma_{t_1}},t_1)}
    e^{-iE_{\vec{P}}^{\sigma_{t_1}}t_1}\Phi_{j,\sigma_{t_1}}^{(t_1)}\, ,
\end{equation}
for each cell  in $\mathscr{G}^{(t_1)}$. 
Our goal is to prove that
\begin{equation}\label{eq-II-95bis}
    \| \, (\ref{eq-II-93bis}) - (\ref{eq-II-93bisbis}) \, \|
	\, \leq \, const\,  \ln t_2 \,  P(t_1,t_2) \,,
\end{equation} 
where
\begin{eqnarray}\label{eq-II-96}
    P(t_1,t_2) & := & \sup_{\vk\in\mathcal{B}_{\sigma_{t_1}}}\Big\| \, (e^{-i(|\vk|t_1-\vk\cdot\vx)}-1)
    W_{\sigma_{t_2}}(\vv_j)W_{\sigma_{t_2}}^{*}(\vnabla
    E^{\sigma_{t_2}}_{\vP})) \, \times
   \\
    &&\quad\quad\quad\quad \times \, e^{i\gamma_{\sigma_{t_2}}(\vv_j,\vnabla
    E_{\vec{P}}^{\sigma_{t_2}},t_1)}e^{-iE_{\vec{P}}^{\sigma_{t_1}}t_1}\Phi_{j,\sigma_{t_1}}^{(t_1)}
    \, \Big\|\, \leq \, \cO( \, t_1^{-\eta}\ln t_2 \, )
    \nonumber
\end{eqnarray}
as $t_1\to+\infty$, for some $\eta>0$,
and for $\beta$ large enough.

Using the identity
\begin{eqnarray}\label{eq-II-94}
    \lefteqn{
	\cW_{\Lambdas,\sigma_{t_2}}(\vv_j,t_1)W_{\sigma_{t_2}}^{*}(\vv_j)
    \, = \, \cW_{\Lambdas,\sigma_{t_1}}(\vv_j,t_1)W_{\sigma_{t_1}}^{*}(\vv_j) \, \times
	}
	\nonumber\\
    & &\times \,
    \exp\Big(\frac{i\alpha}{2}\,\int_{\mathcal{B}_{\sigma_{t_1}}\setminus\mathcal{B}_{\sigma_{t_2}}}
    d^3k \,
    \frac{\vv_j\cdot\vec{\Sigma}_{\vv_j}(\vk)\sin(\vk\cdot\vx-|\vk|t_1)}
    {|\vk|(1-\widehat{k}\cdot\vv_j)}\Big)\times
	\quad\quad\quad\label{eq-II-95}\\
    & &\times\,
    \exp\Big(\alpha^{\frac{1}{2}}\sum_{\lambda}\int_{\mathcal{B}_{\sigma_{t_1}}\setminus
    \mathcal{B}_{\sigma_{t_2}}} \frac{d^3k}{\sqrt{|\vk|}} \frac{\vv_j\cdot\lbrace\veps_{\vk, \lambda}
	b_{\vk,\lambda}^*(e^{-i(|\vk|t_1-\vk\cdot\vx)}-1)-\,h.c.\,\rbrace}{|\vk|(1-\widehat{k}\cdot\vv_j)}
    \Big) \, ,\nonumber\,
\end{eqnarray} 
the difference between (\ref{eq-II-93bis}) and (\ref{eq-II-93bisbis}) is given by
\begin{eqnarray}
    \lefteqn{
    e^{iHt_1}\cW_{\Lambdas,\sigma_{t_1}}(\vv_j,t_1)W_{\sigma_{t_1}}^{*}(\vv_j)\,
    \,
    \exp\Big(\frac{i\alpha}{2}\,\int_{\mathcal{B}_{\sigma_{t_1}}\setminus\mathcal{B}_{\sigma_{t_2}}}
    d^3k \,
    \frac{\vv_j\cdot\vec{\Sigma}_{\vv_j}(\vk)\sin(\vk\cdot\vx-|\vk|t_1)}
    {|\vk|(1-\widehat{k}\cdot\vv_j)}\Big) \, \times
    }
    \nonumber\\
    & & \quad \quad \times \, \Big[ 
    \exp\Big(\alpha^{\frac{1}{2}}\sum_{\lambda}\int_{\mathcal{B}_{\sigma_{t_1}}\setminus
    \mathcal{B}_{\sigma_{t_2}}} \frac{d^3k}{\sqrt{|\vk|}} \frac{\vv_j\cdot\lbrace\veps_{\vk, \lambda}b_{\vk,
    \lambda}^*(e^{-i(|\vk|t_1-\vk\cdot\vx)}-1)-\,h.c.\,\rbrace}{|\vk|(1-\widehat{k}\cdot\vv_j)}
    \Big)
    \, - \, \cI \Big]\times
    \nonumber \\
    & & \quad \quad \quad \quad \quad \quad \times \,
    W_{\sigma_{t_2}}(\vv_j)W_{\sigma_{t_2}}^{*}(\vnabla
    E^{\sigma_{t_2}}_{\vP})\,e^{i\gamma_{\sigma_{t_1}}(\vv_j,\vnabla E_{\vec{P}}^{\sigma_{t_1}},t_1)}
    e^{-iE_{\vec{P}}^{\sigma_{t_1}}t_1}\Phi_{j,\sigma_{t_1}}^{(t_1)}\,  \label{eq-II-96.1}
    \\
    & + &
    e^{iHt_1}\cW_{\Lambdas,\sigma_{t_1}}(\vv_j,t_1)W_{\sigma_{t_1}}^{*}(\vv_j)\,
    \Big[ \,
    \exp\Big(\frac{i\alpha}{2}\,\int_{\mathcal{B}_{\sigma_{t_1}}\setminus\mathcal{B}_{\sigma_{t_2}}}
    d^3k \,
    \frac{\vv_j\cdot\vec{\Sigma}_{\vv_j}(\vk)\sin(\vk\cdot\vx-|\vk|t_1)}
    {|\vk|(1-\widehat{k}\cdot\vv_j)}\Big) \,
    -\cI \Big]\times \nonumber \\
    & &
    \quad \quad \quad \quad \quad \quad \times \,
    W_{\sigma_{t_2}}(\vv_j)W_{\sigma_{t_2}}^{*}(\vnabla
    E^{\sigma_{t_2}}_{\vP})\,e^{i\gamma_{\sigma_{t_1}}(\vv_j,\vnabla E_{\vec{P}}^{\sigma_{t_1}},t_1)}
    e^{-iE_{\vec{P}}^{\sigma_{t_1}}t_1}\Phi_{j,\sigma_{t_1}}^{(t_1)}\,, \label{eq-II-96.1bis}
\end{eqnarray}
where $\cI $ is the identity operator in $\cH$. 

The norm of the vector (\ref{eq-II-96.1}) equals
\begin{eqnarray} 
    \Big\|\,\Big[
    \exp\big(\alpha^{\frac{1}{2}}\sum_{\lambda}\int_{\mathcal{B}_{\sigma_{t_1}}\setminus
    \mathcal{B}_{\sigma_{t_2}}} \frac{d^3k}{\sqrt{|\vk|}} \frac{\vv_j\cdot\lbrace\veps_{\vk, \lambda}
    b_{\vk,\lambda}^*(e^{-i(|\vk|t_1-\vk\cdot\vx)}-\cI)-\,h.c.\,\rbrace}{|\vk|(1-\widehat{k}\cdot\vv_j)}
    \big)-\cI\Big]\times 
    \nonumber \\ 
    \times \,W_{\sigma_{t_2}}(\vv_j)W_{\sigma_{t_2}}^{*}(\vnabla E^{\sigma_{t_2}}_{\vP})
    e^{i\gamma_{\sigma_{t_1}}(\vv_j,\vnabla E_{\vec{P}}^{\sigma_{t_1}},t_1)}
    e^{-iE_{\vec{P}}^{\sigma_{t_1}}t_1}\Phi_{j,\sigma_{t_1}}^{(t_1)}\Big\|\,.
    \quad\quad\quad
    \label{eq-II-96.2}
\end{eqnarray}

\noindent
We now observe that
\begin{itemize}
\item
for $\vk\in\mathcal{B}_{\sigma_{t_1}}$,
\begin{eqnarray}
    \lefteqn{b_{\vk,\lambda}\,W_{\sigma_{t_2}}(\vv_j)W_{\sigma_{t_2}}^{*}(\vnabla
    E^{\sigma_{t_2}}_{\vP}) \,  }\\
    &=&W_{\sigma_{t_2}}(\vv_j)W_{\sigma_{t_2}}^{*}(\vnabla
    E^{\sigma_{t_2}}_{\vP}) \, b_{\vk,\lambda}\\
    & &+ \, W_{\sigma_{t_2}}(\vv_j)W_{\sigma_{t_2}}^{*}(\vnabla
    E^{\sigma_{t_2}}_{\vP}) \, f_{\vk,\lambda}(\vv_j,\vP)\,,
\end{eqnarray}
where
\begin{equation}
    \int_{\mathcal{B}_{\sigma_{t_1}}\setminus
    \mathcal{B}_{\sigma_{t_2}}}\,d^3k\,|f_{\vk,\lambda}(\vv_j,\vP)|^2 \, \leq \, \cO(|\ln\sigma_{t_2}|)
\end{equation}
uniformly in $\vv_j$, and in $\vP\in\cS$, and where $j$ enumerates the cells.
\item
for $\vk\in\mathcal{B}_{\sigma_{t_1}}$,
\begin{equation}
    b_{\vk,\lambda}\,e^{i\gamma_{\sigma_{t_2}}(\vv_j,\vnabla
    E_{\vec{P}}^{\sigma_{t_1}},t_1)}e^{-iE_{\vec{P}}^{\sigma_{t_1}}t_1}\Phi_{j,\sigma_{t_1}}^{(t_1)}=0\,,
\end{equation}
because of the infrared properties of $\Phi_{j,\sigma_{t_1}}^{(t_1)}$.
\end{itemize}

>From the Schwarz inequality, we therefore get
\begin{equation} 
    (\ref{eq-II-96.2}) 
	\, \leq \,
	c \, |\ln \sigma_{t_2}| \, P(t_1,t_2) \,,
\end{equation}
for some finite constant $c$ as claimed in (\ref{eq-II-95bis}), where
\begin{equation}\label{eq-II.96.2.1} 
    P(t_1,t_2) \, = \, \sup_{\vk\in\mathcal{B}_{\sigma_{t_1}}}
    \Big\| \, (e^{-i(|\vk|t_1-\vk\cdot\vx)}-1) \, 
    W_{\sigma_{t_2}}(\vv_j)
    W_{\sigma_{t_2}}^{*}(\vnabla E^{\sigma_{t_2}}_{\vP})
    e^{i\gamma_{\sigma_{t_2}}(\vv_j,\vnabla
    E_{\vec{P}}^{\sigma_{t_2}},t_1)}e^{-iE_{\vec{P}}^{\sigma_{t_2}}t_1}\Phi_{j,\sigma_{t_1}}^{(t_1)}
    \Big\|\,, 
\end{equation} 
as defined in (\ref{eq-II-96}).
To estimate $P(t_1,t_2)$, we regroup the terms inside the norm into
\begin{eqnarray}
    \lefteqn{ (e^{-i(|\vk|t_1-\vk\cdot\vx)}-1)
    W_{\sigma_{t_2}}(\vv_j)
    \, W_{\sigma_{t_2}}^{*}(\vnabla E^{\sigma_{t_2}}_{\vP})
    \, e^{i\gamma_{\sigma_{t_2}}(\vv_j,\vnabla E_{\vec{P}}^{\sigma_{t_2}},t_1)}
    \, e^{-iE_{\vec{P}}^{\sigma_{t_2}}t_1}
    \, \Phi_{j,\sigma_{t_1}}^{(t_1)}
    }
    \nonumber\\
    & & = \; \;
    W_{\sigma_{t_2}}(\vv_j)
    \, W_{\sigma_{t_2}}^{*}(\vnabla E^{\sigma_{t_2}}_{\vP-\vk})
    \, (e^{-i(|\vk|t_1-\vk\cdot\vx)}-\cI) 
    \label{eq-II-96.3} 
    \,e^{i\gamma_{\sigma_{t_2}}(\vv_j,\vnabla E_{\vec{P}}^{\sigma_{t_2}},t_1)}
    \, e^{-iE_{\vec{P}}^{\sigma_{t_2}}t_1}
    \, \Phi_{j,\sigma_{t_1}}^{(t_1)} \quad 
	\\
    & & \; \; + \; W_{\sigma_{t_2}}(\vv_j)
    \, W_{\sigma_{t_2}}^{*}(\vnabla E^{\sigma_{t_2}}_{\vP-\vk})
    \, e^{i\gamma_{\sigma_{t_2}}(\vv_j,\vnabla E_{\vec{P}}^{\sigma_{t_2}},t_1)}
    \, e^{-iE_{\vec{P}}^{\sigma_{t_2}}t_1}
    \, \Phi_{j,\sigma_{t_1}}^{(t_1)}
    \label{eq-II-96.4}\\
    &  & \; \; - \; W_{\sigma_{t_2}}(\vv_j)
    \, W_{\sigma_{t_2}}^{*}(\vnabla E^{\sigma_{t_2}}_{\vP})
    \, e^{i\gamma_{\sigma_{t_2}}(\vv_j,\vnabla E_{\vec{P}}^{\sigma_{t_2}},t_1)}
    \, e^{-iE_{\vec{P}}^{\sigma_{t_2}}t_1}
    \, \Phi_{j,\sigma_{t_1}}^{(t_1)}\,.
    \label{eq-II-96.5}
\end{eqnarray}
We next prove that
\begin{equation}\label{eq-II-97}
    \|(\ref{eq-II-96.3})\|
    \; , \; \;
    \|(\ref{eq-II-96.4})-(\ref{eq-II-96.5})\|
    \, \leq \, \cO( \, (\sigma_{t_1})^{\rho} \, t_1\, \ln t_2  \, )
\end{equation}
for some $\rho>0$. To this end, we use:
\begin{itemize}
\item[i)]
The H\"older regularity of $\Phi_{\vec P}^\sigma$ and $\vnabla E_{\vec P}^\sigma$
described under condition $(\mathscr{I}2)$ in Theorem {\ref{thm-cfp-2}}.
\item[ii)]
The regularity of the phase function
\begin{equation}
    \gamma_{\sigma_{t_2}}(\vv_j,\vnabla
    E_{\vec{P}}^{\sigma_{t_2}},t_1)
\end{equation}
with respect to $\vec{P}\in supp h \subset \mathcal{S}$ expressed in the following estimate, which
is similar to (\ref{eq-A-3}) in Lemma \ref{lemm-A1}: For $\vk \in \cB_{\sigma_{t_1}}$ and $t_1$ large enough
\begin{equation}\label{eq-A-3bis}
    \big| \, \gamma_{\sigma_{t_2}}(\vv_j,\vnabla E_{\vec{P}}^{\sigma_{t_2}},t_1)
    \, - \,
    \gamma_{\sigma_{t_2}}(\vv_j,\vnabla E^{\sigma_{t_2}}_{\vec{P}-\vec{k} },t_1) \, \big|
    \, \leq \, \cO( \, |\vk|^{\frac{1}{4}(1-\delta'')}\,t_1^{(1-\theta)} \, ) \,,
\end{equation}
where $0<\theta(<1)$ can be chosen arbitrarily close to $1$.
\item[iii)]
The estimate
\begin{equation}\label{eq-II-97.1}
	\| \, b_{\vk,\lambda}\Psi_{\vP}^{\sigma} \, \| \, \leq \, C \, \frac{\one_{\sigma,\Lambda}(\vk)}{|\vk|^{3/2}}
\end{equation}
from $(\mathscr{I}5)$ in Theorem {\ref{thm-cfp-2}} for $\vP\in\mathcal{S}$,
which implies
\begin{equation}\label{eq-II-97.2}
	\| \, N_f^{1/2} \, \Psi_{\vP}^{\sigma} \, \|
	\, = \,
	\big(\sum_{\lambda} \, \int d^3k \, \| \,
	b_{\vk,\lambda} \, \Psi_{\vP}^{\sigma} \, \|^2 \big)^{1/2}
	\, \leq \, C \, | \, \ln \sigma \, |^{1/2} \,. 
\end{equation}
Likewise,
\begin{eqnarray}\label{eq-II-97.3}
	\| \, N_f^{1/2} \, \Phi_{\vP}^{\sigma} \, \|
	& = &
	\big( \,\sum_{\lambda} \int_{\cB_{\Lambda}\setminus\cB_\sigma} d^3k \, \big\| \,
	\big( \, b_{\vk,\lambda} \, + \, \cO( |\vk|^{-3/2}) \, \big) \,
	\Psi_{\vP}^{\sigma} \, \big\|^2 \big)^{1/2}
	\nonumber\\
	& \leq & C \, | \, \ln \sigma \, |^{1/2} \,,
\end{eqnarray}
which controls the expected photon number in the states
$\lbrace\Phi_{\vP}^{\sigma_{t_1}}\rbrace$.
As a side remark, we note that the true size is in fact $\cO(1)$, uniformly in $\sigma$,
but the logarithmically divergent bound here is sufficient for our purposes.
\item[iv)]
The cell decomposition $\mathscr{G}^{(t_1)}$ is determined by
$t_1<t_2$. Moreover, since $\beta(>1)$ can be chosen
arbitrarily large and independent of $\epsilon$, $\sigma_{t_1}=t_1^{-\beta}$ can be made
as small as desired.
\end{itemize}

We first prove the bound on $\|(\ref{eq-II-96.3})\|$ stated
in  (\ref{eq-II-97}). To this end, we use
\begin{eqnarray}
    \lefteqn{
    \big( \, e^{-i(|\vk|t_1-\vk\cdot\vx)}-\cI \, \big)
    \, e^{i\gamma_{\sigma_{t_2}}(\vv_j,\vnabla E_{\vec{P}}^{\sigma_{t_1}},t_1)}
    \, e^{-iE_{\vec{P}}^{\sigma_{t_1}}t_1}
    \, \Phi_{j,\sigma_{t_1}}^{(t_1)}
    \,
    }
    \\
    & & = \; \;
    e^{i\gamma_{\sigma_{t_2}}(\vv_j,\vnabla E_{\vec{P}-\vk}^{\sigma_{t_1}},t_1)}
    \, e^{-iE_{\vec{P}-\vk}^{\sigma_{t_1}}t_1}(e^{-i(|\vk|t_1-\vk\cdot\vx)}-\cI)
    \, \Phi_{j,\sigma_{t_1}}^{(t_1)}
    \label{eq-II-97.1.1}\\
    & & \; \; + \;
    e^{i\gamma_{\sigma_{t_2}}(\vv_j,\vnabla E_{\vec{P}-\vk}^{\sigma_{t_1}},t_1)}
    \, e^{-iE_{\vec{P}-\vk}^{\sigma_{t_1}}t_1}
    \, \Phi_{j,\sigma_{t_1}}^{(t_1)}
    \label{eq-II-97.1.2}\\
    & & \; \; - \;
    e^{i\gamma_{\sigma_{t_2}}(\vv_j,\vnabla E_{\vec{P}}^{\sigma_{t_1}},t_1)}
    \, e^{-iE_{\vec{P}}^{\sigma_{t_1}}t_1}
    \, \Phi_{j,\sigma_{t_1}}^{(t_1)} \,.
    \label{eq-II-97.1.3}
\end{eqnarray}
The H\"older regularity of $\Phi_{\vP}^{\sigma_{t_1}}$ from i) yields
\begin{equation}
    	\| \, (\ref{eq-II-97.1.1}) \, \| \,
	\leq \, \cO( \,t_1\,\sigma_{t_1}^{(\frac{1}{4}-\delta')}\,t_1^{-\frac{3\epsilon}{2}} \,)\,,
\end{equation}
where $\delta'$ can be chosen arbitrarily small, and independently of $\epsilon$.
The derivation of
a similar estimate is given in the proof of Theorem {\ref{theo-A'2}} in the Appendix,
starting from (\ref{eq-A-29'}), where we refer for details. 
The H\"older continuity of $E_{\vec{P}}^{\sigma_{t_1}}$ and
$\vnabla E_{\vec{P}}^{\sigma_{t_1}}$, again from i), combined
with  ii), with $\theta$ sufficiently close to $1$, implies that, with $\vk\in\cB_{\sigma_{t_1}}$,
\begin{equation}
    \| \, (\ref{eq-II-97.1.2})-(\ref{eq-II-97.1.3}) \, \| \, \leq \,
    \cO( \, t_1\,\sigma_{t_1}^{\frac{1}{5}}\,t_1^{-\frac{3\epsilon}{2}} \, )\,,
\end{equation}
as desired.
\\

To prove the bound on $\| \, (\ref{eq-II-96.4})-(\ref{eq-II-96.5}) \, \|$
asserted in  (\ref{eq-II-97}), we write
\begin{equation}
    W_{\sigma_{t_2}}^{*}(\vnabla
    E^{\sigma_{t_2}}_{\vP-\vk})-W_{\sigma_{t_2}}^{*}(\vnabla
    E^{\sigma_{t_2}}_{\vP}) \, = \, W_{\sigma_{t_2}}^{*}(\vnabla
    E^{\sigma_{t_2}}_{\vP})(W_{\sigma_{t_2}}^{*}(\vnabla
    E^{\sigma_{t_2}}_{\vP-\vk} \, ; \, \vnabla
    E^{\sigma_{t_2}}_{\vP})-\cI)\,,
\end{equation}
where
\begin{eqnarray*}
	W^*_{\sigma_{t_2}}(\vnabla E^{\sigma_{t_2}}_{\vP-\vk} \, ; \,  \vnabla
	E^{\sigma_{t_2}}_{\vP})  
	\, := \, W_{\sigma_{t_2}}(\vnabla E^{\sigma_{t_2}}_{\vP}) \,  W^*_{\sigma_{t_2}}(\vnabla
	E^{\sigma_{t_2}}_{\vP-\vk}) \,,
\end{eqnarray*}
and apply the Schwarz inequality in the form
\begin{eqnarray}\label{eq-97bisbisbis}
	\lefteqn{
	\Big\| \, (W_{\sigma_{t_2}}^{*}(\vnabla
    E^{\sigma_{t_2}}_{\vP+\vk} \, ; \, \vnabla
    E^{\sigma_{t_2}}_{\vP})-\cI) \, \widetilde \Phi \, \Big\|
    }
    \\
    &\leq& C \, \big( \, \int_{\cB_{\sigma_{t_2}}} \frac{ d^3 q }{|\vq|^3}\, \big)^{1/2}
   \,\sup_{\vP\in supp\, h,\,\vk \in \mathcal{B}_{\sigma_{t_1}}} \big| \, \vnabla E^{\sigma_{t_2}}_{\vP-\vk}-\vnabla E^{\sigma_{t_2}}_{\vP} \, \big|
    \, \| \, N_f^{1/2} \widetilde \Phi\, \|
    \nonumber
\end{eqnarray}
where in our case, $\widetilde\Phi\equiv e^{i\gamma_{\sigma_{t_2}}(\vv_j,\vnabla
    E_{\vec{P}}^{\sigma_{t_1}},t_1)}e^{-iE_{\vec{P}}^{\sigma_{t_1}}t_1}
    \, \Phi_{j,\sigma_{t_1}}^{(t_1)}$.
We have
\begin{equation}\label{eq-97bisbis}
    \| N_f^{1/2} \, \Phi_{\vP}^{\sigma_{t_1}}\|\,\leq\, c \,|\ln\sigma_{t_1}|^{1/2}
    \, \leq \, c' \, ( \, \ln t_1 \, )^{1/2} \,,
\end{equation}
as a consequence of iii). Due to i),
\begin{equation}
    \sup_{\vP\in supp\,h,\,\vk \in \mathcal{B}_{\sigma_{t_1}}} \,
    \big| \, \vnabla E^{\sigma_{t_2}}_{\vP-\vk} \, - \, \vnabla E^{\sigma_{t_2}}_{\vP}  \, \big|
    \, \leq \, \cO( \, \sigma_{t_1}^{\frac{1}{4}-\delta''} \, )
\end{equation}
where $\delta''>0$ is arbitrarily small,
and independent of $\epsilon$  (see (\ref{eq-II-6})).
Therefore,
$$
    \sup_{\vk\in\mathcal{B}_{\sigma_{t_1}}}\,
    \big\| \, (\ref{eq-II-96.4})-(\ref{eq-II-96.5}) \, \big\|\,
    \leq \, \cO((\ln t_{2}) \,(\sigma_{t_1})^{\rho'})
$$
for some $\rho'>0$ which does not depend on $\epsilon$ (recalling that $t_1<t_2$).
\\

We may now return to (\ref{eq-II-95bis}). From iv), and the fact that
the number of cells is $N(t_1)\approx t_1^{3\epsilon}$, summation over all cells yields
\begin{equation}
    \sum_{j=1}^{N(t_1)}\big\| \, (\ref{eq-II-93bis}) \, - \, (\ref{eq-II-93bisbis}) \, \big\|
    \, \leq \, \cO(\frac{\ln(t_2)}{t_1^\rho})
    \label{eq-II-98}
\end{equation}
for some $\rho>0$, provided that $\beta$ is sufficiently large.
This agrees with (\ref{eq:IRchange-ineq-1}).

The sum $\sum_{j=1}^{N(t_1)}\| \, (\ref{eq-II-96.1bis}) \, \|$ can be treated in a similar way.
\\

\noindent
%%%%%%%%%%%%%%%%%%%%%%%%%%%%%%%%%%%%%%%%%%%%%%%%%%%
\emph{\underline{Analysis of Step b)}}
%%%%%%%%%%%%%%%%%%%%%%%%%%%%%%%%%%%%%%%%%%%%%%%%%%%
$\;$\\
\\
To show that the norm difference of the two vectors
corresponding to the change (\ref{eq-II-92}) in (\ref{eq-II-89bis})
is bounded by the r.h.s. of  (\ref{eq-II-72}), we argue similarly as for step {\em a)},
and we shall not reiterate the details.
One again uses properties i) -- iv) as in step {\em a)}. 
\\

%\newpage
\noindent
%%%%%%%%%%%%%%%%%%%%%%%%%%%%%%%%%%%%%%%%%%%%%%%%%%%
\emph{\underline{Analysis of Step c)}}
%%%%%%%%%%%%%%%%%%%%%%%%%%%%%%%%%%%%%%%%%%%%%%%%%%%
$\;$\\
\\
Finally, we prove that the difference of the vectors corresponding to (\ref{eq-II-93}) satisfies
\begin{eqnarray}\label{eq-II-99}
	&&\Big\| \, \sum_{j=1}^{N(t_1)}\cW_{\Lambdas,\sigma_{t_1}}(\vv_j,t_1)
	\big[W|_{\sigma_{t_2}}^{\sigma_{t_1}}(\vv_j) W^{*}|_{\sigma_{t_2}}^{\sigma_{t_1}}
	(\vnabla E^{\sigma_{t_1}}_{\vP})-\cI\big] 
	\, e^{i\gamma_{\sigma_{t_1}}(\vv_j,\vnabla
  	E_{\vec{P}}^{\sigma_{t_1}},t_1)}
	e^{-iE_{\vec{P}}^{\sigma_{t_1}}t_1}\psi_{j,\sigma_{t_1}}^{(t_1)} \, \Big\|^2
	\nonumber\\
	&&\quad\quad\quad\quad\quad\quad\, \leq \, \cO\big((\ln(t_2))^2 /t_1^{\rho}\big) \,,
\end{eqnarray}
where we define
\begin{eqnarray}
    W|_{\sigma_{t_2}}^{\sigma_{t_1}}(\vv_j)
    \, := \, W^{*}_{\sigma_{t_1}}(\vv_j)W_{\sigma_{t_2}}(\vv_j)\,,
    \label{eq-II-100}
\end{eqnarray}
and likewise,
\begin{eqnarray}
    W^{*}|_{\sigma_{t_2}}^{\sigma_{t_1}}(\vnabla E^{\sigma_{t_1}}_{\vP})
    \, := \,
    W_{\sigma_{t_2}}^{*}(\vnabla E^{\sigma_{t_1}}_{\vP})
    \, W_{\sigma_{t_1}}(\vnabla E^{\sigma_{t_1}}_{\vP})\,.
    \label{eq-II-101}
\end{eqnarray}
We separately discuss the diagonal and off-diagonal contributions
to  (\ref{eq-II-99}) from the sum over cells in $\mathscr{G}^{(t_1)}$.
\\

\noindent
$\bullet$ {\em The diagonal terms in  (\ref{eq-II-99}).}
\\
\\
To bound the diagonal terms in (\ref{eq-II-99}),  
we use that,
with $\vv_j\equiv \vnabla E^{\sigma_{t_1}}|_{\vP=\vP^*_j}$,
\begin{equation}
    W|_{\sigma_{t_2}}^{\sigma_{t_1}}(\vv_j \ ; \, \vnabla E^{\sigma_{t_1}}_{\vP})
    \, := \,
    W|_{\sigma_{t_2}}^{\sigma_{t_1}}(\vv_j)
    W^{*}|_{\sigma_{t_2}}^{\sigma_{t_1}}(\vnabla
    E^{\sigma_{t_1}}_{\vP})  \,
\end{equation}
allows for an estimate similar to (\ref{eq-97bisbisbis}),
where we now use that
\begin{equation}\label{eq-II-102}
    \sup_{\vP\in\mathscr{G}^{(t_1)}_{j}}|\vnabla E^{\sigma_{t_1}}_{\vP}-\vv_j|
    \, \leq \, \cO( \, t_1^{-\epsilon(\frac{1}{4}-\delta'')} \, ) \,.
\end{equation}
The latter follows from the H\"older regularity of $\vnabla E^{\sigma}_{\vP}$,
due to condition  $(\mathscr{I}2)$ in Theorem {\ref{thm-cfp-2}}; see (\ref{eq-II-6}). 
Moreover, we use (\ref{eq-97bisbis}) to bound the expected photon number in the states
$\lbrace\Psi_{\vP}^{\sigma_{t_1}}\rbrace$. 

Hereby we find that the sum of diagonal terms can be bounded by
\begin{equation}
    \cO( \, N(t_1)\,\|\psi_{j,\sigma_{t_1}}^{(t_1)}\|^2 \, (\ln t_2)
    \, t_1^{-\epsilon(\frac{1}{4}-\delta'')} \, )
    \, \leq \, \cO( \,  t_1^{-\rho} \, \ln t_2\, )
\end{equation}
for some $\rho>0$, using $N(t_1)=\cO( t_1^{3\epsilon})$,
and $\|\psi_{j,\sigma_{t_1}}^{(t_1)}\|^2=\cO( t_1^{-3\epsilon})$.
\\

\noindent
$\bullet$ {\em The off-diagonal terms in  (\ref{eq-II-99}).}
\\
\\
Next, we bound the off-diagonal terms in (\ref{eq-II-99}), corresponding
to the inner product of vectors supported on cells $j\neq l$
of the partition $\mathscr{G}^{(t_1)}$.
Those are similar to the off-diagonal terms $\widehat{M}_{l,j}^{1}(t,s)$
in (\ref{eq-II-25}) that were discussed in detail previously.
Correspondingly,
we can apply the methods developed in Section {\ref{sec-II.2.1}}, up to
some modifications which we explain now.

Our goal is to prove the asymptotic orthogonality of the off-diagonal terms in  (\ref{eq-II-99}). 

We first of all prove the auxiliary result
\begin{equation}\label{eq-II-103}
    \lim_{s\to+\infty}\| \, \vec{a}_{\sigma_{t_1}}(\overline{\vec{\eta}_{l,j}})(s)\,
    W^{*}|_{\sigma_{t_2}}^{\sigma_{t_1}}(\vnabla
    E^{\sigma_{t_1}}_{\vP})e^{-iH^{\sigma_{t_1}}s}\psi_{j,\sigma_{t_1}}^{(t_1)} \, \| \, = \, 0\,.
\end{equation}
To this end, we compare
\begin{equation}\label{eq-II-104}
    W^{*}|_{\sigma_{t_2}}^{\sigma_{t_1}}(\vnabla
    E^{\sigma_{t_1}}_{\vP})\one_{\mathscr{G}^{(t_1)}_j}(\vP) \,
\end{equation}
(where $\one_{\mathscr{G}^{(t_1)}_j}$ is the characteristic function of the cell
$\mathscr{G}^{(t_1)}_j$) to its discretization:
\begin{itemize}
\item[{\em 1.}]
We pick $\bar{t}$ large enough such that
$\mathscr{G}^{(\bar{t})}$ is a sub-partition of $\mathscr{G}^{(t)}$; in
particular,
$\mathscr{G}^{(t)}_j=\sum_{m(j)=1}^{M}\mathscr{G}^{(\bar{t})}_{m(j)}$,
where $M=\frac{N(\bar{t})}{N(t)}$.
\item[{\em 2.}]
Furthermore, defining $\vec{u}_{m(j)}\, :=\, \vnabla
E^{\sigma_{t_1}}_{\vP^*_{m(j)}}$, where $\vP_{m(j)}^*$ is the center
of the cell $\mathscr{G}^{(\bar{t})}_{m(j)}$, we have, for $\vP\in\mathscr{G}^{(\bar{t})}_{m(j)}$,
\begin{equation}\label{eq-II-105}
    |\vec{u}_{m(j)}-\vnabla E^{\sigma_{t_1}}_{\vP}| \, \leq \,
    C \,\big(\frac{1}{\bar{t}}\big)^{\epsilon\,(\frac{1}{4}-\delta'')}\, ,
\end{equation}
where $C$ is uniform in $t_1$.  
\item[{\em 3.}]
We define
\begin{equation}\label{eq-II-106}
    \mathbb{W}_{\sigma_{t_2}}^{\sigma_{t_1}}(M)\,:=\,
    \sum_{m(j)=1}^{M}W^{*}|_{\sigma_{t_2}}^{\sigma_{t_1}}(\vec{u}_{m(j)})\,
    \one_{\mathscr{G}^{(\bar{t})}_{m(j)}}(\vP)\,
\end{equation}
\noindent
and rewrite the vector
\begin{equation}\label{eq-II-107}
    \vec{a}_{\sigma_{t_1}}(\overline{\vec{\eta}_{l,j}})(s)\,W^{*}|_{\sigma_{t_2}}^{\sigma_{t_1}}
    (\vnabla E^{\sigma_{t_1}}_{\vP})e^{-iH^{\sigma_{t_1}}s}\psi_{j,\sigma_{t_1}}^{(t_1)}
\end{equation}
in (\ref{eq-II-103}) as
\begin{eqnarray}\label{eq-II-108}
    & &\sum_{\lambda}\int_{\mathcal{B}_{\Lambdas }\setminus\mathcal{B}_{\sigma_{t_1}}}
    d^3k \, e^{-i\vk\cdot\vx}\,\big[W^{*}|_{\sigma_{t_2}}^{\sigma_{t_1}}(\vnabla
    E^{\sigma_{t_1}}_{\vP})-\mathbb{W}_{\sigma_{t_2}}^{\sigma_{t_1}}(M)\big]\times\quad\quad\quad\quad\\
    & &\quad\quad\quad\quad\quad\quad\times \, \vec{\eta}_{l,j}(\vk)\cdot\veps_{\vk, \lambda}^{\;*}b_{\vk,
    \lambda}e^{i|\vk|s} \, e^{-iE_{\vP}^{\sigma_{t_1}}s}\psi_{j,\sigma_{t_1}}^{(t_1)}\nonumber\\
    &+&\sum_{m(j)=1}^{M}W^{*}|_{\sigma_{t_2}}^{\sigma_{t_1}}(\vec{u}_{m(j)})
    \sum_{\lambda}\int_{\mathcal{B}_{\Lambdas }\setminus\mathcal{B}_{\sigma_{t_1}}}\,
    d^3k \, \vec{\eta}_{l,j}(\vk)\cdot\veps_{\vk, \lambda}^{\;*} a_{\vk,
    \lambda}e^{i|\vk|s} \, \times
    \nonumber\\
    & &\quad\quad\quad\quad\quad\quad \times \,
    e^{-iE_{\vP}^{\sigma_{t_1}}s}\psi_{m(j),\sigma_{t_1}}^{(\bar{t})}\quad . \quad\,\label{eq-II-109}
\end{eqnarray}
\end{itemize}
We now observe that, at fixed $t_1$, (\ref{eq-II-9}) and the bound (\ref{eq-II-105}) imply that
the vector in (\ref{eq-II-108}) converges to the zero vector as
$\bar{t}\to+\infty$, uniformly in $s$.
Moreover, the norm of the vector in (\ref{eq-II-109}) tends to zero, as
$s\to+\infty$, at fixed $\bar{t}$.
This proves (\ref{eq-II-103}).

The main difference between  (\ref{eq-II-99}) and the similar expression 
in (\ref{eq-II-40}) that is differentiated in $s$
is the operator
\begin{equation}\label{eq-II-111}
    W|_{\sigma_{t_2}}^{\sigma_{t_1}}(\vv_j)W^{*}|_{\sigma_{t_2}}^{\sigma_{t_1}}
    (\vnabla E^{\sigma_{t_1}}_{\vP}) \,,
\end{equation}
which is absent in (\ref{eq-II-40}). To control it,
we first note that the Hamiltonian
\begin{equation}\label{eq-II-112}
\widehat{H}^{\sigma_{t_1}}:=\int^{\oplus}\widehat{H}_{\vP}^{\sigma_{t_1}}d^3P\,,
\end{equation}
where
\begin{equation}\label{eq-II-113}
\widehat{H}_{\vP}^{\sigma_{t_1}}\, := \,
\frac{\big(\vec{P}-\vec{P}^{f}_{>\sigma_{t_1}}
+\alpha^{1/2} \vec{A}^{\sigma_{t_1}} \big)^2}{2}
\; + \;H^{f}_{>\sigma_{t_1}}
\end{equation}
with
\begin{equation}\label{eq-II-114}
    \vec{P}^{f}_{>\sigma_{t_1}}:=\int_{\mathbb{R}^3\setminus\mathcal{B}_{\sigma_{t_1}}}\vk\,
    b^*_{\vk, \lambda}b_{\vk, \lambda}\,d^3k \,,
\end{equation}
and
\begin{equation}\label{eq-II-115}
    H_{>\sigma_{t_1}}^{f}\,:=\int_{\mathbb{R}^3\setminus\mathcal{B}_{\sigma_{t_1}}}|\vk|\,
    b^*_{\vk, \lambda}b_{\vk, \lambda}\,d^3k \,,
\end{equation}
satisfies
\begin{equation}\label{eq-II-116}
	\widehat{H}_{\vP}^{\sigma_{t_1}}\,\Psi_{\vP}^{\sigma_{t_1}}
	\,=\,E_{\vec{P}}^{\sigma_{t_1}}\,\Psi_{\vP}^{\sigma_{t_1}}\,,
\end{equation}
and
\begin{equation}\label{eq-II-117}
[W|_{\sigma_{t_2}}^{\sigma_{t_1}}(\vv_j)W^{*}|_{\sigma_{t_2}}^{\sigma_{t_1}}(\vnabla
E^{\sigma_{t_1}}_{\vP})\,,\,\widehat{H}^{\sigma_{t_1}}]=0\,.
\end{equation}
Using (\ref{eq-II-117}), the vector in (\ref{eq-II-99}) corresponding to the $j$-th cell
can be written as
\begin{equation}\label{eq-II-118}
	e^{i \widehat{H}^{\sigma_{t_1}}s} \, \cW_{\Lambdas,\sigma_{t_1}}(\vv_j,s)
	\, e^{i\gamma_{\sigma_{t_1}}(\vv_j,\nabla E_{\vec{P}}^{\sigma_{t_1}},s)}
	\, e^{-i\widehat{H}^{\sigma_{t_1}}s}
  	\, W|_{\sigma_{t_2}}^{\sigma_{t_1}}(\vv_j \, ; \, \vnabla E^{\sigma_{t_1}}_{\vP})
	\, \psi_{j,\sigma_{t_1}}^{(t)}\,,
\end{equation}
where we recall that
$$
	W|_{\sigma_{t_2}}^{\sigma_{t_1}}(\vv_j \, ; \, \vnabla E^{\sigma_{t_1}}_{\vP})
	\, = \, W|_{\sigma_{t_2}}^{\sigma_{t_1}}(\vv_j)
	W^{*}|_{\sigma_{t_2}}^{\sigma_{t_1}}(\vnabla E^{\sigma_{t_1}}_{\vP}) \,.
$$

Similarly to our strategy in Section {\ref{sec-II.2.1}}, we control
$\widehat{M}_{l,j}^{1}(t,s)$  by integrating a sufficiently strong bound on
$|\frac{d}{ds}(\widehat{M}_{l,j}^{1}(t,s))|$ with respect to
$s$ over $[t,\infty)$; see (\ref{eq-II-37}).
The derivative in $s$ of the $j$-th cell vector has the form
\begin{eqnarray}
    \lefteqn{\frac{d}{ds}\Big( \, e^{i\widehat{H}^{\sigma_{t_1}}s}
    \, \cW_{\Lambdas,\sigma_{t_1}}(\vv_j,s)
    \, e^{i\gamma_{\sigma_{t_1}}(\vv_j,\vnabla E_{\vec{P}}^{\sigma_{t_1}},s)}
    \, e^{-iE_{\vec{P}}^{\sigma_{t_1}}s}
    \, W|_{\sigma_{t_2}}^{\sigma_{t_1}}(\vv_j \, ; \, \vnabla E^{\sigma_{t_1}}_{\vP})
    \, \psi_{j,\sigma_{t_1}}^{(t_1)}\,
    \Big)\, }
    \nonumber\\
    &=&\; \; 
    i \, e^{i\widehat{H}^{\sigma_{t_1}}s}
    \, \cW_{\Lambdas,\sigma_{t_1}}(\vv_j,s)
    \, \alpha\, i[\widehat{H}^{\sigma_{t_1}},\vx] \, \cdot
    \int_{\mathcal{B}_{\Lambdas }\setminus\mathcal{B}_{\sigma_{t_1}}}\vec{\Sigma}_{\vv_j}(\vk)
    \cos(\vk\cdot\vx-|\vk|s) \, d^3k
    \times\nonumber\\
    & &\quad\quad\quad\quad\quad\quad\quad\quad\quad
    \times \,
    e^{-iE_{\vec{P}}^{\sigma_{t_1}}s}
    \, e^{i\gamma_{\sigma_{t_1}}(\vv_j,\nabla E_{\vec{P}}^{\sigma_t},s)}
    \, W|_{\sigma_{t_2}}^{\sigma_{t_1}}(\vv_j \, ; \, \vnabla E^{\sigma_{t_1}}_{\vP})
    \, \psi_{j,\sigma_t}^{(t)}\nonumber\\
    &&+
    \, i \, e^{i\widehat{H}^{\sigma_{t_1}}s}
    \, \cW_{\Lambdas,\sigma_{t_1}}(\vv_j,s) \, \alpha^{2}\,
    \int_{\mathcal{B}_{\Lambdas }\setminus\mathcal{B}_{\sigma_{t_1}}}
    \vec{\Sigma}_{\vv_j}(\vk) \, \cos(\vk\cdot\vx-|\vk|s) \, d^3k \, \cdot
    \nonumber\\
    & &\quad\quad\quad
    \cdot \, \int_{\mathcal{B}_{\Lambdas }\setminus\mathcal{B}_{\sigma_{t_1}}}
    \vec{\Sigma}_{\vv_j}(\vq)\cos(\vq\cdot\vx-|\vq|s)d^3q \, \times
    \label{eq-II-119}\\
    & & \quad\quad\quad\quad\quad\quad\quad\quad\quad
    \times
    \, e^{i\gamma_{\sigma_{t_1}}(\vv_j,\nabla E_{\vec{P}}^{\sigma_t},s)}
    \, e^{-iE_{\vec{P}}^{\sigma_{t_1}}s}
    \, W|_{\sigma_{t_2}}^{\sigma_{t_1}}(\vv_j \, ; \, \vnabla E^{\sigma_{t_1}}_{\vP})
    \, \psi_{j,\sigma_{t_1}}^{(t)} \, \nonumber
    \\
    & &+
    \, i \, e^{i\widehat{H}^{\sigma_{t_1}}s}
    \, \cW_{\Lambdas,\sigma_{t_1}}(\vv_j,s)
    \, \frac{d\gamma_{\sigma_{t_1}}(\vv_j,\nabla E_{\vec{P}}^{\sigma_{t_1}},s)}{ds} \, \times
    \nonumber\\
    & & \quad\quad\quad\quad\quad\quad\quad\quad\quad
    \times
    \, e^{i\gamma_{\sigma_{t_1}}(\vv_j,\vnabla E_{\vec{P}}^{\sigma_{t_1}},s)}
    \, e^{-iE_{\vec{P}}^{\sigma_{t_1}}s}
    \, W|_{\sigma_{t_2}}^{\sigma_{t_1}}(\vv_j \, ; \, \vnabla E^{\sigma_{t_1}}_{\vP})
    \, \psi_{j,\sigma_{t_1}}^{(t_1)}\,.
    \nonumber
\end{eqnarray}
Due to the similarity of this expression with (\ref{eq-II-40}) -- (\ref{eq-II-45}),
we can essentially adopt the analysis presented in Section {\ref{sec-II.2.1}}.
The only difference here is the operator
$\widehat{H}^{\sigma_{t_1}}$ instead of $H^{\sigma_{t_1}}$, and the
additional term involving the commutator
\begin{equation}\label{eq-II-120}
    [\chi_{h}(\frac{\vx}{s})\,,\,W^{*}|_{\sigma_{t_2}}^{\sigma_{t_1}}(\vnabla
    E^{\sigma_{t_1}}_{\vP})]
\end{equation}
applied to the one-particle state
\begin{equation}\label{eq-II-121}
    e^{i\gamma_{\sigma_{t_1}}(\vv_j,\vnabla E_{\vec{P}}^{\sigma_{t_1}},s)}
    \, e^{-iE_{\vec{P}}^{\sigma_{t_1}}s}
    \, \psi_{j,\sigma_{t_1}}^{(t_1)}\,.
\end{equation}
However, the latter tends to zero as $s\to+\infty$, at a rate 
$\cO(\frac{1}{s^{\eta}})$, for some $\epsilon$-independent $\eta>0$.
This follows from the H\"older regularity of $\vnabla E^{\sigma}_{\vP}$
(condition $(\mathscr{I}2)$ in Theorem {\ref{thm-cfp-2}}), and (\ref{eq-II-9}).
Similarly, we treat the commutator (\ref{eq-II-120}) with the infrared tail (\ref{eq-II-52bis}) in place of
$\chi_{h}(\frac{\vx}{s})$ (and with $\widehat{H}^{\sigma_{t_1}}$ replacing $H^{\sigma_{t_1}}$).
It is then straightforward to see that we arrive at (\ref{eq-II-72}).
\QED

%%%%%%%%%%%%%%%%%%%%%%%%%%%%%%%%%%%%%%%%%%%%%%%%%%%%%%%%%%%%%%%%%%%%%%%%%%%%%%%%%%%%%%%%%%%%%%%%%%%%%%%%%%%%%%%%%%%%%%%%%%%%%%%%%%%%%%%%%%%%%%%%%%%%%%%%%%%%%%%%%%%%%%%%%%%%%%%%%%%%%%%%%%%%%%%%%%%%%%%%%%%%%%%%%%%%%%%%%%%%%%%%%%%%%%%%%%%%%%%%%%%%%%%%%%%%Scattering---subspaces%%%%%%%%%%%%%%%%%%%%%%%%%%%%%%%%%%%%%%%%%%%%%%%%%%%%%%%%%%%%%%%%%%%%%%%%%%%%%%%%%%%%%%%%%%%%%%%%%%%%%%%%%%%%%%%%%%%%%%%%%%%%%%%%%%%%%%%%%%%%%%%%%%%%%%%%%%%%%%%%%%%%%%%%%%%%%%%%%%%%%%%%%%%%%%%%%%%%%%%%%%%%%%%%%%%%%%%%%%%%%%%%%%%%%%%%%%%%%%%%%%%%%%%%%%%%%%%%%%%%%%%%%%%%%%%%%%%%%%%%%%%%%%%
\newpage

\section{Scattering subspaces and asymptotic observables} \label{sec-III}
\resetequ

This section is dedicated to the following key constructions in the scattering
theory for an infraparticle with the quantized electromagnetic field:
\\

\begin{itemize}
\item[i)]
We define scattering subspaces $\cH^{out/in}$ which are invariant under space-time
translations, built from vectors $\lbrace\, \Psi_{h,\Lambdas }^{out/in}\,\rbrace$.

To this end, we first define a subspace, $\cHs_{\Lambdas }$,
depending on the choice of a threshold frequency $\Lambdas $ with the following purpose: Apart from
photons with energy smaller than $\Lambdas $, this subspace contains states describing
only a freely moving (asymptotic) electron.

Adding asymptotic photons to the states in
$\cHs_{\Lambdas }$, we define spaces of scattering states of the system,
where the asymptotic electron velocity is restricted to the
region $\lbrace \vnabla E_{\vP} \, : \, |\vP|<\frac{1}{3} \rbrace$ .

We note that the choice of $\cHs_{\Lambdas }$ is not unique,
except for the behavior of the dressing photon cloud in the infrared
limit. It is useful because
\begin{itemize}
\item
in the construction of the spaces of scattering states, we can separate ``hard photons'' from
the photon cloud present in the states
in  $\cHs_{\Lambdas }$, which is not completely removable --
each state in the scattering spaces contains an {\em infinite number} of
asymptotic photons.
\item
from the physical point of view, every experimental setup is limited by
a threshold energy $\Lambdas $ below which photons
cannot  be measured.
\\
\end{itemize}

\item[ii)]
The construction of asymptotic algebras of observables,
$\cA^{out/in}_{ph}$ and $\cA^{out/in}_{el}$, related to
the electromagnetic field and to the electron, respectively.

The asymptotic algebras are
\begin{itemize}
\item
the Weyl algebra, $\cA^{out/in}_{ph}$, associated to the asymptotic
electromagnetic field;
\item
the algebra $\cA^{out/in}_{el}$ generated by smooth functions of compact support
of the asymptotic velocity of the electron.
\end{itemize}

The two algebras $\cA^{out/in}_{ph}$ and $\cA^{out/in}_{el}$ commute.
This is the mathematical counterpart of the
asymptotic decoupling between the photons and the electron. This decoupling
is, however, far from trivial: In fact, in contrast to
a theory with a mass gap or a theory where the interaction with the soft
modes of the field is turned off, the system is characterized by  the emission of soft photons for arbitrarily
long times.

In this respect, the asymptotic convergence of the electron velocity is
a new conceptual result, obtained from the solution of the infraparticle problem in
a concrete model, here non-relativistic QED.
Furthermore, the emission of soft photons for arbitrarily
long times is reflected in the representation of the asymptotic
electromagnetic algebra, which is non-Fock but only locally Fock (see Section {\ref{ssec-III.2}}).
More precisely, the representation
can be decomposed on the spectrum of the asymptotic velocity of the electron; for different values
of the asymptotic velocity,  the representations turn
out to be inequivalent. Only for $\vnabla E_{\vP}=0$, the representation is
Fock, otherwise they are coherent non-Fock. The coherent photon cloud,
labeled by the asymptotic velocity, is the well known Bloch-Nordsieck cloud.
\\
\end{itemize}

All the results and definition clearly hold for both the \emph{out} and the
\emph{in}-states. We shall restrict ourselves to the discussion of
\emph{out}-states.

%%%%%%%%%%%%%%%%%%%%%%%%%%%%%%%%%%%%%%%%%%%%%%%%%%%%%%%%%%%%%%%%%%%%%%%%%%%%%%
\subsection{Scattering subspaces and ``One-particle'' subspaces with counter threshold $\Lambdas $}
%%%%%%%%%%%%%%%%%%%%%%%%%%%%%%%%%%%%%%%%%%%%%%%%%%%%%%%%%%%%%%%%%%%%%%%%%%%%%%
In Section {\ref{sec-II}}, we have constructed a scattering state with electron
wave function $h$, and a dressing cloud exhibiting the correct
behavior in the limit $\vk\to0$, with maximal photon frequency $\Lambdas $.

To construct a space which is invariant under space-time translations,
we may either focus on the vectors
\begin{equation}
	e^{-i\vec{a}\cdot\vec{P}}e^{-iH\tau}\psi_{h,\Lambdas }^{out}\, ,
	\label{eq-III-1}
\end{equation}
or on the vectors obtained from
\begin{equation}
	s-\lim_{t\to+\infty}e^{iHt}\sum_{j=1}^{N(t)}\cW_{\Lambdas,\sigma_{t}}^{\tau,\,\vec{a}}(\vv_j,t)
	e^{i\gamma_{\sigma_{t}}(\vv_j,\vnabla
  	E_{\vec{P}}^{\sigma_{t}},t)}e^{-iE_{\vec{P}}^{\sigma_{t}}t}
	\psi_{j,\sigma_{t}}^{(t)}(\tau,\vec{a})\,,
	\label{eq-III-2}
\end{equation}
where
\begin{eqnarray} 
	\cW_{\Lambdas,\sigma_{t}}^{\tau,\vec{a}}(\vv_j,t) \, 
	\label{eq-III-3} 
	\, := \, \exp\Big( \, \alpha^{\frac{1}{2}}
	\sum_{\lambda}\int_{\mathcal{B}_{\Lambdas }\setminus\mathcal{B}_{\sigma_{t}}}
	\frac{d^3k}{\sqrt{|\vk|}}
	\frac{\vv_j\cdot\lbrace\veps_{\vk, \lambda}
	a_{\vk,\lambda}^*e^{-i|\vk|(t+\tau)}
	e^{-i\vk\cdot\vec{a}}-h.c.\rbrace}{|\vk|(1-\widehat{k}\cdot\vv_j)} \, \Big)\,, 
\end{eqnarray}
and
\begin{eqnarray}
	\psi_{j,\sigma_{t}}^{(t)}(\tau,\vec{a}) \, :=
	\, \int_{\mathscr{G}^{(t)}_j}e^{-i\vec{a}\cdot\vec{P}}e^{-iE_{\vec{P}}^{\sigma_{t}}\tau}\,
	h(\vec{P})\psi_{\vec{P}}^{\sigma_{t}}d^3P \,.
	\label{eq-III-4}
\end{eqnarray}
Using Theorem~\ref{theo-II.2}, one straightforwardly finds that
\begin{eqnarray}
	\lefteqn{e^{-i\vec{a}\cdot\vec{P}}e^{-iH\tau}\psi_{h,\Lambdas }^{out} }
	\label{eq-III-5}
	\\
	&=&s-\lim_{t\to+\infty}e^{-i\vec{a}\cdot\vec{P}}e^{-iH\tau}e^{iHt}\sum_{j=1}^{N(t)}
	\cW_{\Lambdas,\sigma_t}(\vv_j,t)e^{i\gamma_{\sigma_t}(\vv_j,\vnabla
  	E_{\vec{P}}^{\sigma_t},t)}e^{-iE_{\vec{P}}^{\sigma_t}t}\psi_{j,\sigma_t}^{(t)}
  	\quad\quad\quad
	\label{eq-III-6}
	\\
	&=&s-\lim_{t\to+\infty}e^{iHt}\sum_{j=1}^{N(t+\tau)}\cW_{\Lambdas,\sigma_{t+\tau}}^{\tau,\,\vec{a}}(\vv_j,t)
	e^{i\gamma_{\sigma_{t+\tau}}(\vv_j,\vnabla
  	E_{\vec{P}}^{\sigma_{t+\tau}},t+\tau)} \, 
  	\label{eq-III-7} 
	e^{-iE_{\vec{P}}^{\sigma_{t+\tau}}t}
  	\psi_{j,\sigma_{t+\tau}}^{(t+\tau)}(\tau,\vec{a})\,. 
\end{eqnarray}
The two limits (\ref{eq-III-2}) and (\ref{eq-III-7}) coincide; this follows
straightforwardly from the line of analysis presented in the previous section.

Therefore, we can define the ``one-particle'' space corresponding to the frequency
threshold $\Lambdas $ as
\begin{equation}
	\cHs_{\Lambdas }:=\overline{\Big\lbrace
  	\bigvee\,\psi_{h,\Lambdas }^{out/in}(\tau,\vec{a})\,:h(\vP)\in\, C_{0}^{1}
  	(\cS\setminus\mathcal{B}_{r_{\alpha}})\,,\,\tau\in\RR,\,\vec{a}\in\RR^3\Big\rbrace}\,.\label{eq-III-8}
\end{equation}
By construction, $\cHs_{\Lambdas }$ is invariant under space-time
translations.

General scattering states of the system can contain
an arbitrarily large number of ``hard'' photons, i.e., photons with an
energy above a frequency threshold, say for instance $\Lambdas $.
One can construct such states based on $\cHs_{\Lambdas }$
according to the following procedure.

We consider positive energy solutions of the form
\begin{equation}\label{eq-III-10}
	F_{t}(\vy) \, := \, \int \frac{d^3k}{2 \, (2\pi)^{3}\sqrt{|k|}} \, \widehat{F}(\vk) \, e^{-i|k|t+i\vk\cdot\vy}
\end{equation}
of the free wave equation
\begin{equation}\label{eq-III-9}
	\vnabla_{\vy}\cdot\vnabla_{\vy}F_{t}(\vy)-
	\frac{\partial^{2}F_{t}(\vy)}{\partial t^{2}} \, = \, 0 \, ,
\end{equation}
which exhibit fast decay in $|\vy|$ for arbitrary fixed $t$, and where
$\widehat{F}(\vk)\in C_0^{\infty}(\RR^3\backslash\lbrace0\rbrace)$.

We then construct vector-valued test functions
\begin{equation}\label{eq-III-11}
	\vF_{t}(\vy) \, := \, \sum_{\lambda = \pm}\int \frac{d^3k}{2 \, (2\pi)^{3}\sqrt{|k|}} \,
	\veps_{\vk, \lambda}^{\;*}\widehat{F}^{\lambda}(\vk) \, e^{-i|k|t+i\vk \cdot \vy}
\end{equation}
satisfying the wave equation (\ref{eq-III-9}), with
\begin{equation}\label{eq-III.12bis}
	\widehat{\vF}(\vk) \, := \, \sum_{\lambda}\veps_{\vk, \lambda}^{\;*}
	\widehat{F}^{\lambda}(\vk) \; \; \in \; \;
	C_0^{\infty}(\RR^3\backslash\lbrace0\rbrace\, ; \, \CC^3)
\end{equation}

We set
\begin{equation}\label{eq-II-13'}
	\vA(t,\vy) \, := \,  e^{iHt}\vA(\vy)e^{-iHt} \, ;
\end{equation}
here $\vA(\vy)$ is the expression in (\ref{eq-I-14}) with $\Lambda=\infty$.
An asymptotic vector potential is constructed starting from LSZ ($t\to\pm\infty$)
limits of interpolating field operators
\begin{equation}\label{eq-III-12}
	\vA[\vF_{t},t] \, := \, i\int\big(\vA(t,\vy)\cdot\frac{\partial\vF_{t}(\vy)}{\partial t}-
	\frac{\partial\vA(t,\vy)}{\partial t}\cdot\vF_{t}(\vy)\big)d^3y,
\end{equation}
with $\vF_t$ as in (\ref{eq-III-11}) for the positive-energy component,
and with $-\overline{\vF_t}$ for the negative energy component.
We define
\begin{equation}\label{eq-III-14}
	\psi_{h,\vec{F}}^{out/in} \, := \, s-\lim_{t\to +/-\infty}\psi_{h,\vec{F}}(t)\,,
\end{equation}
where
\begin{equation}\label{eq-III-15}
	\psi_{h,\vec{F}}(t) \, := \, e^{i\big(\vA[\vF_{t},t]-\vA[\overline{\vF_t},t]\big)}\psi_{h,\Lambdas }(t)\,.
\end{equation}
Here, $\psi_{h,\Lambdas }(t)$ approximates a vector
$\psi_{h,\Lambdas }^{out}$ in $\cHs_{\Lambdas }$ (we temporarily drop the
 dependence on $(\tau,\vec{a})$ in our notation).
The existence of the limit in (\ref{eq-III-14}) is a straightforward
consequence of standard decay estimates for oscillating integrals under the
assumption in (\ref{eq-III.12bis}), combined with the propagation estimate (\ref{eq-II-49}).

Finally, we can define the scattering subspaces as
\begin{equation}\label{eq-III-17}
	\cH^{out/in} \, := \, \overline{\Big\lbrace \,
  	\bigvee\,\psi_{h,\vec{F}}^{out/in} \, : \,
  	h(\vP)\in\, C_{0}^{1}(\cS\setminus\mathcal{B}_{r_{\alpha}})\,,\,
  	\widehat{\vF}\in\, C_{0}^{\infty}(\RR^3\setminus0\,;\,\CC^3)} \, \Big\rbrace\,.
\end{equation}
$\;$

%\newpage
%%%%%%%%%%%%%%%%%%%%%%%%%%%%%%%%%%%%%%%%%%%%%%%%%%%%%%%%%%%%%%%%%%%%%%%%%%%%%%
%%%%%%%%%%%%%%%%%%%%%%%%%%%%%%%%%%%%%%%%%%%%%%%%%%%%%%%%%%%%%%%%%%%%%%%%%%%%%%%
\subsection{Asymptotic algebras and Bloch-Nordsieck coherent factor}
\label{ssec-III.2}
%%%%%%%%%%%%%%%%%%%%%%%%%%%%%%%%%%%%%%%%%%%%%%%%%%%%%%%%%%%%%%%%%%%%%%%%%%%%%%%
%%%%%%%%%%%%%%%%%%%%%%%%%%%%%%%%%%%%%%%%%%%%%%%%%%%%%%%%%%%%%%%%%%%%%%%%%%%%%%%
We now state some theorems concerning the construction of the asymptotic
algebras. The proofs can be easily derived using the arguments developed in
Sections {\ref{sec-II.2}} and {\ref{sec-II.3}}; for
further details we refer to \cite{Pizzo2005}.
\begin{theorem}
\label{thm-III.1}
The functions $f\in C_{0}^{\infty}(\RR^3)$, of the variable $e^{iHt}\,\frac{\vec{x}}{t}\,e^{-iHt}$, have
strong limits for $t\to\infty$ in $\cH^{out/in}$, namely:
\begin{equation}\label{eq-III-19}
s-\lim_{t\to +/-\infty}e^{iHt}\,f(\frac{\vx}{t})\,e^{-iHt}\psi_{h,\vec{F}}^{out/in}=:\psi_{h\,f_{\vnabla
    E},\vec{F}}^{out/in}
\end{equation}
where $f_{\vnabla E}(\vP):=\lim_{\sigma\to 0}f(\vnabla E_{\vP}^{\sigma})$.
\end{theorem}
The proof is obtained from an adaptation of the proof of Theorem {\ref{theo-A'2}} in
the Appendix.

For the radiation field, we have the following result.
\begin{theorem}
The LSZ Weyl operators
\begin{equation}\label{eq-III-20}
	\Big\lbrace \, e^{i \big(\vA[\vG_{t},t]-\vA[\overline{\vG_t},t] \big)}\,:\,\widehat{G}^{\lambda}(\vk)\in
	L^{2}(\RR^3,(1+|\vk|^{-1})d^3k)\,,\,\lambda=\pm \, \Big\rbrace
\end{equation}
have strong limits in $\cH^{out/in}$:
\begin{equation}\label{eq-III-21}
\cW^{out/in}(\vec{G}) \, := \, s-\lim_{t\to+/-\infty} e^{i \big(\vA[\vG_{t},t]-\vA[\overline{\vG_t},t] \big)}\,.
\end{equation}
The limiting operators are unitary, and have the following properties:
\begin{itemize}
\item[i)]
\begin{equation}\label{eq-III-22}
	\cW^{out/in}(\vec{G})\cW^{out/in}(\vec{G}') \, = \,
	\cW^{out/in}(\vec{G}+\vec{G}')e^{-\frac{\rho(\vec{G},\vec{G}')}{2}}
\end{equation}
where
\begin{equation}
	\rho(\vec{G},\vec{G}')=2iIm(\sum_{\lambda}\int\widehat{G}^{\lambda}(\vk)
	\overline{\widehat{G'}^{\lambda}(\vk)}d^3k) \,.
\end{equation}
$\;$

\item[ii)]
The mapping $\RR\ni s\longrightarrow\cW^{out/in}(s\,\vec{G})$ defines a strongly
continuous, one parameter group of unitary operators.
\\

\item[iii)]
\begin{equation}\label{eq-III-23}
	e^{iH\tau}\cW^{out/in}(\vec{G})e^{-iH\tau}=\cW^{out/in}(\vec{G}_{-\tau})
\end{equation}
where $\vec{G}_{-\tau}$ is a freely evolved, vector-valued test function
in the time $-\tau$.
\\

\end{itemize}
\end{theorem}
Next, we define
\begin{itemize}
\item
$\cA^{out/in}_{el}$ as the norm closure of the (abelian) *algebra generated by the
limits in (\ref{eq-III-19}).

\item
$\cA^{out/in}_{ph}$ as the norm closure of the *algebra generated by the
unitary operators in (\ref{eq-III-21}).
\end{itemize}

>From (\ref{eq-III-22}) and (\ref{eq-III-23}), we conclude
that $\cA^{out/in}_{ph}$ is the Weyl algebra associated to a free radiation
field. Moreover, from straightforward approximation arguments applied to the generators, we can prove
that the two algebras, $\cA^{out/in}_{el}$ and $\cA^{out/in}_{ph}$,  commute.

Moreover, we can next establish key properties of the
representation $\Pi$ of the algebras $\cA^{out/in}_{ph}$ for the
concrete model at hand that confirm structural features
derived in \cite{FrMoSt} under general assumptions.

To study the infrared features of the representation
of  $\cA^{out/in}_{ph}$, it suffices to analyze the expectation of the
generators $\lbrace\,\cW^{out}(\vec{G})\,\rbrace$ of the algebra
with  respect to   arbitrary states of the form
$\psi_{h,\Lambdas }^{out}$,
\begin{eqnarray}
\lefteqn{
    \bra \, \psi_{h,\Lambdas }^{out}\,,\,\cW^{out}(\vec{G}) \, \psi_{h,\Lambdas }^{out} \, \ket \,  }
    \label{eq-III-24}\\
    \lefteqn{= \, \lim_{t\to+\infty}\sum_{j=1,\,l=1}^{N(t)}\Bra \, e^{i\gamma_{\sigma_t}(\vv_{l},\vnabla
    E_{\vec{P}}^{\sigma_t},t)}e^{-iE_{\vec{P}}^{\sigma_t}t}\psi_{l,\sigma_t}^{(t)}\,,}\label{eq-III-25}\\
    & &\,\cW_{\sigma_t}^{*}(\vv_{l},t)
    e^{i\big(\vA[\vG_{t},t]-\vA[\overline{\vG_t},t]\big)}\cW_{\Lambdas,\sigma_t}(\vv_j,t)e^{i\gamma_{\sigma_t}(\vv_j,\vnabla
    E_{\vec{P}}^{\sigma_t},t)}e^{-iE_{\vec{P}}^{\sigma_t}t}\psi_{j,\sigma_t}^{(t)}\, \Ket \,.\nonumber
\end{eqnarray}
In the step passing from (\ref{eq-III-24}) to (\ref{eq-III-25}), we use Theorem~\ref{theo-II.2}.
One infers from the arguments developed in Section {\ref{sec-II.2.1}} that the sum
of the off-diagonal terms, $l\neq j$, vanishes in
the limit. Therefore,
\begin{eqnarray}\label{eq-III-26}
	(\ref{eq-III-25}) & = &
	\lim_{t\to+\infty}\sum_{j=1}^{N(t)}\Bra \, e^{i\gamma_{\sigma_t}(\vv_{j},\vnabla
    E_{\vec{P}}^{\sigma_t},t)}e^{-iE_{\vec{P}}^{\sigma_t}t}\psi_{j,\sigma_t}^{(t)}\,,\\
    & &\quad\quad\quad\,e^{i\big(\vA[\vG_{t},t]-\vA[\overline{\vG_t},t]\big)}
    e^{\varrho_{\vec{v}_j}(\vec{G})}e^{i\gamma_{\sigma_t}(\vv_j,\vnabla
    E_{\vec{P}}^{\sigma_t},t)}e^{-iE_{\vec{P}}^{\sigma_t}t}\psi_{j,\sigma_t}^{(t)}\, \Ket\nonumber\,,
\end{eqnarray}
where
\begin{equation}\label{eq-III-29}
    \varrho_{\vec{u}}(\vec{G})\,:=\,2iRe\big(\alpha^{\frac{1}{2}}\sum_{\lambda}
    \int_{\mathcal{B}_{\Lambdas }}\widehat{G}^{\lambda}(\vk)\frac{\vec{u}\cdot
    \veps_{\vk,\lambda}^{\;*}}{|\vec{k}|^{\frac{3}{2}}(1-\vec{u}\cdot\widehat{k})}d^3k\big)\,.
\end{equation}
After solving an ODE analogous to (\ref{eq-II-29}), we find that the diagonal terms yield
\begin{equation}\label{eq-III-27}
    \bra \, \psi_{h,\Lambdas }^{out}\,,\,\cW^{out}(\vec{G})\psi_{h,\Lambdas }^{out} \, \ket \, = \,
    \int e^{-\frac{C_{\vec{G}}}{2}}e^{\varrho_{\vnabla E_{\vec{P}}}(\vec{G})} \, | h(\vP) |^2 \, d^3P\,,
\end{equation}
where
\begin{equation}\label{eq-III-28}
C_{\vec{G}}\,:=\,\int|\widehat{\vec{G}}(\vk)|^2d^3k\,,
\end{equation}
Here, we also use that $\vv_j\equiv\vnabla E_{\vP_j^*}^{\sigma_t}$, combined with
the convergence $\vnabla E_{\vP}^{\sigma_t}\to\vnabla E_{\vP}$ (as $t\to\infty$
and $\vP\in\mathcal{S}$).

Now, we can
reproduce the following results in \cite{FrMoSt}:
The representation $\Pi(\cA_{ph}^{\;out/in})$ is given by a direct integral on the
spectrum of the operator $\vv_{as}^{\;out/in}$ in $\cH^{out/in}$, defined by
\begin{equation}\label{eq-III-30}
   s- f(\vv_{as}^{\,out/in}) \, := \, \lim_{t\to\,+/-\infty}e^{iHt}f(\frac{\vx}{t})e^{-iHt}
\end{equation}
for any $f\in C_{0}^{\infty}(\RR^3)$,
of mutually inequivalent, irreducible representations.
These representations are coherent non-Fock for values $\vv_{as}^{\;out/in}\neq 0$.
The coherent factors, labeled by $\vv_{as}^{\;out/in}$, are
\begin{equation}\label{eq-III-31}
    \alpha^{\frac{1}{2}}\frac{\vv_{as}^{\;out/in}\cdot
    \veps_{\vk,\lambda}}{|\vec{k}|^{\frac{3}{2}}(1-\vv_{as}^{\;out/in}\cdot\widehat{k})}\,
	\; \; \; \; \; \; \;
	{\rm and }
	\; \; \; \; \; \; \;
    \alpha^{\frac{1}{2}}\frac{\vv_{as}^{\;out/in}\cdot \veps_{\vk,
    \lambda}^{\;*}}{|\vec{k}|^{\frac{3}{2}}(1-\vv_{as}^{\;out/in}\cdot\widehat{k})}\,,
\end{equation}
for the annihilation and the creation part, $a^{out/in}_{\vk,\lambda}$ and
$a^{out/in\,*}_{\vk,\lambda}$, respectively.

The representation $\Pi(\cA_{ph}^{out/in})$ is locally Fock in momentum space. This property is
equivalent to the following one:

For any $\Lambdas >0$,
and $\widehat{\vG^{\Lambdas} }\in\, C_{0}^{\infty}(\RR^3\setminus\mathcal{B}_{\Lambdas }\,;\,\CC^3)$,
the operator
\begin{equation}\label{eq-III-32}
    \vA[-\overline{\vG^{\Lambdas }_t},t]
\end{equation}
annihilates   vectors of the type  $\psi_{h,\Lambdas }^{out}$ in the limit $t\to+\infty$, i.e.,
\begin{equation}\label{eq-III-33}
    \lim_{t\to+\infty}\vA[-\overline{\vG^{\Lambdas }_t},t]\psi_{h,\Lambdas }^{out} \, = \, 0\,.
\end{equation}
To prove this, we first consider Theorem~\ref{theo-II.2}, then
\begin{equation}\label{eq-III-34}
    \lim_{t\to+\infty}\vA[-\overline{\vG^{\Lambdas }_t},t]\psi_{h,\Lambdas }^{out} \, = \,
    \lim_{t\to+\infty}\vA[-\overline{\vG^{\Lambdas }_t},t]\psi_{h,\Lambdas }(t)\,.
\end{equation}
Next, we rewrite the vector
\begin{equation}\label{eq-III-35}
    \vA[-\overline{\vG^{\Lambdas }_t},t]\psi_{h,\Lambdas }(t) \, = \,
    e^{iHt}\vA[-\overline{\vG^{\Lambdas }_t},0]\sum_{j=1}^{N(t)}\cW_{\Lambdas,\sigma_t}(\vv_j,t)
    e^{i\gamma_{\sigma_t}(\vv_j,\vnabla
    E_{\vec{P}}^{\sigma_t},t)}e^{-iE_{\vec{P}}^{\sigma_t}t}\psi_{j,\sigma_t}^{(t)}
\end{equation}
as
\begin{eqnarray}\label{eq-III-36}
    - \int_{t}^{+\infty}\frac{d}{ds}\lbrace\,e^{iHs}\vA[-\overline{\vG^{\Lambdas }_s},0]
    \sum_{j=1}^{N(t)}\cW_{\Lambdas,\sigma_t}(\vv_j,s)e^{i\gamma_{\sigma_t}(\vv_j,\vnabla
    E_{\vec{P}}^{\sigma_t},s)}e^{-iE_{\vec{P}}^{\sigma_t}s}\psi_{j,\sigma_t}^{(t)}\rbrace\,ds \quad \\
    \,+\,\lim_{s\to+\infty}e^{iHs}\sum_{j=1}^{N(t)}\cW_{\Lambdas,\sigma_t}(\vv_j,s)
    \vA[-\overline{\vG^{\Lambdas }_s},0]e^{i\gamma_ {\sigma_t}(\vv_j,\vnabla
    E_{\vec{P}}^{\sigma_t},s)}e^{-iE_{\vec{P}}^{\sigma_t}s}\psi_{j,\sigma_t}^{(t)} \,.
    \quad \quad\label{eq-III-37}
\end{eqnarray}
The integral in  (\ref{eq-III-36}), and the limit in  (\ref{eq-III-37})
exist. To see this, it is enough to follow the procedure in
Section {\ref{sec-II.2.1}}, taking into account that the operator
\begin{equation}
    \vA[-\overline{\vG^{\Lambdas }_s},0]\frac{1}{(H^{\sigma_t}+i)}[H^{\sigma_t},\vx]
\end{equation}
is bounded, uniformly in $t$ and $s$. The limit (\ref{eq-III-37})
vanishes at fixed $t$ because of condition ($\mathscr{I}4$) in Theorem {\ref{thm-cfp-2}}.
Therefore we finally conclude that the limit (\ref{eq-III-34}) vanishes.
\\

\noindent\underline{\em Li\'enard-Wiechert fields generated by the charge}
\\
\\
Now we briefly explain how to obtain the result stated in (\ref{eq-III-72}).
The assertion is obvious for the longitudinal degrees of freedom;
see the definition of $F_{\mu\nu}$ in (\ref{eq-III-72bis}).
For the transverse degrees of freedom, we argue as follows.
Similarly to the treatment of (\ref{eq-III-24}),
we arrive at a sum over the diagonal terms,
\begin{eqnarray}
    \lefteqn{
    \lim_{t\to\pm\infty}\,\Big\langle \, \psi_{h,\vF}^{out/in} \, , \,
    e^{iHt}\int d^3y\,F_{\mu\nu}^{tr}(0,\vy) \,
    \tilde{\delta}_\Lambda(\vy-\vx-\vec{d}) \, e^{-iHt}\psi_{h,\vF}^{out/in}\Big\rangle \,
    }
    \nonumber\\
    & =&\lim_{t\to\pm\infty}\sum_{j=1}^{N(t)}\Big\langle \, \psi_{j,\sigma_t}^{\sigma_t}\, , \,
    \int d^3y\,F_{\mu\nu}^{tr}(0,\vy) \, \tilde{\delta}_\Lambda(\vy-\vx-\vec{d})
    \,\psi_{j,\sigma_t}^{\sigma_t} \, \Big\rangle \, . \nonumber
\end{eqnarray}
%by exploiting the usual decay properties
%in time of solutions $g_{t,\vy}$ of the free wave equation
%whose Fourier transform is in $L^1(\RR^3,d^3k)$, and behaves like $|\vk|^{-1}$ for $\vk\to0$.
Then, one uses the pull-through formula as in Lemma 6.1 in \cite{ChFr},  
and Proposition 5.1 in \cite{ChFr} which identifies the infrared coherent factor by showing that
\begin{equation}
    \Big| \Bra \, \Psi_{\vec{P}}^{\sigma} \, , \, b_{\vk,\lambda} \, \Psi_{\vec{P}}^{\sigma} \, \Ket
    \,+\,\alpha^{\frac{1}{2}}\,\frac{\one_{\sigma,\Lambda}(\vk)}{|\vk|^{\frac{1}{2}}}
    \, \frac{1}{|\vk|-\vk\cdot\vnabla E_{\vP}^{\sigma}}
    \, \veps_{\vk,\lambda}\cdot\vnabla E^{\sigma}_{\vP} \, \Big| \, \leq \, \alpha^{1/2} \, C \, |\vk|^{-1}
\end{equation}
for $\vk\to0$. These ingredients imply that
\begin{eqnarray}
    &| \, \vec{d} \, |^2&\Big|\lim_{t\to\pm\infty}\Big\lbrace\sum_{j=1}^{N(t)}\Big\langle \, \psi_{j,\sigma_t}^{\sigma_t}
    \, , \,\int d^3y\,F_{\mu\nu}^{tr}(0,\vy)
    \, \tilde{\delta}_\Lambda(\vy-\vx-\vec{d})\,\psi_{j,\sigma_t}^{\sigma_t}\Big\rangle -
    \\
    & &\quad\quad-\sum_{j=1}^{N(t)}\int_{\mathscr{G}^{(t)}_j}|h(\vec{P})|^2
    \Big\langle \, \psi_{\vec{P}}^{\sigma_t} \, , \, \psi_{\vec{P}}^{\sigma_t} \, \Big\rangle \,
    \big(F_{\mu\nu}^{\vnabla E_{\vP}^{\sigma_t}}\big)^{tr} (0,\vec{d}) \, d^3P
    \, \Big\rbrace\,\Big| \, \leq \, \cO(|\vec{d}|^{-1/2})\nonumber\,,
\end{eqnarray}
which vanishes in the limit $|\vec d|\rightarrow\infty$, as asserted in (\ref{eq-III-72}).

\newpage
%%%%%%%%%%%%%%%%%%%%%%%%%%%%%%%%%%%%%%%%%%%%%%%%%%%%%%%%%%%%%%%%%%%%%%%%%%%%%%%%%%%%%%%%%%%%%%%%%%%%%%%%%%%%%%%%%%%%%%%%%%%%%%%%%%%%%%%%%%%%%%%%%%%%%%%%%%%%%%%%%%%%%%%%%%%%%%%%%%%%%%%%%%%%%%%%%%%%%%%%%%%%%%%%%%%%%%%%%%%%%%%%%%%%%%%%%%%%%%%%%%%%%%%%%%%%%%%%%%%%%%%%%%%%%%%%%%%%%%%%%%%%%%%%%%%%%%%%%%%%%%%%%%%%%%%%%%%%%%%%%%%%%%% A P P E N D I X %%%%%%%%%%%%%%%%%%%%%%%%%%%%%%%%%%%%%%%%%%%%%%%%%%%%%%%%%%%%%%%%%%%%%%%%%%%%%%%%%%%%%%%%%%%%%%%%%%%%%%%%%%%%%%%%%%%%%%%%%%%%%%%%%%%%%%%%%%%%%%%%%%%%%%%%%%%%%%%%%%%%%%%%%%%%%%%%%%%%%%%%%%%%%%%%%%%%%%%%%%%%%%%%%%%%%%%%%%%%%%%%%%%%%%%%%%%%%%%%%%%%%%%%%%%%%%%%%%%%%%%%%%%%%%%%%%%%%%%%%%%%%%%%%%%%%%%%%%%%%%%%%%%%%%%%%%%%%%%%%%%%%%%%%%%%%%%%%%%%%%%%%%%%%%%%%%%%%%%%%%%%%%%%%%%%%%%
\appendix
\secct{ }
\label{sec-A}
%%%%%%%%%%%%%%%%%%%%%%%%%%%%%%%%%%%%%%%%%%%%%%%%%%%%%%%%%%%%% LEMMA 1 %%%%%%%%%%%%%%%%%%%%%%%%%%%%%%%%%%%%%%%%%%%%%%%%%%%%%%%%%%%%%%%%%%%%%%%%%%%%%%%%%%%%%%
In this part of the Appendix, we present detailed proofs of auxiliary
results used in Section {\ref{sec-II}}.
\\

\begin{lemma}\label{lemm-A1}
The following estimates hold for $\vP\in\mathcal{S}$:
\\

\begin{itemize}
\item[(i)] For $t_2>t_1\gg1$,
\begin{equation}\label{eq-A-1}
|\gamma_{\sigma_{t_2}}(\vv_j,\vnabla
  E_{\vec{P}}^{\sigma_{t_2}},(\sigma_{t_2})^{-\frac{1}{\theta}})-\gamma_{\sigma_{t_2}}(\vv_{l(j)},\vnabla
  E_{\vec{P}}^{\sigma_{t_2}},(\sigma_{t_2})^{-\frac{1}{\theta}})| \, \leq \, \cO(|\vv_j-\vv_{l(j)}|) \, ,
\end{equation}
where $\vv_j\equiv \vnabla
  E_{\vec{P}_j^*}^{\sigma_{t_1}}$ and  $\vv_{\ell(j)}\equiv \vnabla
  E_{\vec{P}_{\ell(j)}^*}^{\sigma_{t_2}}$\,.
\\
	
\item[(ii)] For $t_2>t_1\gg1$,
\begin{eqnarray}\label{eq-A-2} 
    | \, \gamma_{\sigma_{t_2}}(\vv_j,\vnabla
    E_{\vec{P}}^{\sigma_{t_2}},t_1)-\gamma_{\sigma_{t_1}}(\vv_j,\vnabla
    E_{\vec{P}}^{\sigma_{t_1}},t_1) \, | 
	\, \leq \,
    \cO\big(\,[(\sigma_{t_1})^{{\frac{1}{2}(1-\delta)}}\,t_1^{ 1-\theta }+t_1 \,\sigma_{t_1}]\,\big) \, .
    \\ \nonumber
\end{eqnarray}

\item[(iii)]
For $s,t\gg1$ and $\vec{q}\in\lbrace \vec{q}\,:\,|\vec{q}|<s^{(1-\theta)}\rbrace$, 
\begin{equation}\label{eq-A-3}
    | \, \gamma_{\sigma_t}(\vv_j,\vnabla
    E_{\vec{P}}^{\sigma_t},s)-\gamma_{\sigma_t}(\vv_j,\vnabla
    E_{\vec{P}+\frac{\vec{q}}{s}}^{\sigma_t},s) \, |
    \, \leq \, \cO(s^{-\frac{\theta}{4}(1-\delta'')}\,s^{(1-\theta)}) \,,
\end{equation}
whenever $\gamma_{\sigma_t}(\vv_j,\vnabla E_{\vec{P}}^{\sigma_t},s)$ is defined.
\end{itemize}
\end{lemma}

\Proof
\\
The proofs only require the definition of the phase factor, and some
elementary integral estimates, using conditions $(\mathscr{I}1)$ and $(\mathscr{I}2)$
in Theorem {\ref{thm-cfp-2}}. \QED

%\newpage
%%%%%%%%%%%%%%%%%%%%%%%%%%%%%%%%%%%%%%%%%%%%%%%%%%%%%%%%%%%%%%%%%%%%%%%%%%%%%%%%%%%%%%%%%%%%%%%%%%%%%%%%%%%% L E M M A  A.2 %%%%%%%%%%%%%%%%%%%%%%%%%%%%%%%%%%%%%%%%%%%%%%%%%%%%%%%%%%%%%%%%%%%%%%%%%%%%%%%%%%%%%%%%%
\begin{lemma}\label{lemm-A2}
For $s\geq t\gg 1$, the estimates
\begin{equation}\label{eq-A-3'}
    \sup_{\vx\in\RR^3}\,\Big| \, \int_{\mathcal{B}_{\Lambdas }\setminus\mathcal{B}_{\sigma_{t}}}
    \Sigma^{l}_{\vv_j}(\vk)\cos(\vk\cdot\vx-|\vk|s)d^3k\,\Big|
    \, \leq \,\cO(\frac{|\ln\,\sigma_t|}{s})\,,
\end{equation}
\begin{equation}\label{eq-A-4'}
    \sup_{\vx\in\RR^3}\, \Big| \, \int_{\mathcal{B}_{\Lambdas }\setminus\mathcal{B}_{\sigma_{t}^{S}}}
    \Sigma^{l}_{\vv_j}(\vk)\cos(\vk\cdot\vx-|\vk|s)d^3k\,\chi_h(\frac{\vx}{s}) \, \Big|
    \, \leq \, \cO(\frac{t^{\theta}}{s^2})\,,
\end{equation}
hold, 
where
\begin{eqnarray}
    \Sigma^{l}_{\vv_j}(\vk) \, := \, 2 \, \sum_{l'}(\delta_{l,l'}-\frac{k^lk^{l'}}{|\vk|^2})
    v^{l'}_{j}\frac{1}{|\vk|^2(1-\widehat{k}\cdot\vv_j)} \, ,
\end{eqnarray}
and where $\sigma_t:=t^{-\beta}$, $\sigma_\tau^{S}:=\tau^{-\theta}$, with $\beta>1$, $0<\theta<1$.
Moreover, $\chi_h(\vec y)=0$ for $|\vec y|<\nu_{min}$ and $|\vec y|>\nu_{max}$ with
$0<\nu_{min}<\nu_{max}<1$ (see (\ref{eq-II-7})).
\end{lemma}
\noindent
\Proof
\\
To prove the estimate (\ref{eq-A-3'}),
we consider the variable $\vx$ first in the set
$$
    \lbrace \vx\in\RR^3\,:\,|\vx|<(1-\rho)s,\quad0<\rho<1 \rbrace \, .
$$
We denote by $\theta_{\vk}$ the angle between $\vx $ and $\vk$.
Integration with respect to $|\vk|$ yields
\begin{eqnarray}
    \lefteqn{\big|\int_{\mathcal{B}_{\Lambdas }\setminus\mathcal{B}_{\sigma_{t}}}
    \Sigma^{l}_{\vv_j}(\vk)\cos(\vk\cdot\vx-|\vk|s)d^3k\,\big| }\\
    &=
    &\big|\int\,\widehat\Sigma^{l}_{\vv_j}(\widehat k)\frac{\sin(\Lambdas \widehat{k}\cdot\vx-\Lambdas  s)-
    \sin(t^{-\beta}\widehat{k}\cdot\vx-t^{-\beta}s)}{\widehat{k}\cdot\vx-s}d\Omega_{\vk}\,\big|\\
    &\leq &\frac{2}{\rho\,s}\int\,| \widehat{\Sigma}_{\vv_j}(\widehat{k})|d\Omega_{\vk}\,,
\end{eqnarray}
where $\widehat{\Sigma}^{l}_{\vv_j}(\widehat{k}):=|\vk|^2\,\Sigma^{l}_{\vv_j}(\vk)$.

For $\vx$ in the set
$$
    \lbrace \vx\in\RR^3\,:\,|\vx|>(1-\rho)s,\quad0<\rho<1 \rbrace \, ,
$$
we integrate by parts with respect to $\cos\theta_{\vk}$, and observe that the two functions
\begin{equation}
\widehat{\Sigma}^{l}_{\vv_j}(\widehat{k})\quad\text{and}\quad
\frac{d(\widehat{\Sigma}^{l}_{\vv_j}(\widehat{k}))}{d\cos\theta_{\vk}}
\end{equation}
belong to $L^1(S^2\,;\,d\Omega_{\vk})$. This yields
\begin{eqnarray}
    \lefteqn{\int_{\mathcal{B}_{\Lambdas }\setminus\mathcal{B}_{\sigma_{t}}}
    \Sigma^{l}_{\vv_j}(\vk)\cos(\vk\cdot\vx-|\vk|s)d^3k }\\
    &=&\,-\int^{\Lambdas }_{\sigma_{t}}\int\widehat\Sigma^{l}_{\vv_j}(\widehat{k})|_{\theta_{\vk}=\pi}\,
    \frac{\sin(|\vk|\,|\vx|+|\vk|s)}{|\vk|\,|\vx|}\,d|\vk|d\phi_{\vk}\label{eq-A-11'}\\
    &
    &- \int^{\Lambdas }_{\sigma_{t}}\int\widehat\Sigma^{l}_{\vv_j}(\widehat{k})|_{\theta_{\vk}=0}\,
    \frac{\sin(|\vk|\,|\vx|-|\vk|s)}{|\vk|\,|\vx|}\,d|\vk|d\phi_{\vk}\label{eq-A-12'}\\
    &
    &-\int_{\mathcal{B}_{\Lambdas }\setminus\mathcal{B}_{\sigma_{t}}}
    \frac{d(\widehat\Sigma^{l}_{\vv_j}(\widehat{k}))}{d\cos\theta_{\vk}}\,
    \frac{\sin(\vk\cdot\vx-|\vk|s)}{|\vk|\,|\vx|}\,\frac{d^3k}{|\vk|^2}\, .
    \label{eq-A-13'}
\end{eqnarray}
The absolute values of (\ref{eq-A-11'}), (\ref{eq-A-12'}), and (\ref{eq-A-13'}),
are all bounded above by $\cO(\frac{\ln \sigma_t}{s-\rho s})$, as one easily verifies.
This establishes  (\ref{eq-A-3'}), uniformly in $\vx\in\RR^3$.

To prove (\ref{eq-A-4'}), we consider $\vx$ in a set of the form
\begin{equation}
\lbrace
\vx\in\RR^3\,:\,(1-\rho')s>|\vx|>(1-\rho)s,\quad 0<\rho'<\rho<1 \rbrace\,.\label{eq-A-15'}
\end{equation}
We apply integration by parts with respect to $|\vk|$ in (\ref{eq-A-11'}), (\ref{eq-A-12'}),
and (\ref{eq-A-13'}) in the case $\sigma_t^S$. As an example, we get for (\ref{eq-A-11'})
\begin{eqnarray} 
    (\ref{eq-A-11'}) &=&
    \int\widehat\Sigma^{l}_{\vv_j}(\widehat{k})|_{\theta_{\vk}=\pi}\,
    \frac{\cos(\Lambdas \,(|\vx|+s))}{\Lambdas \,(|\vx|+s)\,|\vx|}\,d\phi_{\vk}\\
    & &-\int\widehat\Sigma^{l}_{\vv_j}(\widehat{k})|_{\theta_{\vk}=\pi}\,
    \frac{\cos(t^{-\theta}\,(|\vx|+s))}{t^{-\theta}\,(|\vx|+s)\,|\vx|}\,d\phi_{\vk}\\
    & &+\int^{\Lambdas }_{\sigma_{t}^{S}}\int\widehat\Sigma^{l}_{\vv_j}(\widehat{k})|_{\theta_{\vk}=\pi}\,
    \frac{\cos(|\vk|\,|\vx|+|\vk|s)}{|\vk|^2\,|\vx|(|\vx|+s)}\,d|k|d\phi_{\vk}\,.
\end{eqnarray}
Since $\vx$ is assumed to be an element of (\ref{eq-A-15'}), it follows that the
bound (\ref{eq-A-4'}) holds for (\ref{eq-A-11'}).
In the same manner, one obtains a similar bound for (\ref{eq-A-12'}) and (\ref{eq-A-13'}).
\QED

\begin{theorem}\label{theo-A'2}
For $\theta<1$ sufficiently close to 1, and $s\geq t$, the propagation estimate
\begin{eqnarray}\label{eq-II-49'}
    \lefteqn{
  \Big\| \, \chi_h(\frac{\vx}{s})e^{i\gamma_{\sigma_t}(\vv_j,\vnabla
  E_{\vec{P}}^{\sigma_t},s)}e^{-iE_{\vec{P}}^{\sigma_t}s}\psi_{j,\sigma_t}^{(t)}
  }
  \\
  &&\quad\quad\quad\quad 
  \quad\quad - \, \chi_h(\vnabla E^{\sigma_t}_{\vP})e^{i\gamma_{\sigma_t}(\vv_j,\vnabla
  E_{\vec{P}}^{\sigma_t},s)}e^{-iE_{\vec{P}}^{\sigma_t}s}\psi_{j,\sigma_t}^{(t)}\, \Big\|
  \nonumber 
  \label{eq-A-21'}
  \\
  &&\quad\leq \, c \,  \frac{1}{s^{\nu}}\,\frac{1}{t^{\frac{3\epsilon}{2}}}\,|\ln(\sigma_t)|\, 
\end{eqnarray}
holds, where $\nu>0$ is independent of $\epsilon$.
\end{theorem}
\Proof
\\
Since the detailed proof of a closely related result is given in Theorem A2 of
\cite{Pizzo2005}, we only sketch the argument.

Expressing  $\chi_h$ (which we assume to be real) in terms of its Fourier transform $\widehat{\chi}_h$,
start from the bound 
\begin{eqnarray}
    \lefteqn{\|\int \, d^3q \, \widehat{\chi}_h(\vq)(e^{-i\vq\cdot\vnabla E_{\vP}^{\sigma_t}}
    -e^{-i\vq\cdot\frac{\vx}{s}})
    \, e^{i\gamma_{\sigma_t}(\vv_j,\vnabla
    E_{\vec{P}}^{\sigma_t},s)}e^{-iE_{\vec{P}}^{\sigma_t}s}\psi_{j,\sigma_t}^{(t)}\| }\\
    &\leq& 
    \|\int\,d^3q \, \widehat{\chi}_h(\vq)(e^{-i\vq\cdot\vnabla E_{\vP}^{\sigma_t}}
    -e^{i(E_{\vP}^{\sigma_t}-E_{\vP+\frac{\vq}{s}}^{\sigma_t})s})
    \,e^{i\gamma_{\sigma_t}(\vv_j,\vnabla
    E_{\vec{P}}^{\sigma_t},s)}\,\psi_{j,\sigma_t}^{(t)}\|\quad\quad\quad\label{eq-A-24'}\\
    & &+\|\int \, d^3q\,\widehat{\chi}_h(\vq)
    \, e^{i(E_{\vP}^{\sigma_t}-E_{\vP+\frac{\vec{q}}{s}}^{\sigma_t})s}\,
    (e^{-i\vq\cdot\frac{\vx}{s}}-1)   \,e^{i\gamma_{\sigma_t}(\vv_j,\vnabla
    E_{\vec{P}}^{\sigma_t},s)}\,\psi_{j,\sigma_t}^{(t)}\|\,.\quad\quad\quad
    \label{eq-A-25'}
\end{eqnarray}
We split the integration domains of (\ref{eq-A-24'}) and (\ref{eq-A-25'}) into the two regions
\begin{equation}
I_+:=\lbrace \vq\,:\,|\vq|\,>\,s^{1-\theta}\rbrace\quad \quad \text{and}\quad \quad I_-:=\lbrace
\vq\,:\,|\vq|\,\leq\, s^{1-\theta}\rbrace\,.
\end{equation}
In both (\ref{eq-A-24'}) and (\ref{eq-A-25'}), the contribution to the integral from $I_+$ is controlled by the
decay properties of $\widehat{\chi}(\vq)$, and one easily derives the bound in (\ref{eq-A-21'}).
For the contributions to (\ref{eq-A-24'}) from the integral over $I_-$, the existence of the gradient of the energy,
the H\"older property in $\vP$ of the gradient, and the
decay properties of $\widehat{\chi}(\vq)$ are enough to produce the bound in (\ref{eq-A-21'}).

To control (\ref{eq-A-25'}), we note that the two vectors
\begin{equation}
    \Psi_{\vP-\frac{\vq}{s}}^{\sigma_t}\quad\quad\text{and}\quad\quad
    \widehat{\Psi}_{\vP-\frac{\vq}{s}}^{\sigma_t}:=e^{-i\vq\cdot\frac{\vx}{s}}\,\Psi_{\vP}^{\sigma_t}
\end{equation}
belong to the same fiber space $\cH_{{\vP}-\frac{\vq}{s}}$, and that,
as vectors in Fock space, $\widehat{\Psi}_{\vP-\frac{\vq}{s}}^{\sigma_t}$ and
$\Psi_{\vP}^{\sigma_t}$ coincide, i.e.,
\begin{equation}
I_{\vP-\frac{\vq}{s}}(e^{-i\vq\cdot\frac{\vx}{s}}\,\Psi_{\vP}^{\sigma_t})\,\equiv\,I_{\vP}(\Psi_{\vP}^{\sigma_t})\,.
\end{equation}
We split and estimate  (\ref{eq-A-25'}) by
\begin{eqnarray}\label{eq-A-29'} 
	(\ref{eq-A-25'}) 
&=&\big\|\int_{I_-}\widehat{\chi}_h(\vq)\,\int_{\Gamma_j^{(t)}}e^{i(E_{\vP-\frac{\vq}{s}}^{\sigma_t}
-E_{\vP}^{\sigma_t})s}\,h_{\vP}\,e^{i\gamma_{\sigma_t}(\vv_j,\vnabla
  E_{\vec{P}}^{\sigma_t},s)}\widehat{\Psi}_{\vP-\frac{\vq}{s}}^{\sigma_t}\,d^3Pd^3q\\
& &\quad-\int_{I_-}\widehat{\chi}_h(\vq)\,\int_{\Gamma_j^{(t)}}
e^{i(E_{\vP}^{\sigma_t}-E_{\vP+\frac{\vq}{s}}^{\sigma_t})s}
\,h_{\vP}\,e^{i\gamma_{\sigma_t}(\vv_j,\vnabla
  E_{\vec{P}}^{\sigma_t},s)}\Psi_{\vP}^{\sigma_t}\,d^3Pd^3q\big\|
  \\ 
&\leq&\big\|\int_{I_-}\widehat{\chi}_h(\vq)\,\int_{\Gamma_j^{(t)}}e^{i(E_{\vP-\frac{\vq}{s}}^{\sigma_t}
-E_{\vP}^{\sigma_t})s}\,h_{\vP}\,e^{i\gamma_{\sigma_t}(\vv_j,\vnabla
  E_{\vec{P}}^{\sigma_t},s)}\widehat{\Psi}_{\vP-\frac{\vq}{s}}^{\sigma_t}\,d^3Pd^3q
\label{eq-A-32'}\\
& &\quad-\int_{I_-}\widehat{\chi}_h(\vq)\,\int_{\Gamma_j^{(t)}}e^{i(E_{\vP-\frac{\vq}{s}}^{\sigma_t}
-E_{\vP}^{\sigma_t})s}\,h_{\vP}\,e^{i\gamma_{\sigma_t}(\vv_j,\vnabla
  E_{\vec{P}}^{\sigma_t},s)}\Psi_{\vP-\frac{\vq}{s}}^{\sigma_t}\,d^3Pd^3q\big\|\quad\quad\quad\nonumber\\
& &+ \, \big\|\int_{I_-}\widehat{\chi}_h(\vq)\,\int_{\Gamma_j^{(t)}}e^{i(E_{\vP-\frac{\vq}{s}}^{\sigma_t}
-E_{\vP}^{\sigma_t})s}\,h_{\vP}\,e^{i\gamma_{\sigma_t}(\vv_j,\vnabla
  E_{\vec{P}}^{\sigma_t},s)}\Psi_{\vP-\frac{\vq}{s}}^{\sigma_t}\,d^3Pd^3q\quad\quad\label{eq-A-33'}\\
& &\quad-\int_{I_-}\widehat{\chi}_h(\vq)\,\int_{\Gamma_j^{(t)}}e^{i(E_{\vP-\frac{\vq}{s}}^{\sigma_t}
-E_{\vP}^{\sigma_t})s}\,h_{\vP-\frac{\vq}{s}}\,e^{i\gamma_{\sigma_t}(\vv_j,\vnabla
  E_{\vec{P}-\frac{\vq}{s}}^{\sigma_t},s)}\Psi_{\vP-\frac{\vq}{s}}^{\sigma_t}\,d^3Pd^3q\big\|\quad\quad\nonumber\\
& &+ \, \big\|\,\int_{I_-}\widehat\chi_h(\vq)
 \int_{\Gamma_j^{(t)}}e^{i(E_{\vP-\frac{\vq}{s}}^{\sigma_t}-E_{\vP}^{\sigma_t})s}
\,h_{\vP-\frac{\vq}{s}}\,e^{i\gamma_{\sigma_t}(\vv_j,\vnabla
  E_{\vec{P}-\frac{\vq}{s}}^{\sigma_t},s)}\Psi_{\vP-\frac{\vq}{s}}^{\sigma_t}\,d^3Pd^3q\quad\quad\label{eq-A-34'}\\
& &\quad-\int_{I_-}\widehat{\chi}_h(\vq)\,\int_{\Gamma_j^{(t)}}e^{i(E_{\vP}^{\sigma_t}
-E_{\vP+\frac{\vq}{s}}^{\sigma_t})s}\,h_{\vP }\,e^{i\gamma_{\sigma_t}(\vv_j,\vnabla
  E_{\vec{P} }^{\sigma_t},s)}\Psi_{\vP}^{\sigma_t}\,d^3Pd^3q\,\big\|\,.\nonumber
\end{eqnarray}
The terms (\ref{eq-A-32'}), (\ref{eq-A-33'}), and (\ref{eq-A-34'}) can be bounded by
\begin{equation}
    (\ref{eq-A-32'}) \, \leq \, \int_{I_-}|\widehat{\chi}_h(\vq)|
    \big[\int_{\Gamma_j^{(t)}}\,|h_{\vP}|^2\|I_{\vP}(\Psi_{\vP}^{\sigma_t})-I_{\vP-\frac{\vq}{s}}
    (\Psi_{\vP-\frac{\vq}{s}}^{\sigma_t})\|_{\cF}^2\,d^3P\,\big]^{\frac{1}{2}}d^3q\,,
    \label{eq-A-35'}
\end{equation}
\begin{equation}
    \label{eq-A-36'} 
	(\ref{eq-A-33'}) 
    \, \leq \, \int_{I_-}|\widehat{\chi}_h(\vq)|\big[\int_{\Gamma_j^{(t)}}\,|\Delta_{\frac{\vq}{s}}\,
    ( h_{\vP}e^{i\gamma_{\sigma_t}(\vv_j,\vnabla E_{\vec{P}}^{\sigma_t},s)})|^2
    \, \|I_{\vP-\frac{\vq}{s}}
    (\Psi_{\vP-\frac{\vq}{s}}^{\sigma_t})\|_{\cF}^2\,d^3P\,\big]^{\frac{1}{2}}d^3q\,,
\end{equation}
and
\begin{equation}
    (\ref{eq-A-34'}) \, \leq \, \int_{I_-}|\widehat{\chi}_h(\vq)|\big[\int_{O_{\frac{\vq}{s}}^j}\,
    |h_{\vP}|^2 \, \|I_{\vP}(\Psi_{\vP}^{\sigma_t})\|_{\cF}^2\,d^3P\,\big]^{\frac{1}{2}}d^3q\,,
    \label{eq-A-37'}
\end{equation}
where \begin{equation}
    \Delta_{\frac{\vq}{s}}\,( h_{\vP}e^{i\gamma_{\sigma_t}(\vv_j,\vnabla
    E_{\vec{P}}^{\sigma_t},s)})\,:=\, h_{\vP}e^{i\gamma_{\sigma_t}(\vv_j,\vnabla
    E_{\vec{P}}^{\sigma_t},s)}-h_{\vP-\frac{\vq}{s}}e^{i\gamma_{\sigma_t}(\vv_j,\vnabla
    E_{\vec{P}-\frac{\vq}{s}}^{\sigma_t},s)}\,,
\end{equation}
and $O_{\frac{\vq}{s}}^j:=\,(\Gamma_j^{(t)}\cup\Gamma_j^{(t)\,,\,\frac{\vq}{s}}) \setminus
(\Gamma_j^{(t)}\cap\Gamma_j^{(t)\,,\,\frac{\vq}{s}})$, where  $\Gamma_j^{(t)\,,\,\frac{\vq}{s}}$
is  the translate by $\frac{\vq}{s}$ of the cell $\Gamma_j^{(t)}$.

Using (\ref{eq-A-3}), the $C^1-$regularity of $h_{\vP}$, and the definition of $I_{-}$,
one readily shows that the terms (\ref{eq-A-36'}), (\ref{eq-A-37'}) satisfy the bound (\ref{eq-A-21'}),
as desired.

To estimate (\ref{eq-A-35'}),  we use the inequality
\begin{eqnarray}
    \lefteqn{\|I_{\vP}\big(\Psi_{\vP}^{\sigma_t}\big)-I_{\vP-\frac{\vq}{s}}
    \big(\Psi_{\vP-\frac{\vq}{s}}^{\sigma_t}\big)\|_{\cF} }\\
    &\leq&\|I_{\vP}\big(W_{\sigma_t}(\vnabla
    E_{\vP}^{\sigma_t})\Psi_{\vP}^{\sigma_t}\big)-I_{\vP-\frac{\vq}{s}}\big(W_{\sigma_t}(\vnabla
    E_{\vP-\frac{\vq}{s}}^{\sigma_t})\Psi_{\vP-\frac{\vq}{s}}^{\sigma_t}\big)\|_{\cF}\label{eq-A-39'}
    \\
    & &+\|I_{\vP-\frac{\vq}{s}}\big((W_{\sigma_t}^{*}(\vnabla
    E_{\vP}^{\sigma_t})-W_{\sigma_t}^{*}(\vnabla
    E_{\vP-\frac{\vq}{s}}^{\sigma_t}))\,W_{\sigma_t}(\vnabla
    E_{\vP-\frac{\vq}{s}}^{\sigma_t})\Psi_{\vP-\frac{\vq}{s}}^{\sigma_t}\big)\|_{\cF}\,,\quad\quad
\label{eq-A-40'}
\end{eqnarray}
where it is clear that
\begin{equation}
    W_{\sigma_t}(\vnabla
    E_{\vP}^{\sigma_t})\Psi_{\vP}^{\sigma_t}=\Phi_{\vec{P}}^{\sigma_t}\,.
\end{equation}
Moreover, we use properties ($\mathscr{I}2$), ($\mathscr{I}5$)
in Theorem {\ref{thm-cfp-2}}, where we recall that
\begin{itemize}
\item[($\mathscr{I}2$)]
\emph{H\"older regularity in  $\vec{P}\in\mathcal{S}$ uniformly
in $\sigma \geq 0$ holds in the sense of
\begin{equation}\label{eq-A-42'}
    \|\Phi_{\vec{P}}^{\sigma}-\Phi_{\vec{P}+\Delta\vec{P}}^{\sigma}\|_{\cF} \leq
    C_{\delta'}|\Delta \vec{P}|^{\frac{1}{4}-\delta'}
\end{equation}
and
\begin{equation} \label{eq-A-43'}
    |\vnabla E_{\vec{P}}^{\sigma}-\vnabla E_{\vec{P}+\Delta\vec{P}}^{\sigma}|\leq
    C_{\delta''}|\Delta \vec{P}|^{\frac{1}{4}-\delta''} \, ,
\end{equation}
for any $0<\delta''<\delta'<\frac{1}{4}$, with $\vec{P}\,,\,\vec{P}+\Delta\vec{P} \in
\mathcal{S}$, and where $C_{\delta'}$ and $C_{\delta''}$ are
finite constants depending on $\delta'$ and $\delta''$, respectively}.
\end{itemize}
We can bound (\ref{eq-A-39'}) by use of (\ref{eq-A-42'}).

In order to bound (\ref{eq-A-40'}), we recall the definition of the Weyl operator
\begin{equation}
    W_{\sigma}(\vnabla E_{\vP}^{\sigma}) \, := \, \exp\Big(\, \alpha^{\frac{1}{2}}
    \sum_{\lambda} \int_{\mathcal{B}_{\Lambda}\setminus \mathcal{B}_{\sigma}} d^3k \, 
    \frac{\vnabla E_{\vP}^{\sigma}}{|\vec{k}|^{\frac{3}{2}}\delta_{\vP,\sigma}(\widehat{k})} \cdot
    (\veps_{\vk,\lambda}b_{\vk,\lambda}^{*} - h.c.)\Big) \, ,
\end{equation}
and we note that
\begin{equation}
    (\ref{eq-A-40'}) \, \leq \, c \,
    \big| \, \vnabla E_{\vP}^{\sigma_t} - \vnabla E_{\vP-\frac{\vq}{s}}^{\sigma_t}\, \big|
    \, \Ksp_t \,
    \Big( \, \Ksp_t \,  + \, \Big( \,\sum_{\lambda} \int_{\mathcal{B}_{\Lambda}\setminus \mathcal{B}_{\sigma_t}}
    d^3k \, \| \, b_{\vk,\lambda}\Phi_{\vP-\frac{\vq}{s}}^{\sigma} \, \|_{\cF}^2 \Big)^{\frac12} \, \Big)
\end{equation}
from a simple application of the Schwarz inequality, where
\begin{equation}
	\Ksp_t \, := \, \Big( \int_{\mathcal{B}_{\Lambda}\setminus \mathcal{B}_{\sigma_t}}
    \frac{d^3k}{|\vec{k}|^3}\Big)^{\frac12} 
    \, = \,  \cO( |\ln\sigma_t|^{\frac{1}{2}}) \,.
\end{equation}  
Moreover, we have
\begin{equation}
   \sum_{\lambda} \int_{\mathcal{B}_{\Lambda}\setminus \mathcal{B}_{\sigma_t}} d^3k \,
    \| \, b_{\vk,\lambda}\Phi_{\vP-\frac{\vq}{s}}^{\sigma} \, \|_{\cF}^2
    \, \leq \, c \, |\ln\sigma_t| \, ,
\end{equation}
which is derived similarly as (\ref{eq-II-97.3}).

>From H\"older continuity of $\vnabla E_{\vP}^{\sigma}$ in $\vP$, (\ref{eq-A-43'}),
we obtain a contribution to the upper bound on (\ref{eq-A-40'}) which exhibits
a power law decay in $s$.

We conclude that (\ref{eq-A-35'}) is bounded by (\ref{eq-A-21'}), as claimed.
\QED

%\newpage

\noindent{\bf Remark:}
By a similar procedure, one finds that for $t_2\geq s \geq t_1$,
\begin{eqnarray}
	\lefteqn{
	\Big\| \, \chi_h(\frac{\vx}{s}) \, e^{ i\gamma_{\sigma_{t_2}}(\vv_j,\vnabla E_\vP^{\sigma_{t_2}}, s)}
	\,  e^{ - i E_\vP^{\sigma_{t_2}} s } \,
	\psi_{j,\sigma_{t_2}}^{(t_1)}
	} 
	\nonumber\\
	&& - \,  \chi_h(\vnabla E_\vP^{\sigma_{t_2}}) \, e^{ i\gamma_{\sigma_{t_2}}(\vv_j,\vnabla E_\vP^{\sigma_{t_2}}, s)}
	\, e^{ - i E_\vP^{\sigma_{t_2}} s } \,	 \psi_{j,\sigma_{t_2}}^{(t_1)} \, \Big\| 
	\, \leq \, c \, \frac{1}{s^\nu} \, \frac{\ln(\sigma_{t_2})}{t_1^{3/2}} \,.
\end{eqnarray}
Analogous extensions hold for the estimates in the next theorem.
\\

\begin{theorem}\label{theo-A4}
Both
\begin{equation}\label{eq-A-4}
    \Big\| \, \int_t^{+\infty}
    e^{i H^{\sigma_t}s} \mathcal{W}_{\Lambdas,\sigma_t}(\vec v_j,s)
    \Big[\cJ|_{\sigma_{t}}^{\sigma^{S}_s}(s)\chi_h(\frac{\vx}{s})-
    \frac{d\widehat{\gamma}_{\sigma_t}(\vv_j,\frac{\vx}{s},s)}{ds}\Big]
    \,e^{i\gamma_{\sigma_t}(\vv_j,\vnabla
    E_{\vec{P}}^{\sigma_t},s)}e^{-iE_{\vec{P}}^{\sigma_t}s}(E_{\vec{P}}^{\sigma_{t}}+i) \, \psi_{j,\sigma_t}^{(t)}ds
    \, \Big\|\,
\end{equation}
and
\begin{eqnarray}\label{eq-A-4bis}
    &\Big\| &
    \int_t^{+\infty}e^{i H^{\sigma_t}s}
    \mathcal{W}_{\Lambdas,\sigma_t}(\vec v_j,s)
    \Big\lbrace\,\frac{d\widehat{\gamma}_{\sigma_t}(\vv_j,\frac{\vx}{s},s)}{ds}\,
    e^{i\gamma_{\sigma_t}(\vv_j,\vnabla
    E_{\vec{P}}^{\sigma_t},s)}e^{-iE_{\vec{P}}^{\sigma_t}s}(E_{\vec{P}}^{\sigma_{t}}+i)
    \, \psi_{j,\sigma_t}^{(t)}\\
    & &\quad \quad \,-\,\frac{d\gamma_{\sigma_t}(\vv_j,\vnabla
    E_{\vec{P}}^{\sigma_t},s)}{ds}\, e^{i\gamma_{\sigma_t}(\vv_j,\vnabla
    E_{\vec{P}}^{\sigma_t},s)}e^{-iE_{\vec{P}}^{\sigma_t}s}(E_{\vec{P}}^{\sigma_{t}}+i)
    \, \psi_{j,\sigma_t}^{(t)}\Big\rbrace\,ds\,\, \Big\|\, \nonumber
\end{eqnarray}
are bounded by
\begin{equation}\label{eq-A-4bisbis}
     \frac{1}{t^{\eta}}\,|\ln(\sigma_t)|^2\,t^{-\frac{3\epsilon}{2}} \, ,
\end{equation}
where $\eta>0$ is $\epsilon$-independent.
$\cJ|_{\sigma_{t}}^{\sigma^{S}_s}(s)$,
$\frac{d\widehat{\gamma}_{\sigma_t}(\vv_j,\frac{\vx}{s},s)}{ds}$, and $\frac{d\gamma_{\sigma_t}(\vv_j,\vnabla
  E_{\vec{P}}^{\sigma_t},s)}{ds}$ are defined in
(\ref{eq-II-52}), (\ref{eq-II-52bis}), and (\ref{eq-II-27}) -- (\ref{eq-II-28}), respectively.

\end{theorem}

\Proof
\\
We recall from (\ref{eq-II-52}) that for $\sigma_s^S\geq\sigma_t$,
\begin{eqnarray}
    \cJ|_{\sigma_{t}}^{\sigma^{S}_s}(s)&=&\alpha\,i[H^{\sigma_t},\vx]\cdot
    \int_{\mathcal{B}_{\sigma_s^{S}}\setminus\mathcal{B}_{\sigma_t}}
    \vec{\Sigma}_{\vv_j}(\vk) \, \frac{1}{H^{\sigma_t}+i} \, \cos(\vk\cdot\vx-|\vk|s)d^3k\, ,
    \nonumber
\end{eqnarray}
where  $\sigma_{s}^S:=\frac{1}{s^{\theta}}$ is the slow cut-off, and from (\ref{eq-II-52bis})
\begin{eqnarray}
    \frac{d\hat \gamma_{\sigma_t}(\vec v_j,\frac{\vec x}{s}, s)}{ds} &:=&
    \alpha \, e^{-iH^{\sigma_t} s} \, \frac{1}{H^{\sigma_t}+i} \,
    \frac{d\{ e^{iH^{\sigma_t} s} \vec x_h(s) e^{-iH^{\sigma_t} s} \} }{ds} e^{iH^{\sigma_t} s} \, \cdot
    \nonumber\\
    &&\cdot \int_{\mathcal{B}_{\sigma_s^{S}}\setminus\mathcal{B}_{\sigma_t}}
    \vec{\Sigma}_{\vv_j}(\vk) \, \cos(\vk\cdot\vnabla E_\vP^{\sigma_t}s-|\vk|s)d^3k \,.
\end{eqnarray}
For $s$ such that $\sigma_s^S\leq\sigma_t$ the expressions (\ref{eq-A-4}) 
and (\ref{eq-A-4bis}) are identically zero.
By unitarity of $e^{iH^{\sigma_t}s}$ and $\cW_{\sigma_t}(\vv_j,s)$,
we can replace the part in the integrand of (\ref{eq-A-4}) proportional to
$\cJ|_{\sigma_{t}}^{\sigma^{S}_s}(s)$ by
\begin{eqnarray}\label{eq-A-5}
    & &e^{iH^{\sigma_t}s}\cW_{\Lambdas,\sigma_t}(\vv_j,s)\alpha\,i[H^{\sigma_{t}}\,,\,\vx]
    \, \frac{1}{H^{\sigma_t}+i} \, \chi_h(\frac{\vx}{s}) \, \cdot
     \,  \quad\\
    & &\quad \quad \cdot
    \int_{\mathcal{B}_{\sigma_s^S}\setminus\mathcal{B}_{\sigma_t}}d^3k \, \vec{\Sigma}_{\vv_j}(\vk) 
    \, \cos(\vk\cdot\vnabla
    E_{\vec{P}}^{\sigma_t}s-|\vk|s) \, e^{i\gamma_{\sigma_t}(\vv_j,\vnabla
    E_{\vec{P}}^{\sigma_t},s)} \, e^{-iE_{\vec{P}}^{\sigma_{t}}s}(E_{\vec{P}}^{\sigma_t}+i)\psi_{j,\sigma_t}^{(t)}
    \, , \nonumber
\end{eqnarray}
up to a term which yields an integral bounded in norm by
\begin{equation}\label{eq-A-5bis}
    \frac{1}{t^{\eta}}\,|\ln(\sigma_t)|^2\,t^{-\frac{3\epsilon}{2}}\,.
\end{equation}
To justify this step, we exploit the fact that the operator
\begin{equation}\label{eq-A-6}
    i[H^{\sigma_{t}}\,,\,\vx]\,\frac{1}{H^{\sigma_t}+i}
\end{equation}
is bounded. Moreover, we are applying the propagation estimate
\begin{eqnarray}\label{eq-A-7}
    &&\Big\| \, \Big\{ \int_{\mathcal{B}_{\sigma_s^S}\setminus\mathcal{B}_{\sigma_t}}
    \vec{\Sigma}_{\vv_j}(\vk)\cos(\vk\cdot\vx-|\vk|s)d^3k
    \\
    &&\quad\quad - \, \int_{\mathcal{B}_{\sigma_s^S}\setminus\mathcal{B}_{\sigma_t}}
    \vec{\Sigma}_{\vv_j}(\vk)\cos(\vk\cdot\vnabla
    E_{\vec{P}}^{\sigma_t}s-|\vk|s)d^3k \, \Big\}
    \,  e^{i \gamma_{\sigma_t}(\vec v_j, \vnabla E^{\sigma_t}_{\vP} , s)} \, e^{-iE^{\sigma_t}_{\vP} s}
    (E_{\vP}^{\sigma_t} + i) \, \psi_{j,\sigma_t}^{(t)} \, \Big\|
    \nonumber\\
    &&\quad \quad \quad \quad \, \leq \, 
    \, c \,  \frac{1}{s^{1+\nu}}\,\frac{1}{t^{\frac{3\epsilon}{2}}}\,|\ln(\sigma_t)|\,
    \,,
    \nonumber
\end{eqnarray}
for some $\nu>0$,
which is similar to  (\ref{eq-II-49'}).
To obtain the upper bound, we exploit the fact that
due to the slow cut-off $\sigma_s^S=s^{-\theta}$, $\theta>0$, in  $\cJ|_{\sigma_{t}}^{\sigma^{S}_s}(s)$,
the upper integration bound in the radial
part of the momentum variables vanishes in the limit $s\to\infty$.
We note that we have to assume $\theta<1$ as required in (\ref{eq-II-51}),
in order to use the result in Lemma \ref{lemm-A2}.

Next, we approximate (\ref{eq-A-5}) by
\begin{eqnarray}\label{eq-A-8}
    & &e^{iH^{\sigma_t}s}\cW_{\Lambdas,\sigma_t}(\vv_j,s)\alpha\,e^{-iH^{\sigma_t}s}
    \frac{1}{H^{\sigma_t}+i}\frac{d\vec{x}(s)}{ds}\, \chi_h(\frac{\vx(s)}{s})\,\cdot \quad\\
    & &\quad
    \cdot \int_{\mathcal{B}_{\sigma_s^S}\setminus \mathcal{B}_{\sigma_t}}
    d^3k\, \vec{\Sigma}_{\vv_j}(\vk) \cos(\vk\cdot\vnabla
    E_{\vec{P}}^{\sigma_t}s-|\vk|s) \, (E_{\vec{P}}^{\sigma_t}+i)
    e^{i\gamma_{\sigma_t}(\vv_j,\vnabla E_{\vec{P}}^{\sigma_t},s)}\psi_{j,\sigma_t}^{(t)}\nonumber
\end{eqnarray}
where $\vx(s):=e^{iH^{\sigma_t}s}\vx e^{-iH^{\sigma_t}s}$.
To pass from (\ref{eq-A-5}) to (\ref{eq-A-8}), we have used
\begin{equation}\label{eq-A-9}
	\frac{d\vec x(s)}{ds} \, \frac{1}{H^{\sigma_t}+i} \, = \,
	\frac{1}{H^{\sigma_t}+i} \, \frac{d\vec x(s)}{ds} \, - \,  
    \frac{1}{H^{\sigma_t}+i} \,
    \frac{d[\vec x(s) \, , \, H^{\sigma_t}]}{ds} 
    \,\frac{1}{H^{\sigma_t}+i} \, ,
\end{equation}
and we have noticed that the term containing
\begin{equation}\label{eq-A-10}
	\frac{1}{H^{\sigma_t}+i}\,\frac{d[\vec x(s) \, , \, H^{\sigma_t}]}{ds}\,\frac{1}{H^{\sigma_t}+i}
\end{equation}
can be neglected because an integration by parts shows that the corresponding integral
is  bounded in norm by $\frac{1}{t^{\nu}}\,|\ln(\sigma_t)|^2\,t^{-\frac{3\epsilon}{2}}$
for some $\nu>0$ and $\epsilon$-independent.
This uses
\begin{equation}\label{eq-A-11}
    \sup_{\vP\in\mathcal{S}}\Big|\int_{\mathcal{B}_{\sigma_s^S}\setminus \mathcal{B}_{\sigma_t}}
    d^3k \, \vec{\Sigma}_{\vv_j}(\vk)\,\cos(\vk\cdot\vnabla
    E_{\vec{P}}^{\sigma_t}s-|\vk|s) \, \Big|
    \, \leq \, \cO\big(\frac{|\ln\big(\sigma_t)|}{s}\big)\,
\end{equation}
and
\begin{equation}\label{eq-A-11b}
    \sup_{\vP\in\mathcal{S}}\Big| \,
    \frac{d}{ds} \, \int_{\mathcal{B}_{\sigma_s^S}\setminus\mathcal{B}_{\sigma_t}}
    d^3k \, \vec{\Sigma}_{\vv_j}(\vk)\cos(\vk\cdot\vnabla
    E_{\vec{P}}^{\sigma_t}s-|\vk|s) \, \Big|
    \, \leq \, \cO\big(\frac{1}{s^{1+\theta}}\big)\, ,
\end{equation}
which can be derived as in Lemma {\ref{lemm-A2}}.

To bound the integral corresponding to (\ref{eq-A-8}), we note that up to a
term whose integral is bounded in norm by (\ref{eq-A-4bisbis}), one can replace
$\frac{d\vec{x}(s)}{ds} \, \chi_h(\frac{\vx(s)}{s})$ by
\begin{equation}\label{eq-A-12}
    \frac{d}{ds}\Big(  \, e^{iH^{\sigma_t}s}\vx_h (s)e^{-iH^{\sigma_t}s} \, \Big) \, ,
\end{equation}
where $\vx_h(s):=\vx\chi_h(\frac{\vx}{s})$, with $\chi_h(\vec{y})$ defined as in Section
{\ref{sec-II.2.1}}. This is possible because
\begin{eqnarray}
    \lefteqn{
    \frac{d}{ds}\Big( \, e^{iH^{\sigma_t}s}\vx_h(s)e^{-iH^{\sigma_t}s} \, \Big)
    }
    \label{eq-A-13}\\
    & =&
    - \, e^{iH^{\sigma_t}s}\vx\,[\frac{\vx}{s^2}\cdot\vec{\nabla}\chi_h(\frac{\vx}{s})]e^{-iH^{\sigma_t}s}
    \label{eq-A-14}\\
    & &+ \, e^{iH^{\sigma_t}s} 
    i[H^{\sigma_{t}}\,,\,\vx] 
    \chi_h(\frac{\vx}{s})e^{-iH^{\sigma_t}s}
    \label{eq-A-15}\\
    & &
    + \, e^{iH^{\sigma_t}s}\frac{\vx}{s}\,\big[ \vec{\nabla}
    \chi_h(\frac{\vx}{s})
    \cdot \frac{i[H^{\sigma_{t}}\,,\,\vx]}{2}\big]e^{-iH^{\sigma_t}s} 
    \label{eq-A-16}\\ 
    & &+ \, e^{iH^{\sigma_t}s}\frac{\vx}{s}\,\big[\frac{i[H^{\sigma_{t}}\,,\,\vx]}{2}\cdot\vec{\nabla}
    \chi_h(\frac{\vx}{s})\big]e^{-iH^{\sigma_t}s}\,,\label{eq-A-18}
\end{eqnarray}
where 
(\ref{eq-A-15}) 
corresponds to $\frac{d\vec{x}(s)}{ds} \, \chi_h(\frac{\vx(s)}{s})$. 
Moreover, we use the fact that the vector operator
$\frac{1}{H^{\sigma_t}+i}\,i[H^{\sigma_{t}}\,,\,\vx]$ is bounded,
and apply the propagation estimate (\ref{eq-II-49'}) to
$\frac{x^i x^j}{s^2}\nabla^j\chi_h(\frac{\vx}{s})$
and to
$\frac{x^i}{s}\nabla^j\chi_h(\frac{\vx}{s})$ 
with appropriate modifications (see (\ref{eq-A-23}) and recall that
$\vnabla \chi_h(\vnabla E_{\vP}^{\sigma_t})=0$ for $\vP\in{\rm supp}\, h$).

We observe that
\begin{eqnarray}\label{eq-A-8bis}
    & &e^{iH^{\sigma_t}s} \, \cW_{\Lambdas,\sigma_t}(\vv_j,s) \, \alpha \, e^{-iH^{\sigma_t}s}
    \, \frac{1}{H^{\sigma_t}+i}
    \, \frac{d ( \, e^{iH^{\sigma_t}s} \vec{x}_{h}(s) e^{-iH^{\sigma_t}s} \, ) }{ds}
    \, \cdot \quad\\
    & &\quad \quad
    \cdot \int_{\mathcal{B}_{\sigma_s^S}\setminus \mathcal{B}_{\sigma_t}}
    d^3k \, \vec{\Sigma}_{\vv_j}(\vk)
    \, \cos(\vk\cdot\vnabla E_{\vec{P}}^{\sigma_t}s-|\vk|s)
    \, (E_{\vec{P}}^{\sigma_t}+i)
    \, e^{i\gamma_{\sigma_t}(\vv_j,\vnabla E_{\vec{P}}^{\sigma_t},s)}
    \, \psi_{j,\sigma_t}^{(t)}\nonumber
\end{eqnarray}
corresponds to
\begin{eqnarray}\label{eq-A-8bisbis} 
    \, e^{iH^{\sigma_t}s}\cW_{\Lambdas,\sigma_t}(\vv_j,s)
    \, \Big[\frac{d \hat \gamma_{\sigma_t}(\vec v_j , \frac{\vec x}{s} , s)}{ds}\Big]
    \, e^{i\gamma_{\sigma_t}(\vv_j,\vnabla E_{\vec{P}}^{\sigma_t},s)}
    \, \psi_{j,\sigma_t}^{(t)} \,.
\end{eqnarray}
This immediately implies (\ref{eq-A-4}).

To prove (\ref{eq-A-4bis}), we need to control the integral
\begin{eqnarray}
    & &\int_{t}^{\bar{s}}e^{iH^{\sigma_t}s}\cW_{\Lambdas,\sigma_t}(\vv_j,s)
    e^{-iH^{\sigma_t}s}\frac{\alpha}{H^{\sigma_t}+i}
    \frac{d ( \, e^{iH^{\sigma_t}s} \vec{x}_{h}(s) e^{-iH^{\sigma_t}s} \, ) }{ds}\cdot
    \quad\quad\label{eq-A-19}\\
    & &\quad \cdot \,
    \int_{\mathcal{B}_{\sigma_s^S}\setminus\mathcal{B}_{\sigma_t}}d^3k \, \vec{\Sigma}_{\vv_j}(\vk)
    \cos(\vk\cdot\vnabla
    E_{\vec{P}}^{\sigma_t}s-|\vk|s)
    \, e^{i\gamma_{\sigma_t}(\vv_j,\vnabla E_{\vec{P}}^{\sigma_t},s)}
    \, (E_{\vec{P}}^{\sigma_t}+i)\psi_{j,\sigma_t}^{(t)}ds\nonumber
\end{eqnarray}
for $\bar{s}\to+\infty$.
An integration by parts with respect to $s$ yields
\begin{eqnarray}
    &
    &e^{iH^{\sigma_t}s}\cW_{\Lambdas,\sigma_t}(\vv_j,s)\frac{\alpha}{H^{\sigma_t}+i}
    \vec{x}_{h}(s)\cdot\int_{\mathcal{B}_{\sigma_s^S}\setminus\mathcal{B}_{\sigma_t}}d^3k \,
    \vec{\Sigma}_{\vv_j}(\vk)\times\label{eq-A-20}\\
    & &\quad\quad\quad\quad\quad\quad\times
    \cos(\vk\cdot\vnabla E_{\vec{P}}^{\sigma_t}s-|\vk|s)\,E_{\vec{P}}^{\sigma_t}\,e^{-iE_{\vec{P}}^{\sigma_t}s}
    \, e^{i\gamma_{\sigma_t}(\vv_j,\vnabla E_{\vec{P}}^{\sigma_t},s)}
    \, (E_{\vec{P}}^{\sigma_t}+i)\psi_{j,\sigma_t}^{(t)}\big|_{t}^{\bar{s}}\nonumber\\
    &&-\int_{t}^{\bar{s}}\Big\lbrace \, \frac{d}{ds} ( \,
    e^{iH^{\sigma_t}s}\cW_{\Lambdas,\sigma_t}(\vv_j,s)e^{-iH^{\sigma_t}s} \, ) \, \Big\rbrace \,
    e^{iH^{\sigma_t}s}\frac{\alpha}{H^{\sigma_t}+i}\times\label{eq-A-21}\\
    & &\quad\quad\quad\quad\quad\quad
    \times\vec{x}_h(s)\cdot\int_{\mathcal{B}_{\sigma_s^S}\setminus\mathcal{B}_{\sigma_t}}
    d^3k \, \vec{\Sigma}_{\vv_j}(\vk)\cos(\vk\cdot\vnabla
    E_{\vec{P}}^{\sigma_t}s-|\vk|s)
    \nonumber\\
    &&\quad\quad\quad\quad\quad\quad\quad\quad\quad\quad\quad\quad\times\,
    e^{-iE_{\vec{P}}^{\sigma_t}s}
    \, e^{i\gamma_{\sigma_t}(\vv_j,\vnabla E_{\vec{P}}^{\sigma_t},s)}
    \, (E_{\vec{P}}^{\sigma_t}+i)\psi_{j,\sigma_t}^{(t)}ds\nonumber\\
    &&-\int_{t}^{\bar{s}}e^{iH^{\sigma_t}s}\cW_{\Lambdas,\sigma_t}(\vv_j,s)\frac{\alpha}{H^{\sigma_t}+i}\times\label{eq-A-22}\\
    & &\quad\quad\quad\quad\quad\quad
    \times\vec{x}_h(s)\cdot\Big\lbrace \, \frac{d}{ds} \, \Big[
    \int_{\mathcal{B}_{\sigma_s^S}\setminus\mathcal{B}_{\sigma_t}}
    d^3k \, \vec{\Sigma}_{\vv_j}(\vk)\cos(\vk\cdot\vnabla
    E_{\vec{P}}^{\sigma_t}s-|\vk|s) 
    \, e^{i\gamma_{\sigma_t}(\vv_j,\vnabla E_{\vec{P}}^{\sigma_t},s)}
    \Big] \,  \Big\rbrace
    \nonumber\\
    &&\quad\quad\quad\quad\quad\quad\quad\quad\quad\quad\quad\quad\times\,
    e^{-iE_{\vec{P}}^{\sigma_t}s}(E_{\vec{P}}^{\sigma_t}+i)\psi_{j,\sigma_t}^{(t)}ds\nonumber.
\end{eqnarray}
Here, we notice that
\begin{equation}\label{eq-A-23}
	\vec{x}_h(s)=\vec{x}\chi_h(\frac{\vx}{s})=-is\int\vec{\nabla}\widehat{\chi_h}(\vec{q})
	e^{-i\vec{q}\cdot\frac{\vx}{s}}d^3q\,.
\end{equation}
Furthermore, the operator
\begin{equation}\label{eq-A-24}
	-i\int\vec{\nabla}\widehat{\chi_h}(\vec{q})e^{-i\vec{q}\cdot\frac{\vx}{s}}d^3q
\end{equation}
tends  to
\begin{equation}\label{eq-A-25}
	-i\int\vec{\nabla}\widehat{\chi_h}(\vec{q})e^{-i\vec{q}\cdot\vnabla
	E_{\vec{P}}^{\sigma_t}} d^3q
\end{equation}
for $s\to \infty$, if it is applied to the vectors
\begin{equation}\label{eq-A-26}
	e^{i\gamma_{\sigma_t}(\vv_j,\vnabla
  	E_{\vec{P}}^{\sigma_t},s)}e^{-iE_{\vec{P}}^{\sigma_t}s}\psi_{j,\sigma_t}^{(t)}\,,
\end{equation}
or
\begin{equation}\label{eq-A-27}
    \int_{\mathcal{B}_{\sigma_s^S}\setminus\mathcal{B}_{\sigma_t}}
    d^3k \, \vec{\Sigma}_{\vv_j}(\vk)\cos(\vk\cdot\vnabla
    E_{\vec{P}}^{\sigma_t}s-|\vk|s)e^{i\gamma_{\sigma_t}(\vv_j,\vnabla
    E_{\vec{P}}^{\sigma_t},s)}e^{-iE_{\vec{P}}^{\sigma_t}s}\psi_{j,\sigma_t}^{(t)}\,,
\end{equation}
or
\begin{equation}\label{eq-A-28}
    \Big\lbrace \frac{d}{ds}\Big[
    \int_{\mathcal{B}_{\sigma_s^S}\setminus\mathcal{B}_{\sigma_t}}
    d^3k \, \vec{\Sigma}(\vk,\vv_j)\cos(\vk\cdot\vnabla
    E_{\vec{P}}^{\sigma_t}s-|\vk|s)
    \, e^{i\gamma_{\sigma_t}(\vv_j,\vnabla E_{\vec{P}}^{\sigma_t},s)}
    \Big]
    \Big\rbrace 
    \, e^{-iE_{\vec{P}}^{\sigma_t}s}\psi_{j,\sigma_t}^{(t)}\,.
\end{equation}
The rate of convergence of the corresponding expression in (\ref{eq-A-4bis}) is bounded by (\ref{eq-A-4bisbis}).

Therefore, we can replace expressions
(\ref{eq-A-20}),(\ref{eq-A-21}), and (\ref{eq-A-22}) by
\begin{eqnarray}
    & &e^{iH^{\sigma_t}s}\cW_{\Lambdas,\sigma_t}(\vv_j,s)e^{-iH^{\sigma_t}s}\alpha\,s\vnabla
    E_{\vec{P}}^{\sigma_t}\cdot\label{eq-A-29}\\
    & &\quad\quad \cdot \int_{\mathcal{B}_{\sigma_s^S}\setminus\mathcal{B}_{\sigma_t}}
    d^3k \,
    \vec{\Sigma}_{\vv_j}(\vk) \cos(\vk\cdot\vnabla
    E_{\vec{P}}^{\sigma_t}s-|\vk|s)e^{i\gamma_{\sigma_t}(\vv_j,\vnabla
    E_{\vec{P}}^{\sigma_t},s)}e^{-iE_{\vec{P}}^{\sigma_t}s}\psi_{j,\sigma_t}^{(t)}\Big|_{t}^{\bar{s}}\nonumber\\
    & &-\int_{t}^{\bar{s}}ds \, \Big\lbrace \, \frac{d}{ds}
    ( \, e^{iH^{\sigma_t}s}\cW_{\Lambdas,\sigma_t}(\vv_j,s)e^{-iH^{\sigma_t}s} \, ) \Big\rbrace
    \alpha\,s\vnabla
    E_{\vec{P}}^{\sigma_t}\cdot\label{eq-A-30}\\
    & &\quad\quad \cdot \int_{\mathcal{B}_{\sigma_s^S}\setminus\mathcal{B}_{\sigma_t}}
    \vec{\Sigma}_{\vv_j}(\vk) \cos(\vk\cdot\vnabla
    E_{\vec{P}}^{\sigma_t}s-|\vk|s)d^3ke^{i\gamma_{\sigma_t}(\vv_j,\vnabla
    E_{\vec{P}}^{\sigma_t},s)}e^{-iE_{\vec{P}}^{\sigma_t}s}\psi_{j,\sigma_t}^{(t)}\nonumber\\
    & &-\int_{t}^{\bar{s}}e^{iH^{\sigma_t}s}\cW_{\Lambdas,\sigma_t}(\vv_j,s)e^{-iH^{\sigma_t}s}\alpha\,s\vnabla
    E_{\vec{P}}^{\sigma_t}\cdot
    \label{eq-A-31}\\
    & &\quad\quad \cdot \,
    \Big\{\frac{d}{ds} \Big[ \int_{\mathcal{B}_{\sigma_s^S}\setminus\mathcal{B}_{\sigma_t}}
    \vec{\Sigma}_{\vv_j}(\vk)\cos(\vk\cdot\vnabla
    E_{\vec{P}}^{\sigma_t}s-|\vk|s)d^3k 
    \, e^{i\gamma_{\sigma_t}(\vv_j,\vnabla E_{\vec{P}}^{\sigma_t},s)}
    \Big]\Big\} \, \times
    \nonumber\\
    &&\quad\quad\quad\quad\quad\quad\quad\quad\quad\quad
    \times 
    \, e^{-iE_{\vec{P}}^{\sigma_t}s}\psi_{j,\sigma_t}^{(t)}ds \, .\nonumber
\end{eqnarray}
Recalling the definition of the phase factor, the sum of the expressions
(\ref{eq-A-29}), (\ref{eq-A-30}), and (\ref{eq-A-31}) can be written compactly as
\begin{equation}\label{eq-A-32}
    \int_{t}^{\bar{s}} ds \,   \, e^{iH^{\sigma_t}s}
    \, \cW_{\Lambdas,\sigma_t}(\vv_j,s)
    \, \frac{d\gamma_{\sigma_t}(\vv_j,\nabla E_{\vec{P}},s)}{ds}
    \, e^{i\gamma_{\sigma_t}(\vv_j,\vnabla E_{\vec{P}}^{\sigma_t},s)}
    \, e^{-iE_{\vec{P}}^{\sigma_t}s} \, \psi_{j,\sigma_t}^{(t)} \, ,
\end{equation}
after an integration by parts.

This implies the asserted bound for (\ref{eq-A-4}).
\QED
\\

\subsubsection*{Acknowledgements}

The authors gratefully acknowledge the support and hospitality of the
Erwin Schr\"odinger Institute (ESI) in Vienna in June 2006,
where this collaboration was initiated.
T.C. was supported by NSF grants DMS-0524909 and DMS-0704031.

%%%%%%%%%%%%%%%%%%%%%%%%%%%%%%%%%%%%%%%%%%%%%%%%%
%\bibliography{/home/vbach/BIB/volle}
%\end{document}
%%%%%%%%%%%%%%%%%%%%%%%%%%%%%%%%%%%%%%%%%%%%%%%%%

\parindent=0pt

\end{document}